\documentclass[longauth]{aaEC}

\usepackage{graphicx}
\usepackage{natbib}
\usepackage{scalerel}
\usepackage{multirow} 
\usepackage[table]{xcolor}
\usepackage{csquotes}
\usepackage{placeins}
\usepackage{lscape}
\usepackage{lastpage}

\usepackage{txfonts}

\usepackage[pdfencoding=auto,psdextra]{hyperref}
\hypersetup{
    colorlinks=true,
    linkcolor=blue,
    filecolor=magenta,      
    urlcolor=blue,
    citecolor=blue
}
\makeatletter
\renewcommand*\aa@pageof{, page \thepage{} of \pageref*{LastPage}}
\makeatother

\usepackage[utf8]{inputenc}

\usepackage[switch, modulo]{lineno}
\nolinenumbers
\usepackage{pdflscape}
\usepackage{euclid}
\defcitealias{Lim2015}{L15}

\begin{document}

\title{\Euclid: Early Release Observations -- The star cluster systems of the Local Group dwarf galaxies IC\,10 and NGC\,6822\thanks{This paper is published on
     behalf of the Euclid Consortium}}

\newcommand{\orcid}[1]{} 	   
\author{J.~M.~Howell\orcid{0009-0002-2242-6515}\thanks{\email{jess.howell@ed.ac.uk}}\inst{\ref{aff1}}
\and A.~M.~N.~Ferguson\inst{\ref{aff1}}
\and S.~S.~Larsen\orcid{0000-0003-0069-1203}\inst{\ref{aff2}}
\and A.~Lan\c{c}on\orcid{0000-0002-7214-8296}\inst{\ref{aff3}}
\and F.~Annibali\inst{\ref{aff4}}
\and J.-C.~Cuillandre\orcid{0000-0002-3263-8645}\inst{\ref{aff5}}
\and L.~K.~Hunt\orcid{0000-0001-9162-2371}\inst{\ref{aff6}}
\and D.~Mart{\'\i}nez-Delgado\inst{\ref{aff7}}
\and D.~Massari\orcid{0000-0001-8892-4301}\inst{\ref{aff4}}
\and T.~Saifollahi\orcid{0000-0002-9554-7660}\inst{\ref{aff3}}
\and K.~Voggel\orcid{0000-0001-6215-0950}\inst{\ref{aff3}}
\and B.~Altieri\orcid{0000-0003-3936-0284}\inst{\ref{aff8}}
\and S.~Andreon\orcid{0000-0002-2041-8784}\inst{\ref{aff9}}
\and N.~Auricchio\orcid{0000-0003-4444-8651}\inst{\ref{aff4}}
\and C.~Baccigalupi\orcid{0000-0002-8211-1630}\inst{\ref{aff10},\ref{aff11},\ref{aff12},\ref{aff13}}
\and M.~Baldi\orcid{0000-0003-4145-1943}\inst{\ref{aff14},\ref{aff4},\ref{aff15}}
\and S.~Bardelli\orcid{0000-0002-8900-0298}\inst{\ref{aff4}}
\and A.~Biviano\orcid{0000-0002-0857-0732}\inst{\ref{aff11},\ref{aff10}}
\and E.~Branchini\orcid{0000-0002-0808-6908}\inst{\ref{aff16},\ref{aff17},\ref{aff9}}
\and M.~Brescia\orcid{0000-0001-9506-5680}\inst{\ref{aff18},\ref{aff19}}
\and J.~Brinchmann\orcid{0000-0003-4359-8797}\inst{\ref{aff20},\ref{aff21},\ref{aff22}}
\and S.~Camera\orcid{0000-0003-3399-3574}\inst{\ref{aff23},\ref{aff24},\ref{aff25}}
\and G.~Ca\~nas-Herrera\orcid{0000-0003-2796-2149}\inst{\ref{aff26},\ref{aff27}}
\and G.~P.~Candini\orcid{0000-0001-9481-8206}\inst{\ref{aff28}}
\and V.~Capobianco\orcid{0000-0002-3309-7692}\inst{\ref{aff25}}
\and C.~Carbone\orcid{0000-0003-0125-3563}\inst{\ref{aff29}}
\and J.~Carretero\orcid{0000-0002-3130-0204}\inst{\ref{aff30},\ref{aff31}}
\and M.~Castellano\orcid{0000-0001-9875-8263}\inst{\ref{aff32}}
\and G.~Castignani\orcid{0000-0001-6831-0687}\inst{\ref{aff4}}
\and S.~Cavuoti\orcid{0000-0002-3787-4196}\inst{\ref{aff19},\ref{aff33}}
\and A.~Cimatti\inst{\ref{aff34}}
\and C.~Colodro-Conde\inst{\ref{aff35}}
\and G.~Congedo\orcid{0000-0003-2508-0046}\inst{\ref{aff1}}
\and C.~J.~Conselice\orcid{0000-0003-1949-7638}\inst{\ref{aff36}}
\and L.~Conversi\orcid{0000-0002-6710-8476}\inst{\ref{aff37},\ref{aff8}}
\and Y.~Copin\orcid{0000-0002-5317-7518}\inst{\ref{aff38}}
\and F.~Courbin\orcid{0000-0003-0758-6510}\inst{\ref{aff39},\ref{aff40},\ref{aff41}}
\and H.~M.~Courtois\orcid{0000-0003-0509-1776}\inst{\ref{aff42}}
\and M.~Cropper\orcid{0000-0003-4571-9468}\inst{\ref{aff28}}
\and A.~Da~Silva\orcid{0000-0002-6385-1609}\inst{\ref{aff43},\ref{aff44}}
\and H.~Degaudenzi\orcid{0000-0002-5887-6799}\inst{\ref{aff45}}
\and G.~De~Lucia\orcid{0000-0002-6220-9104}\inst{\ref{aff11}}
\and F.~Dubath\orcid{0000-0002-6533-2810}\inst{\ref{aff45}}
\and C.~A.~J.~Duncan\orcid{0009-0003-3573-0791}\inst{\ref{aff1}}
\and X.~Dupac\inst{\ref{aff8}}
\and S.~Dusini\orcid{0000-0002-1128-0664}\inst{\ref{aff46}}
\and S.~Escoffier\orcid{0000-0002-2847-7498}\inst{\ref{aff47}}
\and M.~Farina\orcid{0000-0002-3089-7846}\inst{\ref{aff48}}
\and R.~Farinelli\inst{\ref{aff4}}
\and F.~Faustini\orcid{0000-0001-6274-5145}\inst{\ref{aff32},\ref{aff49}}
\and S.~Ferriol\inst{\ref{aff38}}
\and F.~Finelli\orcid{0000-0002-6694-3269}\inst{\ref{aff4},\ref{aff50}}
\and M.~Frailis\orcid{0000-0002-7400-2135}\inst{\ref{aff11}}
\and E.~Franceschi\orcid{0000-0002-0585-6591}\inst{\ref{aff4}}
\and M.~Fumana\orcid{0000-0001-6787-5950}\inst{\ref{aff29}}
\and S.~Galeotta\orcid{0000-0002-3748-5115}\inst{\ref{aff11}}
\and K.~George\orcid{0000-0002-1734-8455}\inst{\ref{aff51}}
\and B.~Gillis\orcid{0000-0002-4478-1270}\inst{\ref{aff1}}
\and C.~Giocoli\orcid{0000-0002-9590-7961}\inst{\ref{aff4},\ref{aff15}}
\and J.~Gracia-Carpio\inst{\ref{aff52}}
\and A.~Grazian\orcid{0000-0002-5688-0663}\inst{\ref{aff53}}
\and F.~Grupp\inst{\ref{aff52},\ref{aff54}}
\and S.~V.~H.~Haugan\orcid{0000-0001-9648-7260}\inst{\ref{aff55}}
\and H.~Hoekstra\orcid{0000-0002-0641-3231}\inst{\ref{aff27}}
\and W.~Holmes\inst{\ref{aff56}}
\and F.~Hormuth\inst{\ref{aff57}}
\and A.~Hornstrup\orcid{0000-0002-3363-0936}\inst{\ref{aff58},\ref{aff59}}
\and K.~Jahnke\orcid{0000-0003-3804-2137}\inst{\ref{aff60}}
\and M.~Jhabvala\inst{\ref{aff61}}
\and E.~Keih\"anen\orcid{0000-0003-1804-7715}\inst{\ref{aff62}}
\and S.~Kermiche\orcid{0000-0002-0302-5735}\inst{\ref{aff47}}
\and B.~Kubik\orcid{0009-0006-5823-4880}\inst{\ref{aff38}}
\and M.~K\"ummel\orcid{0000-0003-2791-2117}\inst{\ref{aff54}}
\and M.~Kunz\orcid{0000-0002-3052-7394}\inst{\ref{aff63}}
\and H.~Kurki-Suonio\orcid{0000-0002-4618-3063}\inst{\ref{aff64},\ref{aff65}}
\and A.~M.~C.~Le~Brun\orcid{0000-0002-0936-4594}\inst{\ref{aff66}}
\and D.~Le~Mignant\orcid{0000-0002-5339-5515}\inst{\ref{aff67}}
\and S.~Ligori\orcid{0000-0003-4172-4606}\inst{\ref{aff25}}
\and P.~B.~Lilje\orcid{0000-0003-4324-7794}\inst{\ref{aff55}}
\and V.~Lindholm\orcid{0000-0003-2317-5471}\inst{\ref{aff64},\ref{aff65}}
\and I.~Lloro\orcid{0000-0001-5966-1434}\inst{\ref{aff68}}
\and G.~Mainetti\orcid{0000-0003-2384-2377}\inst{\ref{aff69}}
\and D.~Maino\inst{\ref{aff70},\ref{aff29},\ref{aff71}}
\and E.~Maiorano\orcid{0000-0003-2593-4355}\inst{\ref{aff4}}
\and O.~Mansutti\orcid{0000-0001-5758-4658}\inst{\ref{aff11}}
\and O.~Marggraf\orcid{0000-0001-7242-3852}\inst{\ref{aff72}}
\and M.~Martinelli\orcid{0000-0002-6943-7732}\inst{\ref{aff32},\ref{aff73}}
\and N.~Martinet\orcid{0000-0003-2786-7790}\inst{\ref{aff67}}
\and F.~Marulli\orcid{0000-0002-8850-0303}\inst{\ref{aff74},\ref{aff4},\ref{aff15}}
\and R.~J.~Massey\orcid{0000-0002-6085-3780}\inst{\ref{aff75}}
\and E.~Medinaceli\orcid{0000-0002-4040-7783}\inst{\ref{aff4}}
\and S.~Mei\orcid{0000-0002-2849-559X}\inst{\ref{aff76},\ref{aff77}}
\and M.~Melchior\inst{\ref{aff78}}
\and Y.~Mellier\inst{\ref{aff79},\ref{aff80}}
\and M.~Meneghetti\orcid{0000-0003-1225-7084}\inst{\ref{aff4},\ref{aff15}}
\and E.~Merlin\orcid{0000-0001-6870-8900}\inst{\ref{aff32}}
\and G.~Meylan\inst{\ref{aff81}}
\and A.~Mora\orcid{0000-0002-1922-8529}\inst{\ref{aff82}}
\and M.~Moresco\orcid{0000-0002-7616-7136}\inst{\ref{aff74},\ref{aff4}}
\and L.~Moscardini\orcid{0000-0002-3473-6716}\inst{\ref{aff74},\ref{aff4},\ref{aff15}}
\and R.~Nakajima\orcid{0009-0009-1213-7040}\inst{\ref{aff72}}
\and C.~Neissner\orcid{0000-0001-8524-4968}\inst{\ref{aff83},\ref{aff31}}
\and S.-M.~Niemi\orcid{0009-0005-0247-0086}\inst{\ref{aff26}}
\and C.~Padilla\orcid{0000-0001-7951-0166}\inst{\ref{aff83}}
\and S.~Paltani\orcid{0000-0002-8108-9179}\inst{\ref{aff45}}
\and F.~Pasian\orcid{0000-0002-4869-3227}\inst{\ref{aff11}}
\and K.~Pedersen\inst{\ref{aff84}}
\and W.~J.~Percival\orcid{0000-0002-0644-5727}\inst{\ref{aff85},\ref{aff86},\ref{aff87}}
\and V.~Pettorino\orcid{0000-0002-4203-9320}\inst{\ref{aff26}}
\and S.~Pires\orcid{0000-0002-0249-2104}\inst{\ref{aff5}}
\and G.~Polenta\orcid{0000-0003-4067-9196}\inst{\ref{aff49}}
\and M.~Poncet\inst{\ref{aff88}}
\and L.~A.~Popa\inst{\ref{aff89}}
\and F.~Raison\orcid{0000-0002-7819-6918}\inst{\ref{aff52}}
\and A.~Renzi\orcid{0000-0001-9856-1970}\inst{\ref{aff90},\ref{aff46}}
\and J.~Rhodes\orcid{0000-0002-4485-8549}\inst{\ref{aff56}}
\and G.~Riccio\inst{\ref{aff19}}
\and E.~Romelli\orcid{0000-0003-3069-9222}\inst{\ref{aff11}}
\and M.~Roncarelli\orcid{0000-0001-9587-7822}\inst{\ref{aff4}}
\and R.~Saglia\orcid{0000-0003-0378-7032}\inst{\ref{aff54},\ref{aff52}}
\and Z.~Sakr\orcid{0000-0002-4823-3757}\inst{\ref{aff91},\ref{aff92},\ref{aff93}}
\and D.~Sapone\orcid{0000-0001-7089-4503}\inst{\ref{aff94}}
\and B.~Sartoris\orcid{0000-0003-1337-5269}\inst{\ref{aff54},\ref{aff11}}
\and M.~Schirmer\orcid{0000-0003-2568-9994}\inst{\ref{aff60}}
\and P.~Schneider\orcid{0000-0001-8561-2679}\inst{\ref{aff72}}
\and T.~Schrabback\orcid{0000-0002-6987-7834}\inst{\ref{aff95}}
\and A.~Secroun\orcid{0000-0003-0505-3710}\inst{\ref{aff47}}
\and G.~Seidel\orcid{0000-0003-2907-353X}\inst{\ref{aff60}}
\and S.~Serrano\orcid{0000-0002-0211-2861}\inst{\ref{aff96},\ref{aff97},\ref{aff98}}
\and P.~Simon\inst{\ref{aff72}}
\and C.~Sirignano\orcid{0000-0002-0995-7146}\inst{\ref{aff90},\ref{aff46}}
\and G.~Sirri\orcid{0000-0003-2626-2853}\inst{\ref{aff15}}
\and J.~Skottfelt\orcid{0000-0003-1310-8283}\inst{\ref{aff99}}
\and L.~Stanco\orcid{0000-0002-9706-5104}\inst{\ref{aff46}}
\and J.~Steinwagner\orcid{0000-0001-7443-1047}\inst{\ref{aff52}}
\and P.~Tallada-Cresp\'{i}\orcid{0000-0002-1336-8328}\inst{\ref{aff30},\ref{aff31}}
\and A.~N.~Taylor\inst{\ref{aff1}}
\and H.~I.~Teplitz\orcid{0000-0002-7064-5424}\inst{\ref{aff100}}
\and I.~Tereno\orcid{0000-0002-4537-6218}\inst{\ref{aff43},\ref{aff101}}
\and S.~Toft\orcid{0000-0003-3631-7176}\inst{\ref{aff102},\ref{aff103}}
\and R.~Toledo-Moreo\orcid{0000-0002-2997-4859}\inst{\ref{aff104}}
\and F.~Torradeflot\orcid{0000-0003-1160-1517}\inst{\ref{aff31},\ref{aff30}}
\and I.~Tutusaus\orcid{0000-0002-3199-0399}\inst{\ref{aff92}}
\and L.~Valenziano\orcid{0000-0002-1170-0104}\inst{\ref{aff4},\ref{aff50}}
\and J.~Valiviita\orcid{0000-0001-6225-3693}\inst{\ref{aff64},\ref{aff65}}
\and T.~Vassallo\orcid{0000-0001-6512-6358}\inst{\ref{aff11},\ref{aff51}}
\and Y.~Wang\orcid{0000-0002-4749-2984}\inst{\ref{aff100}}
\and J.~Weller\orcid{0000-0002-8282-2010}\inst{\ref{aff54},\ref{aff52}}
\and G.~Zamorani\orcid{0000-0002-2318-301X}\inst{\ref{aff4}}
\and I.~A.~Zinchenko\orcid{0000-0002-2944-2449}\inst{\ref{aff105}}
\and J.~Mart\'{i}n-Fleitas\orcid{0000-0002-8594-569X}\inst{\ref{aff106}}
\and V.~Scottez\orcid{0009-0008-3864-940X}\inst{\ref{aff79},\ref{aff107}}}

\institute{Institute for Astronomy, University of Edinburgh, Royal Observatory, Blackford Hill, Edinburgh EH9 3HJ, UK\label{aff1}
\and
Department of Astrophysics/IMAPP, Radboud University, PO Box 9010, 6500 GL Nijmegen, The Netherlands\label{aff2}
\and
Universit\'e de Strasbourg, CNRS, Observatoire astronomique de Strasbourg, UMR 7550, 67000 Strasbourg, France\label{aff3}
\and
INAF-Osservatorio di Astrofisica e Scienza dello Spazio di Bologna, Via Piero Gobetti 93/3, 40129 Bologna, Italy\label{aff4}
\and
Universit\'e Paris-Saclay, Universit\'e Paris Cit\'e, CEA, CNRS, AIM, 91191, Gif-sur-Yvette, France\label{aff5}
\and
INAF-Osservatorio Astrofisico di Arcetri, Largo E. Fermi 5, 50125, Firenze, Italy\label{aff6}
\and
Centro de Estudios de F\'isica del Cosmos de Arag\'on (CEFCA), Plaza San Juan, 1, planta 2, 44001, Teruel, Spain\label{aff7}
\and
ESAC/ESA, Camino Bajo del Castillo, s/n., Urb. Villafranca del Castillo, 28692 Villanueva de la Ca\~nada, Madrid, Spain\label{aff8}
\and
INAF-Osservatorio Astronomico di Brera, Via Brera 28, 20122 Milano, Italy\label{aff9}
\and
IFPU, Institute for Fundamental Physics of the Universe, via Beirut 2, 34151 Trieste, Italy\label{aff10}
\and
INAF-Osservatorio Astronomico di Trieste, Via G. B. Tiepolo 11, 34143 Trieste, Italy\label{aff11}
\and
INFN, Sezione di Trieste, Via Valerio 2, 34127 Trieste TS, Italy\label{aff12}
\and
SISSA, International School for Advanced Studies, Via Bonomea 265, 34136 Trieste TS, Italy\label{aff13}
\and
Dipartimento di Fisica e Astronomia, Universit\`a di Bologna, Via Gobetti 93/2, 40129 Bologna, Italy\label{aff14}
\and
INFN-Sezione di Bologna, Viale Berti Pichat 6/2, 40127 Bologna, Italy\label{aff15}
\and
Dipartimento di Fisica, Universit\`a di Genova, Via Dodecaneso 33, 16146, Genova, Italy\label{aff16}
\and
INFN-Sezione di Genova, Via Dodecaneso 33, 16146, Genova, Italy\label{aff17}
\and
Department of Physics "E. Pancini", University Federico II, Via Cinthia 6, 80126, Napoli, Italy\label{aff18}
\and
INAF-Osservatorio Astronomico di Capodimonte, Via Moiariello 16, 80131 Napoli, Italy\label{aff19}
\and
Instituto de Astrof\'isica e Ci\^encias do Espa\c{c}o, Universidade do Porto, CAUP, Rua das Estrelas, PT4150-762 Porto, Portugal\label{aff20}
\and
Faculdade de Ci\^encias da Universidade do Porto, Rua do Campo de Alegre, 4150-007 Porto, Portugal\label{aff21}
\and
European Southern Observatory, Karl-Schwarzschild-Str.~2, 85748 Garching, Germany\label{aff22}
\and
Dipartimento di Fisica, Universit\`a degli Studi di Torino, Via P. Giuria 1, 10125 Torino, Italy\label{aff23}
\and
INFN-Sezione di Torino, Via P. Giuria 1, 10125 Torino, Italy\label{aff24}
\and
INAF-Osservatorio Astrofisico di Torino, Via Osservatorio 20, 10025 Pino Torinese (TO), Italy\label{aff25}
\and
European Space Agency/ESTEC, Keplerlaan 1, 2201 AZ Noordwijk, The Netherlands\label{aff26}
\and
Leiden Observatory, Leiden University, Einsteinweg 55, 2333 CC Leiden, The Netherlands\label{aff27}
\and
Mullard Space Science Laboratory, University College London, Holmbury St Mary, Dorking, Surrey RH5 6NT, UK\label{aff28}
\and
INAF-IASF Milano, Via Alfonso Corti 12, 20133 Milano, Italy\label{aff29}
\and
Centro de Investigaciones Energ\'eticas, Medioambientales y Tecnol\'ogicas (CIEMAT), Avenida Complutense 40, 28040 Madrid, Spain\label{aff30}
\and
Port d'Informaci\'{o} Cient\'{i}fica, Campus UAB, C. Albareda s/n, 08193 Bellaterra (Barcelona), Spain\label{aff31}
\and
INAF-Osservatorio Astronomico di Roma, Via Frascati 33, 00078 Monteporzio Catone, Italy\label{aff32}
\and
INFN section of Naples, Via Cinthia 6, 80126, Napoli, Italy\label{aff33}
\and
Dipartimento di Fisica e Astronomia "Augusto Righi" - Alma Mater Studiorum Universit\`a di Bologna, Viale Berti Pichat 6/2, 40127 Bologna, Italy\label{aff34}
\and
Instituto de Astrof\'{\i}sica de Canarias, V\'{\i}a L\'actea, 38205 La Laguna, Tenerife, Spain\label{aff35}
\and
Jodrell Bank Centre for Astrophysics, Department of Physics and Astronomy, University of Manchester, Oxford Road, Manchester M13 9PL, UK\label{aff36}
\and
European Space Agency/ESRIN, Largo Galileo Galilei 1, 00044 Frascati, Roma, Italy\label{aff37}
\and
Universit\'e Claude Bernard Lyon 1, CNRS/IN2P3, IP2I Lyon, UMR 5822, Villeurbanne, F-69100, France\label{aff38}
\and
Institut de Ci\`{e}ncies del Cosmos (ICCUB), Universitat de Barcelona (IEEC-UB), Mart\'{i} i Franqu\`{e}s 1, 08028 Barcelona, Spain\label{aff39}
\and
Instituci\'o Catalana de Recerca i Estudis Avan\c{c}ats (ICREA), Passeig de Llu\'{\i}s Companys 23, 08010 Barcelona, Spain\label{aff40}
\and
Institut de Ciencies de l'Espai (IEEC-CSIC), Campus UAB, Carrer de Can Magrans, s/n Cerdanyola del Vall\'es, 08193 Barcelona, Spain\label{aff41}
\and
UCB Lyon 1, CNRS/IN2P3, IUF, IP2I Lyon, 4 rue Enrico Fermi, 69622 Villeurbanne, France\label{aff42}
\and
Departamento de F\'isica, Faculdade de Ci\^encias, Universidade de Lisboa, Edif\'icio C8, Campo Grande, PT1749-016 Lisboa, Portugal\label{aff43}
\and
Instituto de Astrof\'isica e Ci\^encias do Espa\c{c}o, Faculdade de Ci\^encias, Universidade de Lisboa, Campo Grande, 1749-016 Lisboa, Portugal\label{aff44}
\and
Department of Astronomy, University of Geneva, ch. d'Ecogia 16, 1290 Versoix, Switzerland\label{aff45}
\and
INFN-Padova, Via Marzolo 8, 35131 Padova, Italy\label{aff46}
\and
Aix-Marseille Universit\'e, CNRS/IN2P3, CPPM, Marseille, France\label{aff47}
\and
INAF-Istituto di Astrofisica e Planetologia Spaziali, via del Fosso del Cavaliere, 100, 00100 Roma, Italy\label{aff48}
\and
Space Science Data Center, Italian Space Agency, via del Politecnico snc, 00133 Roma, Italy\label{aff49}
\and
INFN-Bologna, Via Irnerio 46, 40126 Bologna, Italy\label{aff50}
\and
University Observatory, LMU Faculty of Physics, Scheinerstrasse 1, 81679 Munich, Germany\label{aff51}
\and
Max Planck Institute for Extraterrestrial Physics, Giessenbachstr. 1, 85748 Garching, Germany\label{aff52}
\and
INAF-Osservatorio Astronomico di Padova, Via dell'Osservatorio 5, 35122 Padova, Italy\label{aff53}
\and
Universit\"ats-Sternwarte M\"unchen, Fakult\"at f\"ur Physik, Ludwig-Maximilians-Universit\"at M\"unchen, Scheinerstrasse 1, 81679 M\"unchen, Germany\label{aff54}
\and
Institute of Theoretical Astrophysics, University of Oslo, P.O. Box 1029 Blindern, 0315 Oslo, Norway\label{aff55}
\and
Jet Propulsion Laboratory, California Institute of Technology, 4800 Oak Grove Drive, Pasadena, CA, 91109, USA\label{aff56}
\and
Felix Hormuth Engineering, Goethestr. 17, 69181 Leimen, Germany\label{aff57}
\and
Technical University of Denmark, Elektrovej 327, 2800 Kgs. Lyngby, Denmark\label{aff58}
\and
Cosmic Dawn Center (DAWN), Denmark\label{aff59}
\and
Max-Planck-Institut f\"ur Astronomie, K\"onigstuhl 17, 69117 Heidelberg, Germany\label{aff60}
\and
NASA Goddard Space Flight Center, Greenbelt, MD 20771, USA\label{aff61}
\and
Department of Physics and Helsinki Institute of Physics, Gustaf H\"allstr\"omin katu 2, 00014 University of Helsinki, Finland\label{aff62}
\and
Universit\'e de Gen\`eve, D\'epartement de Physique Th\'eorique and Centre for Astroparticle Physics, 24 quai Ernest-Ansermet, CH-1211 Gen\`eve 4, Switzerland\label{aff63}
\and
Department of Physics, P.O. Box 64, 00014 University of Helsinki, Finland\label{aff64}
\and
Helsinki Institute of Physics, Gustaf H{\"a}llstr{\"o}min katu 2, University of Helsinki, Helsinki, Finland\label{aff65}
\and
Laboratoire d'etude de l'Univers et des phenomenes eXtremes, Observatoire de Paris, Universit\'e PSL, Sorbonne Universit\'e, CNRS, 92190 Meudon, France\label{aff66}
\and
Aix-Marseille Universit\'e, CNRS, CNES, LAM, Marseille, France\label{aff67}
\and
SKA Observatory, Jodrell Bank, Lower Withington, Macclesfield, Cheshire SK11 9FT, UK\label{aff68}
\and
Centre de Calcul de l'IN2P3/CNRS, 21 avenue Pierre de Coubertin 69627 Villeurbanne Cedex, France\label{aff69}
\and
Dipartimento di Fisica "Aldo Pontremoli", Universit\`a degli Studi di Milano, Via Celoria 16, 20133 Milano, Italy\label{aff70}
\and
INFN-Sezione di Milano, Via Celoria 16, 20133 Milano, Italy\label{aff71}
\and
Universit\"at Bonn, Argelander-Institut f\"ur Astronomie, Auf dem H\"ugel 71, 53121 Bonn, Germany\label{aff72}
\and
INFN-Sezione di Roma, Piazzale Aldo Moro, 2 - c/o Dipartimento di Fisica, Edificio G. Marconi, 00185 Roma, Italy\label{aff73}
\and
Dipartimento di Fisica e Astronomia "Augusto Righi" - Alma Mater Studiorum Universit\`a di Bologna, via Piero Gobetti 93/2, 40129 Bologna, Italy\label{aff74}
\and
Department of Physics, Institute for Computational Cosmology, Durham University, South Road, Durham, DH1 3LE, UK\label{aff75}
\and
Universit\'e Paris Cit\'e, CNRS, Astroparticule et Cosmologie, 75013 Paris, France\label{aff76}
\and
CNRS-UCB International Research Laboratory, Centre Pierre Bin\'etruy, IRL2007, CPB-IN2P3, Berkeley, USA\label{aff77}
\and
University of Applied Sciences and Arts of Northwestern Switzerland, School of Engineering, 5210 Windisch, Switzerland\label{aff78}
\and
Institut d'Astrophysique de Paris, 98bis Boulevard Arago, 75014, Paris, France\label{aff79}
\and
Institut d'Astrophysique de Paris, UMR 7095, CNRS, and Sorbonne Universit\'e, 98 bis boulevard Arago, 75014 Paris, France\label{aff80}
\and
Institute of Physics, Laboratory of Astrophysics, Ecole Polytechnique F\'ed\'erale de Lausanne (EPFL), Observatoire de Sauverny, 1290 Versoix, Switzerland\label{aff81}
\and
Telespazio UK S.L. for European Space Agency (ESA), Camino bajo del Castillo, s/n, Urbanizacion Villafranca del Castillo, Villanueva de la Ca\~nada, 28692 Madrid, Spain\label{aff82}
\and
Institut de F\'{i}sica d'Altes Energies (IFAE), The Barcelona Institute of Science and Technology, Campus UAB, 08193 Bellaterra (Barcelona), Spain\label{aff83}
\and
DARK, Niels Bohr Institute, University of Copenhagen, Jagtvej 155, 2200 Copenhagen, Denmark\label{aff84}
\and
Waterloo Centre for Astrophysics, University of Waterloo, Waterloo, Ontario N2L 3G1, Canada\label{aff85}
\and
Department of Physics and Astronomy, University of Waterloo, Waterloo, Ontario N2L 3G1, Canada\label{aff86}
\and
Perimeter Institute for Theoretical Physics, Waterloo, Ontario N2L 2Y5, Canada\label{aff87}
\and
Centre National d'Etudes Spatiales -- Centre spatial de Toulouse, 18 avenue Edouard Belin, 31401 Toulouse Cedex 9, France\label{aff88}
\and
Institute of Space Science, Str. Atomistilor, nr. 409 M\u{a}gurele, Ilfov, 077125, Romania\label{aff89}
\and
Dipartimento di Fisica e Astronomia "G. Galilei", Universit\`a di Padova, Via Marzolo 8, 35131 Padova, Italy\label{aff90}
\and
Institut f\"ur Theoretische Physik, University of Heidelberg, Philosophenweg 16, 69120 Heidelberg, Germany\label{aff91}
\and
Institut de Recherche en Astrophysique et Plan\'etologie (IRAP), Universit\'e de Toulouse, CNRS, UPS, CNES, 14 Av. Edouard Belin, 31400 Toulouse, France\label{aff92}
\and
Universit\'e St Joseph; Faculty of Sciences, Beirut, Lebanon\label{aff93}
\and
Departamento de F\'isica, FCFM, Universidad de Chile, Blanco Encalada 2008, Santiago, Chile\label{aff94}
\and
Universit\"at Innsbruck, Institut f\"ur Astro- und Teilchenphysik, Technikerstr. 25/8, 6020 Innsbruck, Austria\label{aff95}
\and
Institut d'Estudis Espacials de Catalunya (IEEC),  Edifici RDIT, Campus UPC, 08860 Castelldefels, Barcelona, Spain\label{aff96}
\and
Satlantis, University Science Park, Sede Bld 48940, Leioa-Bilbao, Spain\label{aff97}
\and
Institute of Space Sciences (ICE, CSIC), Campus UAB, Carrer de Can Magrans, s/n, 08193 Barcelona, Spain\label{aff98}
\and
Centre for Electronic Imaging, Open University, Walton Hall, Milton Keynes, MK7~6AA, UK\label{aff99}
\and
Infrared Processing and Analysis Center, California Institute of Technology, Pasadena, CA 91125, USA\label{aff100}
\and
Instituto de Astrof\'isica e Ci\^encias do Espa\c{c}o, Faculdade de Ci\^encias, Universidade de Lisboa, Tapada da Ajuda, 1349-018 Lisboa, Portugal\label{aff101}
\and
Cosmic Dawn Center (DAWN)\label{aff102}
\and
Niels Bohr Institute, University of Copenhagen, Jagtvej 128, 2200 Copenhagen, Denmark\label{aff103}
\and
Universidad Polit\'ecnica de Cartagena, Departamento de Electr\'onica y Tecnolog\'ia de Computadoras,  Plaza del Hospital 1, 30202 Cartagena, Spain\label{aff104}
\and
Astronomisches Rechen-Institut, Zentrum f\"ur Astronomie der Universit\"at Heidelberg, M\"onchhofstr. 12-14, 69120 Heidelberg, Germany\label{aff105}
\and
Aurora Technology for European Space Agency (ESA), Camino bajo del Castillo, s/n, Urbanizacion Villafranca del Castillo, Villanueva de la Ca\~nada, 28692 Madrid, Spain\label{aff106}
\and
ICL, Junia, Universit\'e Catholique de Lille, LITL, 59000 Lille, France\label{aff107}}

   \abstract{Star clusters are valuable indicators of galaxy evolution, offering insights into the buildup of stellar populations across cosmic time. Understanding the intrinsic star cluster populations of dwarf galaxies is particularly important given these systems' role in the hierarchical growth of larger systems.  We use data from \Euclid's Early Release Observation programme to study star clusters in two star-forming dwarf irregular galaxies in the Local Group, NGC\,6822 and IC\,10 [$M_{\star}\sim$ (1--4)\,$\times10^8~M_{\odot}$]. With \Euclid, star clusters are resolved into individual stars across the main bodies and haloes of both galaxies. Through visual inspection of the \IE images, we uncover 30 new star cluster candidates in NGC\,6822 and 16 in IC\,10, ranging from compact to diffuse extended clusters. We compile and re-evaluate previously identified literature candidates, resulting in final combined catalogues of 52 (NGC\,6822) and 71 (IC\,10) cluster candidates with confidence-based classifications. We present homogeneous photometry in \IE, \YE, \JE, and \HE, and in archival {\it UBVRI} data, alongside size measurements and properties derived from the spectral energy distribution fitting code \texttt{BAGPIPES}. Through synthetic cluster injection, we conclude our sample is ${\sim}50\%$ complete to $M \lesssim 10^3~{M}_{\odot}$ for ages $\lesssim100\,\rm{Myr}$, and to $M \lesssim 2\times10^4~{M}_{\odot}$ for ages of ${\sim}10\,\rm{Gyr}$. We find that IC\,10 has more young clusters than NGC\,6822, and its young clusters extend to higher masses, consistent with its starburst nature.  We find several old massive ($\gtrsim10^5\,M_{\odot}$) clusters in both dwarfs, including an exceptional cluster in NGC\,6822's outskirts with a mass of $1.3 \times 10^6\,M_{\odot}$, nearly twice as massive as any other old cluster in either galaxy.  In NGC\,6822, we also identify a previously undetected, old, and extended cluster ($R_{\mathrm{h}} = 12.4 \pm 0.11\,\mathrm{pc}$). Using well-defined criteria, we identify 11 candidate GCs in NGC\,6822 and nine in IC\,10.  Both galaxies have high specific frequencies ($S_{\rm N}$) for their luminosities but remain consistent with the known GC scaling relationships in the low-luminosity regime.}

\keywords{Galaxies: dwarf, Galaxies: star clusters: general, Galaxies: Local Group, Galaxies: individual: NGC\,6822, IC\,10}

   \titlerunning{\Euclid\/: Star cluster systems of IC~10 and NGC~6822}
   \authorrunning{J.~M.~Howell et al.}
   
   \maketitle

\section{\label{sc:Intro}Introduction}

Star clusters are present in nearly all galaxies and can be observed across most of cosmic time \citep[e.g.][]{Forbes2018a, Adamo2024}. They are likely the sites of most, if not all, star formation in the Universe, even if they do not always survive for long as coherent or bound groupings of stars.  In the Local Volume, star clusters are observed with a remarkable range of properties, from massive and young clusters \citep[e.g.][]{Portegies2010} to ancient, faint, and extended ones   \citep[e.g.][]{Huxor2005, Crnojevic2016}. The origin of this diversity is not fully understood but likely reflects the complex interplay between birth conditions, stellar feedback, and dynamical evolution \citep[e.g.][]{Krumholz2019, Lahen2025}.  
Nearby galaxies offer key advantages for studying star clusters, as their proximity allows resolution into individual stars, facilitating both identification and characterisation \citep[e.g.][]{Huxor2014, Johnson2017, Jang2012}. Furthermore, using resolved stars, star clusters can be observed to lower luminosities and surface densities than is possible in more distant systems, yielding insight into small-scale star-formation events and the bottom end of the cluster mass function. 

Comprehensive views of star cluster populations in nearby galaxies are still relatively rare. Historically, many extragalactic star cluster studies have been conducted with the \emph{Hubble} Space Telescope (HST), which captures only a small field of view (FoV) in a single pointing. Studies often choose to focus on the bright inner regions of galaxies where recent star formation is concentrated and hence the number of young clusters is highest \citep[e.g.][]{Adamo2017}.  Even large HST imaging campaigns, such as the PHAT and PHATTER surveys of M31 and M33, capture star clusters only in the main disc \citep{Johnson2015,Johnson2022}. This biased view may have important consequences for our overall understanding of cluster populations, since old-aged objects, including ancient globular clusters (GCs), are often more widely distributed in and around galaxies than young and intermediate-age ones \citep[e.g.][]{Rhode2004, Huxor2014, Veljanoski2015}. 

While young to intermediate-age star clusters are of great interest for understanding star-formation processes and calibrating stellar evolution and population models \citep[e.g.][]{Grasha2019,Renzini1988,Dotter2007}, old clusters provide probes of galaxy assembly histories \citep[e.g.][]{Brodie2006}.  Evidence suggests many GCs are accreted through mergers, and their identification and analysis offer insights into the timeline and nature of hierarchical galaxy growth \citep[e.g.][]{Cote1998, Mackey2010}. In the Milky Way (MW), searches for accreted GCs focus on cluster kinematics, chemistry and ages \citep[e.g.][]{Massari2019,Monty2024}. In contrast, outside the MW, accreted GCs can often be recognised more directly through their association to tidal streams \citep[e.g.][]{Foster2014,Mackey2019}. The GC populations of dwarf galaxies ($M_{\star} \lesssim 10^9~M_{\odot}$) are of particular interest not only for understanding how they might contribute to the build-up of more massive galaxy GC systems via accretion events, but also for understanding whether these small systems have acquired GCs from merger and accretion events themselves \citep{Deason2014}. 

In this paper, we use state-of-the-art data from the \emph{Euclid} satellite \citep{EuclidSkyOverview} to conduct an in-depth study of the star cluster populations in two Local Group (LG) dwarf irregular (dIrr) galaxies, NGC\,6822 and IC\,10. Aside from the Magellanic Clouds, these are the closest gas-rich dwarf galaxies to us and, with stellar masses of ${\sim}$(1--4)\,$\times 10^8~ M_{\odot}$ \citep{Pace2024}, they serve as valuable analogues to the star cluster-forming low-mass galaxies that are now being identified at high redshift \citep[e.g.][]{Mowla2024}.  
\emph{Euclid}'s spectacular capabilities allow us to conduct a comprehensive and systematic survey of star clusters in these systems that is unprecedented in its combination of depth, areal coverage, and spatial resolution.  In particular, the pixel scale of \emph{Euclid} corresponds to sub-parsec scales at the distances of these galaxies, while the FoV captures $>$6~kpc on a side. Additionally, \emph{Euclid} extends star cluster studies in these systems to near-infrared (NIR) wavelengths. 
We summarise the properties of the two galaxies in Table~\ref{table:gal_properties} and briefly review their properties below. 

\begin{figure*}
\centering
\includegraphics[width=\columnwidth*2]{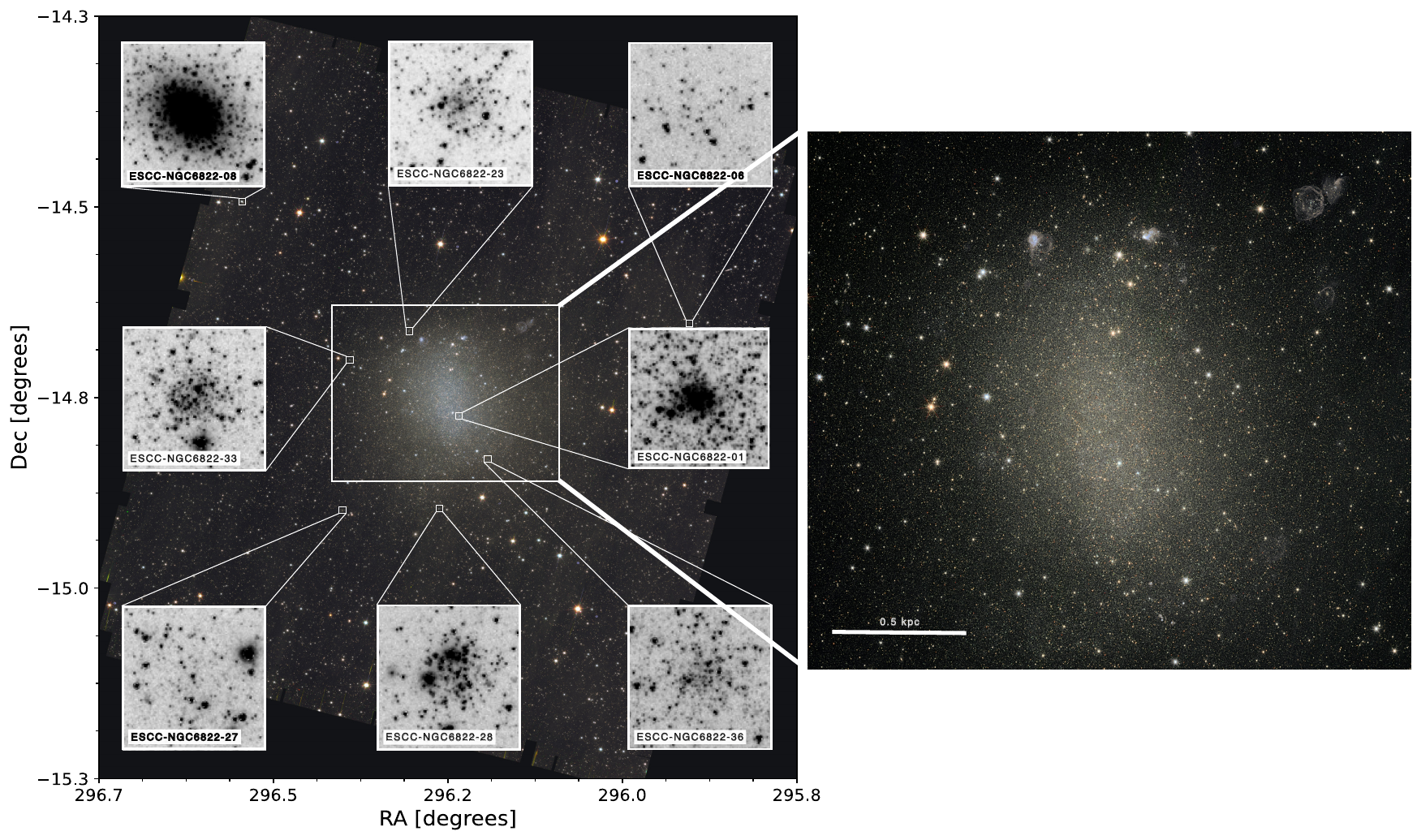}
\caption{Examples of old and young clusters across NGC\,6822, both previously found and new from this study, showcased alongside an RGB image of the \Euclid data displaying its full FoV. \IE, \YE, and \HE are the blue, green, and red channels, respectively. Each cutout is ${\sim}44$ pc on the side in the \IE band, displayed in log scale. The zoom-in region is roughly $16\farcm5 \times 13\farcm5$ ($2.5\,\text{kpc} \times 2 \,\text{kpc}$). The top row includes the GCs SC7 (ESCC-NGC6822-08) and SC5 (ESCC-NGC6822-06), the middle row shows the cluster Hubble-VI (ESCC-NGC6822-01), and the bottom row features a newly identified extended cluster candidate, ESCC-NGC6822-27.}
\label{fig:example_clusts}
\end{figure*}
\begin{figure*}
    \centering
    \includegraphics[width=\columnwidth*2]{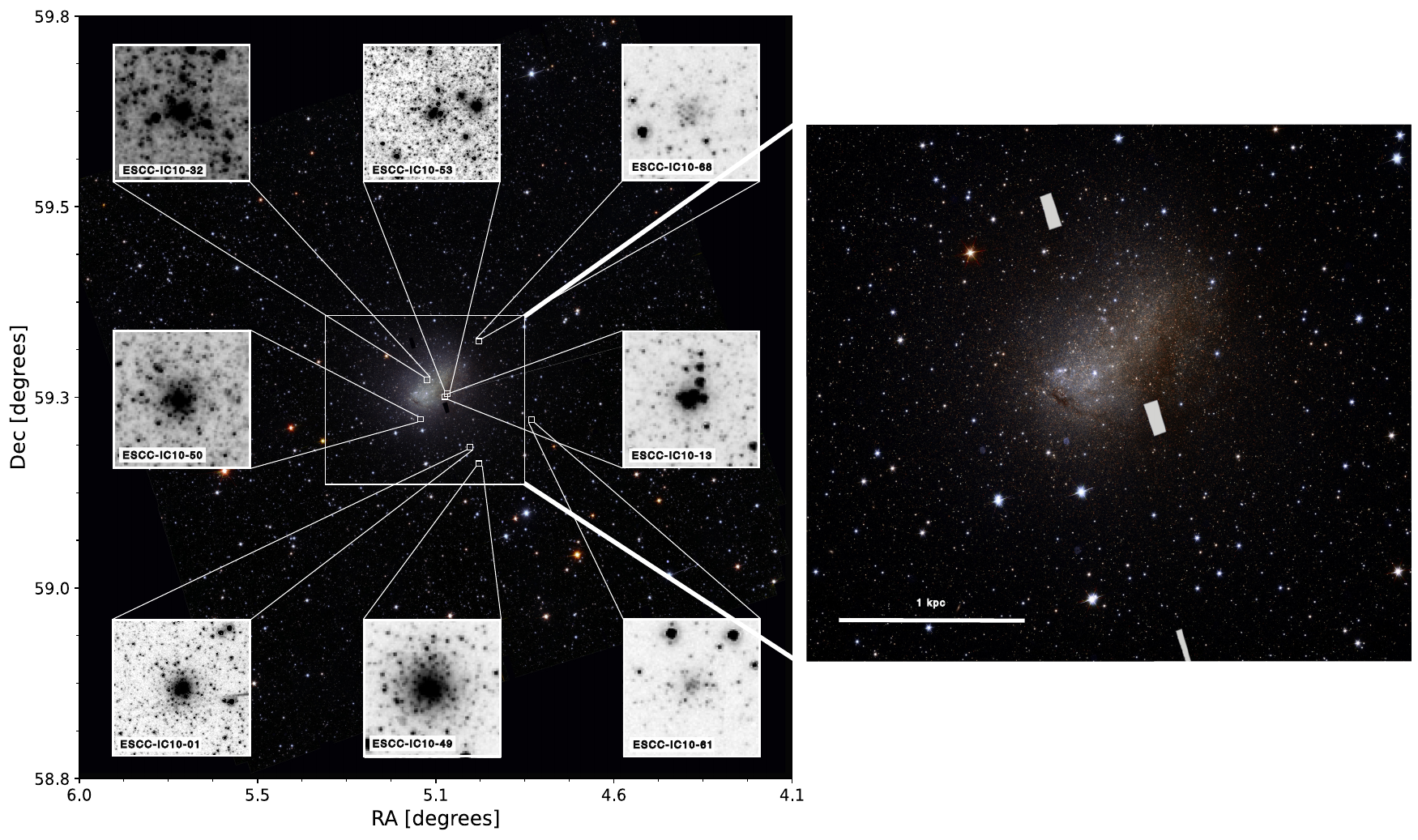}
    \caption{Same as Fig.~\ref{fig:example_clusts} for IC\,10. Each cutout is 63 pc on the side in the \IE band, displayed in asinh scale. The zoom-in region is roughly $15\farcm5 \times 12\farcm9$ ($3.25\,\text{kpc} \times 2.7 \,\text{kpc}$). The sample includes some clusters identified in the halo by \citet{Lim2015}, specifically ESCC-IC10-49 and ESCC-IC10-50.}
    \label{fig:example_clusts_IC10}
\end{figure*}
\section{Overview of the two galaxies}
\subsection{NGC 6822}
\label{ngc6822_info}
First identified by \citet{Barnard1884}, NGC\,6822 is the third nearest dIrr galaxy after the Large and Small Magellanic Clouds and has a distance of 510\,kpc determined from the tip of the red giant branch (RGB, \citealt{Fusco2012}). It has many luminous \ion{H}{ii} regions \citep[e.g.][]{Hodge1988, Efremova2011} and studies suggest it has experienced an increase in its global star-formation rate (SFR) over the last few hundred Myr \citep{Hodge1980, Gallart1996,Fusco2014,Khatamsaz2024}. NGC\,6822's young stellar component is predominantly concentrated within a central bar structure, aligned along the north–south direction \citep{Hodge1977}. In contrast, the older and intermediate-age populations exhibit elliptical distributions and extend to larger radii \citep[e.g.][]{Letarte2002,Battinelli2006, Tantalo2022}. In addition, NGC\,6822 contains a significant quantity of \ion{H}{i} gas which lies in a disc that is misaligned to both the inner bar structure and the extended spheroid \citep{deBlok2000, Weldrake2003}. Notably, this \ion{H}{i} disc rotates perpendicular to the older stellar component and hence constitutes a dynamically decoupled structure \citep{Demers2006}. In spite of the complex stellar structure of NGC\,6822, there is no compelling evidence for a recent merger or interaction \citep{Zhang2021}. In fact,  most studies suggest that the galaxy is not even bound to the MW at the present epoch, although it may have entered within its virial radius in the past \citep[e.g.][]{Battaglia2022, Bennet2024}.

Previous estimates of reddening towards and internal to NGC\,6822 have been fairly consistent. \citet{Massey1995} determined a colour excess of $E(B-V) = 0.26$ near the edges of the galaxy through spectroscopic samples of hot, luminous stars. Similarly, \citet{Gallart1996} found $E(B-V) = 0.24 \pm 0.03$, using a multiwavelength fit to the {\it B, V, R,} and {\it I} magnitudes of Cepheids, and through observations obtained with HST, \citet{Fusco2012} estimated a value of $0.30 \pm 0.032$. These values are consistent with the dust maps of \citet{schlegel1998}, which return foreground values between $E(B-V) = 0.18$ and $0.26$. Some studies suggest the reddening varies between the galaxy's centre and edges. \citet{Massey1995} estimated central reddening at $E(B-V) = 0.45$, while others reported slightly lower values of 0.35--0.37 \citep{Rich2014, Fusco2012, Gieren2006}. However, \citet{Massey2007} noted that the \citet{Massey1995} sample likely had above-average reddening, potentially inflating the estimate.

The metallicity of NGC\,6822's young population is roughly similar to that of the Small Magellanic Cloud (hereafter SMC; $Z \approx 0.2$--0.3\,$\rm Z_{\odot}$). \citet{Venn2001} found a mean iron abundance of $\mathrm{[Fe/H]} = -0.49 \pm 0.22$ using two A-type supergiant spectra. Similarly, \citet{Lee2006} obtained an oxygen abundance of $\mathrm{[O/H]} = -0.55$ from \ion{H}{ii} region spectra. More recently, \citet{Patrick2015} found a mean metallicity of $\mathrm{[Fe/H]} = -0.52 \pm 0.21$ by using NIR spectra of 11 red supergiants. The metallicity of NGC\,6822's older population was found to be $\mathrm{[Fe/
H]} = -1.29 \pm 0.07$ using asymptotic giant branch stars \citep{Sibbons2012,HirschauerA2020} and $\mathrm{[Fe/H]} = -1.05 \pm 0.49$ from the spectra of RGB stars  \citep{Kirby2013}. In contrast, \citet{Swan2016} reported a higher mean metallicity of $\mathrm{[Fe/H]} = -0.84 \pm 0.04$ from \ion{Ca}{ii} spectroscopy of RGB stars, with dispersion $\sigma = 0.31$\,dex.

NGC\,6822's star cluster system has been well-studied, dating back to a century ago, although studies have been carried out in a somewhat piecemeal fashion. Using photographic plates, \citet{Hubble1925} identified five potential clusters, and \citet{Hodge1977} later confirmed four (Hubble-IV, VI, VII, and VIII) of these as genuine clusters, and identified 26 additional potential clusters in the main body. Follow-up studies of the \citet{Hubble1925} clusters using HST and ground-based spectroscopy \citep{wyder2000, Chandar2000, Larsen2022, Hwang2014} confirmed Hubble-VII as a GC, Hubble-VI and VIII as likely young or intermediate-age clusters, and Hubble-IV to be associated with an \ion{H}{ii} region. \citet{Krienke2004} further identified nine new cluster candidates in the central regions using HST imaging. In the remote halo, \citet{Hwang2011} identified four GCs using Canada--France--Hawaii telescope (CFHT) MegaCam data ($3^{\circ} \times \,3^{\circ}$), and \citet{Huxor2013} later discovered three more GCs from an independent analysis of the same data, plus additional coverage. \citet{Veljanoski2015} presented a homogeneous photometric and spectroscopic analysis of NGC\,6288's GC system and used it to infer a dynamical mass of (3--4)\,$\times10^9 M_{\odot}$  within 11~kpc, implying the system is very dark-matter dominated. Particularly striking is the linear alignment of the outermost GCs in NGC\,6822, which extends over ${\sim}17$\,kpc at the distance of the galaxy. 
\begin{table}
\centering
\small
\caption{Galaxy properties.}
\setlength{\tabcolsep}{3.25pt}
\renewcommand{\arraystretch}{1.4}  
\label{table:gal_properties}
\begin{tabular}{lcccccc} 
\hline\hline
 Galaxy & $D^a$ & $E(B-V)^b$ & $M_{\rm V} ^c$ & $b^d$ & $R_{\mathrm{h}}^e$ & SFR$^f$\\
    & kpc & \dots & \dots & deg & arcmin & $M_\odot \,\rm {yr}^{-1}$ \\
\hline
IC\,10 & $720_{-70}^{+90}$ & ${\sim}$0.7--1.15 & $-15.1$ & $-3.3$ & 2.7 & 0.3 \\
NGC\,6822 & $510 \pm 12$ & 0.24--0.45 & $-15.2$ & $-18.4$ & 12.0 & 0.02 \\
\hline
\end{tabular}
\raggedright
\\
$^a$ Distances from \citet{Gerbrandt2015} and \citet{Fusco2012}.\\
$^b$ Total line-of-sight reddening measurements, as detailed in text.\\
$^c$ Absolute magnitudes taken from \citet{McConnachie2012} and \citet{Sanna2010}.\\
$^d$ Galactic latitudes from NASA/IPAC Extragalactic
Database (NED).\\
$^e$ Half-light radii \citep{McConnachie2012,Higgs2021}.\\
$^f$ SFRs determined from SED fitting through CIGALE, and adjusted to our adopted distances \citep{Nersesian2019, Hunt2024}.
\end{table}
\subsection{IC 10}
\label{ic10_info}
Housing a substantial population of Wolf--Rayet stars, IC\,10 is presently undergoing a strong burst of star formation and is arguably the closest example of a starburst galaxy \citep{Richer2001}. IC\,10's low Galactic latitude and formidable foreground extinction have hindered detailed study in the past and it has received comparatively less attention than other dwarf galaxies in the LG.   Similar to NGC\,6822, IC\,10's youngest populations are the most centrally concentrated and the older stars are distributed much more broadly \citep{Demers2004,Sanna2010, Gerbrandt2015}. It has been suggested that the geometric centre of IC\,10's younger populations is offset by a few hundred parsecs from its oldest population \citep{Gerbrandt2015}.  \citet{Nidever2013} argue that a recent interaction might have occurred based on the existence of an extended \ion{H}{i} feature, but there are no clear signs of stellar tidal features or distortions in the outer regions of the galaxy \citep{Demers2004,Sanna2010}.  IC\,10 is a remote satellite of M31 that likely underwent a moderately close pericentric passage (${\sim}130$~kpc) roughly 1--2 Gyr ago \citep{Bennet2024}. 

There is considerable uncertainty in the total reddening towards IC\,10. \citet{Massey2007} measured the total reddening value to be $E(B-V) = 0.81$ using the location of the plume of blue supergiants from {\it UBVRI} wide-field imaging ($37\arcminute \times 37\arcminute$). This aligns with the values of $0.78 \pm 0.06$ and $0.77 \pm 0.07$ reported by \citet{Sanna2008} and \citet{Richer2001}, based on deep HST imaging of field stars and optical spectroscopy, respectively. It also agrees well with the average reddening value of $0.79$ from \citet{Demers2004}, measured using Carbon stars over a $42\arcminute \times 28\arcminute$ field. However, their values vary across the field by as much as 0.30 magnitudes. A few higher values have also been reported, such as \citet{Kim2009}, who found the total reddening to be $E(B-V) = 0.98 \pm 0.06$ through deep \emph{JHK} photometry of the central $4\arcminute \times 7\arcminute$ field and more recently, using NIR and mid-IR data, \citet{DellAgli2018} found a value of $E(B-V) = 1.14$, also restricted to a FoV that excludes the halo. The low Galactic latitude of IC\,10 places it in the regime in which the foreground reddening maps of \citet{schlegel1998} are known to be inaccurate. With the \citet{schlafly2011} recalibration, the dust maps return foreground values between 0.77 and 1.58, considerably larger than some of the total line-of-sight estimates in the literature. 

Similarly to NGC\,6822, IC\,10's metallicity has been estimated to be  $Z \approx 0.2-0.3\,\rm Z_{\odot}$, based on its \ion{H}{ii} regions \citep{Skillman1989,Garnett1990,Lee2003}. IC\,10's older population has received less attention but using RGB stars, \citet{Tikhonov2009} measured a photometric metallicity of $\mathrm{[Fe/H]} = -1.28$.

\citet{Karachentsev1993} first identified seven star cluster candidates in IC\,10 from ground-based {\it B}- and 
{\it V}-band images. \citet{Hunter2001} later found additional candidates using HST imaging, focused on the central and eastern parts of the galaxy. \citet{Tikhonov2009} conducted a more extensive search across the central body and parts of the halo using a combination of HST data and ground-based observations from the 6m BTA and 1m Zeiss telescopes, compiling a list of 57 candidates, subsuming those from \citet{Karachentsev1993} and some from \citet{Hunter2001}. \citet{Sharina2010} identified 17 new objects in archival HST data, while \citet{Lim2015}, hereafter \citetalias{Lim2015}, found five additional cluster candidates in the remote halo using Subaru {\it R}-band imaging with 0\farcs7 seeing. As with NGC\,6822, the previous cluster work has involved an amalgamation of patchy high-resolution HST coverage, wide-field low-resolution coverage, and varying search sensitivities across the galaxy. 

\section{Data}
We combined the Euclid Early Release Observations (ERO, \citealt{Cuillandre2024})\footnote{\cite{EROcite}} with the NOAO Local Group Galaxy Survey (LGGS, \citealt{Massey2007}) to provide broad wavelength coverage from $U$ to \HE, which we used for both star cluster identification and detailed characterisation of their properties.

\subsection{Euclid Early Release Observations} 
As described in \citet{Hunt2024}, \emph{Euclid} observed NGC\,6822 and IC\,10 with one and two Reference Observation Sequences respectively (ROS, see \citealt{Scaramella-EP1}) using the Visible and Near Infrared Spectrometer (VIS and NISP) instruments \citep{EuclidSkyVIS,EuclidSkyNISP}. One ROS consists of four dithered images per band (4 repetitions of 566\,s for \IE and 87.2\,s for \YE, \JE, and \HE). The \emph{Euclid} images cover a ${\sim}0.67\,\rm{deg}^{2}$ area centred on each galaxy, with the \IE band providing a 0\farcs1 pixel scale, while the \YE, \JE, and \HE bands each have a 0\farcs3 pixel scale. Additionally, the point-spread function (PSF) full width at half maximum (FWHM) in the stacked images is  ${\sim}1.6$ pixels in each band \citep{Cuillandre2024}.

All the images have been processed using a bespoke ERO reduction pipeline, as detailed in \citet{Cuillandre2024}. In brief, this involved the removal of instrumental signatures, astrometric calibration, image stacking, photometric calibration onto the AB magnitude system, and the production of science-ready catalogues. Two versions of the image stacks are available: a stack that is optimised for studying diffuse extended emission, in which the background is preserved, and a stack that is optimised for detection and analysis of compact sources, in which the background has been modelled and subtracted.  After some experimentation, we opted to use the extended-emission images for cluster identification and the compact-sources images for photometry and size measurements.  Figures~\ref{fig:example_clusts} and \ref{fig:example_clusts_IC10} show full and zoom-in RGB images of NGC\,6822 and IC\,10, respectively, together with cutouts of some known and new clusters (see Sect. \ref{cluster_ids}).

\subsection{Local Group Galaxy Survey}
\label{lggs}

The LGGS imaged NGC\,6822 and IC\,10 using the CTIO Blanco and Kitt Peak Mayall 4m telescopes, respectively, in each of the {\it UBVRI} and $\rm H\alpha$ filters  \citep{Massey2007}. The two cameras are nearly identical, consisting of a $2 \times 4$ array of $2048 \times 4096$ SITe CCDs, yielding images that cover $37\arcminute \times 37\arcminute$ per pointing with a pixel scale of $0\farcs27$.  The seeing was measured to be $\approx$\,0\farcs9 in each case. A single pointing was obtained for each galaxy, with NGC\,6822 centred in the FoV but IC\,10 offset somewhat to avoid a ghost image from the prime focus corrector. As a result, while the \emph{Euclid} and LGGS images cover much of the same area in NGC\,6822, the \emph{Euclid} IC\,10 images extend slightly more toward the north-east and LGGS images extend more to the south-west.  Nonetheless, the combined fields of view cover all previously identified clusters in IC\,10 and all except for four in NGC\,6822, three of which lie outside both the LGGS and {\it Euclid} footprints, and one outside only the LGGS footprint (see Sect. \ref{litsearch}). From Table 14 of \citet{Massey2007}, the median magnitude uncertainties in the LGGS survey reach $\sigma_V \approx 0.20$, corresponding to a $5\sigma$ point-source detection, at $V \approx 25.0$ in IC\,10 and $V \approx 24.5$ in NGC\,6822. In comparison, \citet{Cuillandre2024} report a $5\sigma$ point-source detection limit in \IE of  ${\sim}27.0$. While \Euclid reaches fainter sources, the LGGS photometry is sufficiently deep and precise for the clusters in our sample, enabling a reliable combination of the two datasets.

To ensure that the LGGS images had a consistent astrometric solution to those from \Euclid, we ran them through  \texttt{Astrometry.net} \citep{Lang2010}. With the improved astrometric calibration, the positions of bright stars showed a median offset of 0\farcs006 in RA and 0\farcs05 in Dec between the \YE and {\it V} images for IC\,10, and 0\farcs006 in RA and 0\farcs08 in Dec for NGC\,6822. 

\section{Cluster identification}
\label{cluster_ids}
\begin{table*}
\centering
\caption{NGC\,6822 cluster candidate list.}
\label{table:ngc6822_search_results}
\begin{tabular}{c c c c c c c c}
\hline\hline
  \multicolumn{1}{c}{Euclid ID} &
  \multicolumn{1}{c}{RA (deg)}&
  \multicolumn{1}{c}{Dec (deg)} &
  \multicolumn{1}{c}{Class} &
  \multicolumn{1}{c}{Recovered?} &
  \multicolumn{1}{c}{Previous ID} &
  \multicolumn{1}{c}{Ref} \\
\hline
  ESCC-NGC6822-01 & 296.223235 & $-14.819755$ & 1 & Yes & Hubble-VI & H25 \\
  ESCC-NGC6822-02 & 296.231932 & $-14.815552$ & 1 & Yes & Hubble-VII & H25 \\
  ESCC-NGC6822-03 & 296.242518 & $-14.720538$ & 1 & Yes & Hubble-VIII & H25 \\
\hline\end{tabular}
\tablefoot{Full list of NGC\,6822 clusters identified in our blind search (classifications 1--5), alongside previously reported candidates (classifications 1--6), both those we recover and those we do not. Previous IDs take the form: Hubble-n (\citealt{Hubble1925}, H25), SCn \citep{Hwang2011,Huxor2013}, Cn \citep{Hodge1977} and KHn \citep{Krienke2004}. Previously reported candidates are listed first, and new clusters are labelled from ESCC-NGC6822-21 onwards. We provide only the first three rows here; the complete table is provided in electronic form at the CDS.}
\end{table*}

\subsection{Blind cluster search}
\label{blind}
At the distances of IC\,10 and NGC\,6822, luminous young and evolved stars are resolved in \Euclid images \citep[Annibali et al., in prep]{Hunt2024}. Compact star clusters appear as central diffuse light concentrations with resolved stellar outskirts, while faint and/or extended star clusters are fully resolved, albeit with irregular shapes. While either type can be readily detected in the sparse outer regions of galaxies,  they can be very hard to distinguish against the field populations in the more crowded regions. Figures~\ref{fig:example_clusts} and \ref{fig:example_clusts_IC10} show examples of the diverse morphologies and brightnesses of the star cluster candidates as seen by \Euclid. Their complex appearance, ranging from partially to fully resolved, and projecting on top of widely varying stellar fields, has previously been recognised as a significant challenge for automated or semi-automated methods of star cluster detection in LG galaxies \citep[e.g.][]{Huxor2008, Huxor2014, Johnson2015}.  As advocated by previous work, we decided that visually inspecting the full survey area was the most efficient and least biased star cluster detection method.

The initial search was conducted by one of us (JH) on the \IE images as these offer the highest spatial resolution compared to the NIR filters, and a greater sensitivity to various cluster ages due to the broad filter bandwidth. The extended-emission images were preferred as they help to highlight instances where there is significant coincident nebular emission. We excluded cases where an object is so heavily embedded in nebular emission that it becomes difficult to discern whether it is a distinct star cluster or a sparser OB association. We also note that very red clusters (heavily dust reddened or embedded) may be more clearly distinguished in the NIR and therefore underrepresented by our search. The search was conducted blindly, without prior knowledge of existing cluster locations, though we inspected cutouts of a few already confirmed clusters before starting, to familiarise ourselves with the likely range of morphologies. This was helpful for learning how to identify faint clusters and for calibrating our expectations for the appearance of star clusters at these distances, and with \emph{Euclid} resolution. Once the cluster candidates were identified in \IE, they were inspected across the other eight bands: {\it U, B, V, R,} and {\it I}, as well as \YE, \JE, \HE. This aided in distinguishing between clusters and background galaxies, as the galaxies tend to drop out in the shorter-wavelength passbands.  Objects exhibiting this behaviour, or showing other indications of a not being a star cluster, were immediately discarded.   Remaining candidates were labelled as either {\lq{likely}\rq} or {\lq{maybe}\rq} and were passed for further classification by two additional team members (AF and SL). By the end of this process, each object initially classed as maybe or likely also had two further (independent) classification attempts into the categories of likely, maybe, and {\lq{unlikely}\rq}. The final classification of a given star cluster candidate was arrived at as follows: 

\begin{itemize}
    \item Three likely classifications == 1;
    \item Two likely and one maybe == 2;
    \item One likely and two maybe == 3;
    \item Three maybe == 4; 
    \item Any combination with unlikely == 5.
\end{itemize}
 
Tables~\ref{table:ngc6822_search_results} and~\ref{table:ic10_search_results} contain the results of our blind star cluster search, detailing the positions and classifications of all cluster candidates with final classifications 1--5, and indicating whether or not the candidate was previously known (see Sect. \ref{litsearch}). Candidates are labelled Euclid Star Cluster Candidate (ESCC) - galaxy - \emph{n}.

\begin{table*}
\centering
\caption{IC\,10 cluster candidate list.}
\label{table:ic10_search_results}
\begin{tabular}{c c c c c c c}
\hline\hline
  \multicolumn{1}{c}{Euclid ID} &
  \multicolumn{1}{c}{RA (deg)} &
  \multicolumn{1}{c}{Dec (deg)} &
  \multicolumn{1}{c}{Class} &
  \multicolumn{1}{c}{Recovered?} &
  \multicolumn{1}{c}{Previous ID} &
  \multicolumn{1}{c}{Ref} \\
\hline
  ESCC-IC10-01 & 4.974880 & 59.224041 & 1 &  Yes & TG1 & TG09 \\
  ESCC-IC10-02 & 4.990859 & 59.331172 & 3 &  Yes & TG2 & TG09 \\
  \dots & 4.997314 & 59.326613 & 4 & No & TG3 & TG09 \\
\hline\end{tabular}
\tablefoot{Same as Table~\ref{table:ngc6822_search_results} but for IC\,10. Previous IDs take the form TG-n \citep{Tikhonov2009,Karachentsev1993}, Ln \citep{Lim2015}, dn or Sn \citep{Sharina2010} and Hn-n \citep{Hunter2001}.
Previously reported candidates are listed first and new clusters are from ESCC-IC10-58 onwards. Literature candidates that were not recovered in our blind search do not have a \emph{Euclid} ID; only their previously assigned IDs are listed. The full table is available at the CDS.}
\end{table*}

\subsection{Literature sample compilation}
\label{litsearch}

As mentioned before, previous studies have searched for star clusters in NGC\,6822 and IC\,10 and we sought to identify which of the candidates uncovered in our \Euclid blind search were already known. To this end, we undertook a thorough literature source compilation.  The process was mostly straightforward for IC\,10, due to the list of 66 cluster candidates already compiled from various sources \citep{Karachentsev1993,Hunter2001,Sharina2010} by \citetalias{Lim2015}. However, we discovered that 21 previously identified cluster candidates were missing from the list by \citetalias{Lim2015} (four from \citealt{Hunter2001} and 17 from \citealt{Sharina2010}), which we included in our compilation.

In the case of NGC\,6822, we did not have the benefit of a recent compilation and hence had to tabulate directly the cluster candidates presented in multiple works spanning several decades \citep{Hubble1925,Hodge1977,Krienke2004,Hwang2011,Huxor2013}.\footnote{We note that \citet{Gouliermis2010} studied hierarchical clustering of blue stars in NGC\,6822 using ground-based imaging. Using a nearest-neighbour density method, they defined clusters as overdensities of point sources that are $3\sigma$ above the average stellar density. However, apart from C24 from \citet{Hodge1977}, these clusters are not visually apparent in the \emph{Euclid} image. Their clusters span a large range of sizes (10--120~pc), with most of them having half-light radii $>30$~pc, which is considerably larger than the objects we identify as star cluster candidates in this paper. Indeed, combined with their stellar masses of $10^3$--$10^4 \, M_{\odot}$, these objects are more akin to low-mass unbound stellar associations. We therefore exclude the \citet{Gouliermis2010} candidates from our final lists and do not aim for completeness in very low-mass young clusters.} In older studies, cluster candidates generally lacked published coordinates \citep{Hodge1977} or had coordinates that did not always correspond to a visible overdensity in the \emph{Euclid} images \citep{Krienke2004}. In such cases, positions were determined on a best-effort basis, either by updating the coordinates to match the nearest cluster candidate to the provided coordinates or, for \citet{Hodge1977}, the closest cluster candidate to the position highlighted in their finding chart. 

In total, we compiled a list of 87 previously identified cluster candidates within IC\,10 and 43 within NGC\,6822 (noting an additional three NGC\,6822 clusters outside of the \emph{Euclid} footprint).  Cross-matching the list of candidates from our blind search with those from the literature, we found that we recovered many but not all of the sources identified in earlier studies, as well as adding many new sources not previously known.  For completeness, all three inspectors examined the positions of the literature candidates and classified them using the same system described above.   Reassuringly, all the literature candidates within the IC\,10 and NGC\,6822 footprints that we classified as classes 1 and 2 already appeared in our blind search list. On the other hand, several literature candidates were missed. In some cases, the reported positions did not correspond to any obvious stellar overdensity in the \Euclid \IE images. In other cases, the
reported positions clearly corresponded to single stars, nebulae, or background galaxies; such objects were assigned to class 6, indicating sources that are definitely not star clusters. In NGC\,6822, our blind search missed 23 out of 43 literature candidates. Of these, one was classified as class 3 and one as class 4, while 14 were assigned to class 5 and four to class 6. As a result, the majority were deemed unconvincing as cluster candidates. In IC\,10, we missed 30 out of 87 previous candidates. Of these, one was classified as class 3, six as class 4, 18 as class 5, and five as class 6. Notably, of the five candidate clusters in the remote halo of IC\,10 found by \citetalias{Lim2015} using Subaru {\it R}-band imaging, three were recovered in our blind search and are confirmed as genuine clusters (ESCC-IC10-49, 51 and 52; 49 and 51 are displayed in Fig.~\ref{fig:example_clusts_IC10}), while the remaining two (L64 and L65) are clearly background galaxies and have been assigned to class 6. While it is unlikely that many class 5 sources are genuine clusters, we note that objects previously identified as faint clusters at UV wavelengths may appear less convincing in the \IE filter due to its longer wavelength. Similarly, it is possible that some faint sources previously identified at HST resolution may be less obvious at the slightly lower resolution of \Euclid. However, since class 5 sources do not appear cluster-like in the \Euclid data, we do not analyse them further in this work. For completeness and to facilitate future studies of these cluster systems, we include all literature candidates, whether recovered or not, and regardless of our classification, in Tables~\ref{table:ngc6822_search_results} and~\ref{table:ic10_search_results}, retaining their original identifiers (based on the earliest published ID). These objects can be identified as having an entry in the column {\lq{Recovered?}\rq}. Thumbnails of the previously-identified candidates that we reclassify as class 5 and 6 are included in the Appendix (Figs.~\ref{fig:6822_compilation} and \ref{fig:ic10_compilation}), alongside all other previously identified and newly discovered clusters. 

The outcome of our combined \emph{Euclid} blind and literature searches can be summarised as follows:

\begin{itemize}
    \item Class 1: 30 in NGC\,6822 (15 previously reported), 29 in IC\,10 (23 previously reported).
    \item Class 2: 10 in NGC\,6822 (five previously reported), 24 in IC\,10 (23 previously reported).
    \item Class 3: Seven in NGC\,6822 (one previously reported), seven in IC\,10 (three previously reported).
    \item Class 4: Five in NGC\,6822 (one previously reported), 11 in IC\,10 (six previously reported)
    \item Class 5: 22 in NGC\,6822 (17 previously reported), 37 in IC\,10 (27 previously reported).
    \item Class 6: four in NGC\,6822 (all previously reported), five in IC\,10 (all previously reported).
    \end{itemize}
    
In summary, our blind search identified 30 new star cluster candidates of classes 1--4 in NGC\,6822 and 16 in IC\,10. Incorporating the 22 and 55 previously reported literature candidates with classifications 1--4 (of which 20 and 48 were independently recovered in our blind search), the total number of candidates in these classes increases to 52 in NGC\,6822 and 71 in IC\,10. For clarity, we refer to our final list of class 1--4 cluster candidates simply as {\lq{clusters}\rq} throughout the remainder of the paper.

\section{Analysis}
\subsection{Cluster photometry} 
\label{photometry}

Some preliminary steps were taken before performing photometry. Firstly, we needed to determine the photometric zeropoints for the LGGS images. These were derived through the 1,000 brightest and unsaturated stars (with photometric errors less than 0.05 in all bands) from \citet{Massey2007}, who provided measurements on the Vega system. The aperture size for measuring total instrumental magnitudes was determined using the average flattening point of the growth curves of the 20 brightest stars (approximately 5.5 pixels for IC\,10, and 7 pixels for NGC\,6822). The zeropoints were then calculated by taking the median of the sigma-clipped differences between the \citet{Massey2007} published magnitudes and our derived instrumental aperture values. The standard deviations of the zero points (ranging from 0.01 to 0.06 magnitudes) were combined with our photometric uncertainties.  It should be noted that while the LGGS photometry is on the Vega system, the \Euclid photometry is on the AB system \citep{Cuillandre2024}.

The next stage was to mask bright stars and contaminating objects in the vicinity of the clusters. As the clusters themselves are largely resolved into individual stars, and given that they lie within highly resolved regions of galaxies, deciding how much to mask was not straightforward.  In the end, we took a conservative approach and decided to mask only the most obvious contaminants.  Cluster centres were determined on the \Euclid \IE images, as these are the deepest and have the highest spatial resolution. For centrally concentrated objects, the centres were derived using a centre-of-mass algorithm within a $2\arcsec$ box, starting from a best-guess point based on visual inspection.  For more diffuse objects, or for those which were impacted by masking, we determined the centres manually. Since the astrometric calibration was consistent across all {\it Euclid} and LGGS images, the \IE centres could be adopted universally. 

We determined the magnitudes of the clusters via aperture photometry across the nine photometric bands: {\it U, B, V, R, I,} \YE, \JE, \HE, and \IE. The measurements were made using \texttt{PhotUtils}, an \texttt{Astropy} package for the detection and photometry of astronomical sources \citep{Bradley2023}. We largely followed the methodology outlined in \citetalias{Lim2015}, who also conducted photometry for a sample of IC\,10 clusters using the LGGS images.  Specifically, we adopted their aperture radii of 6 pixels (corresponding to ${\sim}1\farcs6$, or 5.6 parsecs at the distance of IC\,10) to derive the instrumental aperture magnitudes in the {\it UBVRI} images. This radius was found to be a good compromise between capturing sufficient flux and minimising the influence of contaminating objects. For background subtraction, we used annuli spanning the radial range 20--30 pixels (corresponding to ${\sim}5\farcs4$--$8\farcs1$ or 18.8--28.3 parsec) with three $\sigma$ sigma-clipping. In NGC\,6822, we adopted the same physical-sized apertures, with the 5.6\,pc aperture corresponding to ${\sim}2\farcs3$ (8 pixels), and background annuli spanning 28–42 pixels.  Apertures and background annuli corresponding to the same physical size were used for extracting photometry from the \Euclid \IE\YE\JE\HE images. Colours were derived by subtracting the small-aperture magnitudes from each other (1\farcs6 and 2\farcs3 apertures in IC\,10 and NGC\,6822, respectively).

\begin{figure}
    \centering
    \includegraphics[width=\columnwidth]{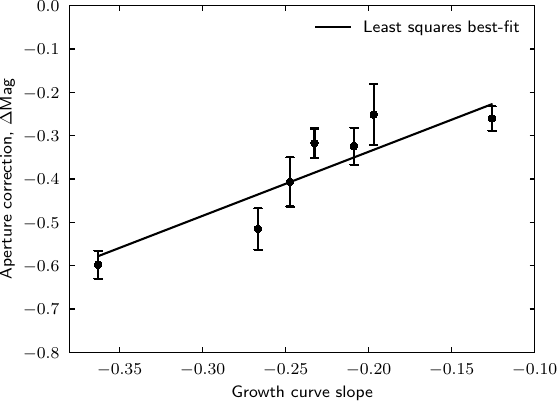}
    \caption{{\it V}-band aperture correction relationship for clusters in IC\,10, derived using bright and isolated clusters.}
    \label{fig:apcorr}
\end{figure}

To go from aperture magnitudes to total magnitudes, aperture corrections are required. For point sources, the correction can be determined in a straightforward manner from the instrumental PSF, while for extended sources, we must also consider the intrinsic radial profile.  Unfortunately, this represents a major source of uncertainty in star cluster studies \citep[e.g.][]{Chandar2010} and methods to deal with this have been widely discussed in the literature \citep[e.g.][]{Cookd2019, Adamoa2015,Deger2022}. A common approach is to derive a relationship between the slope of the enclosed flux growth curve at a given radius and the aperture correction, which can be calibrated using either isolated real clusters or injected synthetic ones.  We again followed the approach of \citetalias{Lim2015}, who sampled the slope of the growth curve at radii where it is typically at its steepest, and calibrated the relationship using a small sample of relatively bright and isolated star clusters in each galaxy (clusters ESCC-IC10-01, -03, -17, -45, -49, -50, and -52, and ESCC-NGC6822-05, -07, -08, -15, and -28). 

Given the uncertainties introduced by applying aperture corrections, we opted to calculate these only for the {\it V} and \YE filters. Total magnitudes in other passbands were then obtained by combining the small-aperture colours with the aperture-corrected {\it V} or \YE magnitudes, a valid approach considering that star clusters should not exhibit strong colour gradients. In this way, all bands are placed on the same total-magnitude scale without requiring uncertain, band-by-band aperture corrections. For the IC\,10 {\it V}-band photometry, the slope of the growth curves was calculated using the difference between the magnitudes calculated within 1\farcs6 and 2\farcs7 aperture radii, and the aperture corrections were calculated from the magnitude difference between 1\farcs6 and 5\farcs4 aperture radii. Figure\,\ref{fig:apcorr} illustrates a linear least squares fit to the data. For NGC\,6822, we used apertures of 2\farcs3 and 3\farcs8 for the slope of the growth curves and 2\farcs3 and 7\farcs5 for the aperture corrections, corresponding to the same physical radii as in IC\,10. The same-sized physical apertures were used for calculating the \Euclid \YE aperture corrections. To prevent over-correction, we limited the maximum aperture correction to be the largest one observed in the isolated clusters, which avoided extrapolating beyond the linear relationship shown in Fig.\,\ref{fig:apcorr}. Thus, the maximum aperture corrections applied were {\it V} = $-1.2$ and $-0.6$, and \YE = $-1.4$ and $-0.8$ for NGC\,6822 and IC\,10, respectively. The mean corrections were somewhat smaller, {\it V} = $-0.76$ and $-0.44$, and \YE = $-0.78$ and $-0.33$ for NGC\,6822 and IC\,10, respectively. Mean aperture correction uncertainties, derived from the slope and intercept uncertainties of the relationship, were approximately 0.02–0.05 magnitudes and were combined in quadrature with photometric errors from Poisson noise, sky background uncertainty, and zeropoint uncertainties. 

Tables~\ref{Table:6822_phot} and \ref{Table:ic10_phot} present the photometric measurements for all candidates with classes 1--4 and are available at the CDS. In NGC\,6822, the resulting median photometric uncertainties in the {\it V}- and \YE-bands are 0.02 and 0.04 for the brightest 50\% of clusters, increasing to 0.08 and 0.13 among the faintest half. In IC\,10, the corresponding photometric uncertainties are 0.05 and 0.06 for the brighter clusters, rising to 0.08 and 0.22 in the fainter population. Some clusters were particularly faint in some photometric bands or partially obscured by nearby bright stars, resulting in large photometric errors. These clusters are excluded from our analysis and flagged in our photometry tables with an asterisk. 

\subsection{SED fitting}
\label{sed_fitting_method}
To derive the physical properties (e.g. ages, masses, metallicities) as well as the line-of-sight extinction values of the clusters, we compared our photometric {\it UBVRI} + \YE\JE\HE measurements to spectral energy distributions (SEDs) predicted by the 2016 version of \citet{Bruzual2003} models, which are updated using the stellar evolutionary tracks of \citet{BressanA2012} and \citet{Marigo2013}. We did not include \IE in SED fitting, as it offers no additional information compared to the optical bands. The stellar population models were constructed and fit using the \texttt{BAGPIPES} code \citep{Carnall2018}. Prior to fitting, \Euclid AB magnitudes and {\it UBVRI} Vega magnitudes were converted to flux using the zero points provided by the SVO filter profile service. The \texttt{BAGPIPES} code assumes a fully sampled initial mass function from \citet{Kroupa2001} and we assumed a single burst of star formation can approximate all clusters. \texttt{BAGPIPES} uses Bayesian inference to derive the properties of the clusters, constructing a probability distribution based upon the $\chi^2$ value for each model fit (the likelihood), assuming
uncertainties are Gaussian and independent. Priors for each parameter can also be set, which we discuss below and detail in Table~\ref{table:bagpipes_params}. The posterior distribution was sampled using the \texttt{MULTINEST} nested sampling algorithm \citep{Skilling2006,Feroz2008,Feroz2009} via the \texttt{PYMULTINEST} interface \citep{Buchner2014}. The median of the posterior distribution (50\textsuperscript{th} percentile) was used to determine the best estimate of each cluster property, with the 16\textsuperscript{th} and 84\textsuperscript{th} percentiles defining the uncertainties. To deal with dust, we adopted a \citet{Cardelli1989} extinction law, an $R_V$ value of 3.1 and allowed the total line-of-sight extinction in the {\it V}-band, $A_V$, to vary according to the range of $E(B-V)$ values in the literature (detailed in Sects.~\ref{ngc6822_info} and~\ref{ic10_info}). We did not correct for foreground extinction before fitting, as \texttt{BAGPIPES} assumes a foreground screen approximation, and applying a single extinction law consistently avoids introducing additional assumptions about the foreground $A_V$ and the intrinsic source spectrum used to compute $A_\lambda$ values. Although the internal extinction law in dwarf galaxies may differ from that of the MW, the uncertainties involved in estimating foreground extinction and correcting for it separately are significant enough that we opt to model the total extinction with a single component. Within a central 3\arcminute\, radius of IC\,10, we imposed an $E(B-V)$ prior with a minimum of 0.8, reflecting the higher extinction values reported in the literature. Beyond this radius, the prior was allowed to extend down to 0.65. For NGC\,6822, the $E(B-V)$ prior ranged from 0.2 to 0.45 in the central regions (approximately within a radius of 6\farcm5), and was capped at 0.35 beyond this, consistent with previous studies. Additionally, we applied high-metallicity, low-age priors to five clusters in NGC\,6822 and seven in IC\,10 that appeared as prominent $\rm H\alpha$ sources in continuum-subtracted images that we constructed from the LGGS images.  Specifically, for these clusters we limited the fits to metallicity values of 0.2--0.3$ \rm\,Z_{\odot}$ for NGC\,6822 and 0.2--0.5 $\rm\,Z_{\odot}$ for IC\,10, and ages of 0--10\,Myr.  We note that since most of the clusters to be fit did not have strong coincident $\rm H\alpha$ emission, we did not include a nebular component in the fits. Tables~\ref{Table:6822_SED} and \ref{Table:ic10_SED} present the \texttt{BAGPIPES} SED-fitting results (including ages, initial mass formed, living stellar masses, $Z/Z_{\odot}$ and $A_v$) for all candidates with classes 1--4 and are available at the CDS. While \texttt{BAGPIPES} provides both the mass formed and the living stellar mass, it does not have an option for including remnants. However, using FSPS models \citep{Conroy2009x, Conroy2010x}, we approximate that the inclusion of the latter contributes an additional ${\sim}1$--5\% in present-day mass. 

\begin{table}
\centering
\caption{\texttt{BAGPIPES} priors.}
\begin{tabular}{l c c}
\hline\hline
Free parameter & NGC\,6822 & IC\,10 \\
\hline
log$_{10}(M/M_{\odot})^a$  & (0.1,7) & (0.1,7) \\
Age [Gyr]  & (0.001,14) & (0.001,14) \\
$Z/Z_{\odot}$ & (0.02,0.3) & (0.05,0.5)\\
$E(B-V)^b$ & (0.2,0.45) & (0.65,1.2) \\
\hline
\end{tabular}
\tablefoot{Uniform priors used within the \texttt{BAGPIPES} SED-fitting routine. $^a$ Mass priors are set in initial mass formed. $^b$Maximum and minimum allowed line-of-sight reddening values. Note that inner and outer galaxy regions have slightly different priors, as detailed in the main text; these represent the overall limits.  Metallicity priors are on a scale where $Z_{\odot} = 0.02$, consistent with \texttt{BAGPIPES}.}
\label{table:bagpipes_params}
\end{table}

\subsection{Half-light radii}
\label{halflight}

We empirically calculated the cluster half-light radii, $R_{\mathrm{h}}$, from the growth curves of \IE magnitude versus radius by identifying the radius where the magnitude is 0.75 fainter than the plateau value. To do this, we first smooth the growth curve using a Savitzky--Golay filter and obtain the derivative to find the flattening point. The growth curves for isolated clusters flatten out at the cluster's boundary. However, in crowded fields, nearby stars contaminate the growth curves, an effect that is especially apparent for low-mass clusters. For these objects, we apply additional masking and identify the local minimum of the derivative in the region where the overall shape of the growth curve plateaus, intentionally disregarding other minima that may appear farther out. Alternative methods for determining cluster sizes involve fitting profiles to the data. However, in the study by \citetalias{Lim2015}, many half-light radii remained undetermined via this method due to the irregular profiles of the clusters. Consequently, we opted for this empirical approach for a more complete analysis.  Our measured half-light radii are also included in Tables~\ref{Table:6822_phot} and \ref{Table:ic10_phot}.

\section{Validation and completeness}
Before proceeding to present our results, we discuss the artificial cluster tests we conducted to assess the completeness of our sample and the accuracy of our measurements. These tests involved creating synthetic clusters using the Padova stellar isochrones \citep{BressanA2012,chen2015,pastorelli2020} and the assumption of a Moffat profile (\citealt{Elson1987}). Moffat profiles are generally used to describe young clusters, while King profiles \citep{king1966} are typical for GCs. However, for the purposes of visual detection, we found that a Moffat distribution could adequately represent the various clusters visible in our images, most of which are not particularly old (see Sect.~\ref{agesmasses}). 

\begin{figure}
\centering
\includegraphics[width=\columnwidth]{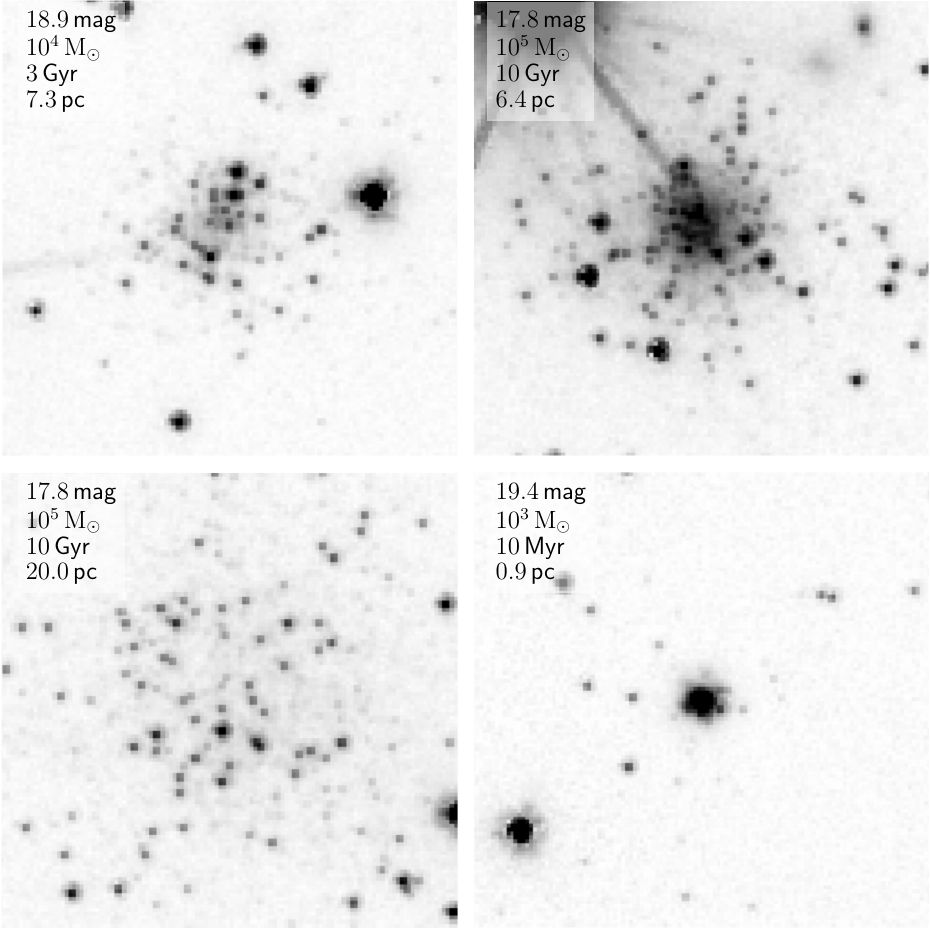}
\caption{Synthetic clusters displayed in asinh scale, placed within the outermost regions of the IC\,10 \IE image. Each cutout is 12 arcseconds on the side with apparent \IE magnitude, age, and half-light radii displayed.}
\label{fig:synth_clusts}
\end{figure}

\begin{figure*}
\centering
\includegraphics[width=\columnwidth*2]{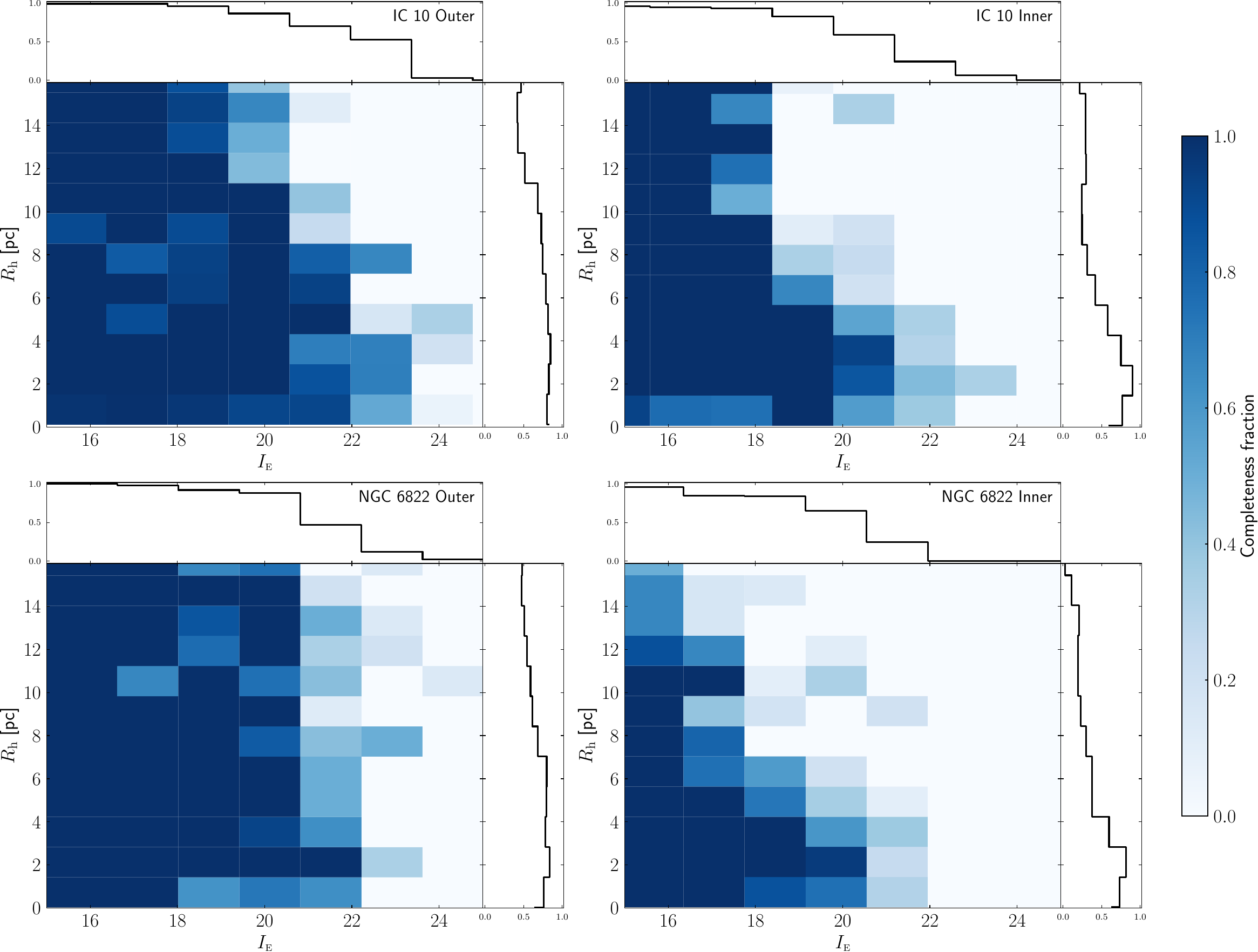}
\caption{Cluster completeness derived from artificial cluster tests for inner and outer regions of each galaxy, as a function of \IE magnitude and half-light radius.}
\label{fig:completeness_fracs}
\end{figure*}

The following steps were taken to build a sample of synthetic clusters.
\begin{enumerate}
  \item Clusters were generated across a grid of ages (10 Myr to 10 Gyr) and stellar masses ($10$ to $10^6\, M_\odot$). Metallicity was fixed at $Z = 0.004$, and a Kroupa initial mass function (IMF) was assumed \citep{Kroupa2001}. 
  \item Moffat profile parameters, core width (0.1--15\,pc) and power index (1.25--1.75), were drawn from uniform distributions, corresponding to half-light radii ranging from 0.1 to 58\,pc. Stars were spatially distributed within a cluster based on the weighting of the Moffat profile and convolved with the \IE PSF. 
  \item Stellar magnitudes were adjusted by the distance modulus of each galaxy and reddened using \citet{schlegel1998} dust maps, re-calibrated via \citet{schlafly2011}, using the \texttt{dustmaps} package \citep{Green2018}.
\end{enumerate}

Example synthetic clusters, spanning a range of physical properties, are demonstrated in Fig.\,\ref{fig:synth_clusts}.  To make the analysis more tractable, we focused on inserting clusters into a representative inner and an outer region of each galaxy, selected from the full \Euclid FoV. The inner regions cover the dense central areas of the galaxies, defined approximately by a box of width 15\arcminute\,centred on NGC\,6822 and 12\arcminute\, in IC\,10. The outer regions, located beyond these boxes, probe the sparser halo environments.  In all, ${\sim}1500$ clusters were inserted at random positions in each region (around 6000 clusters in total). We then performed a search for the synthetic clusters and recorded which ones were successfully recovered.

Figure\,\ref{fig:completeness_fracs} shows the completeness fraction, defined as the recovery rate, versus the magnitude and half-light radius for each region and galaxy. The full two-dimensional map is shown alongside the projected completeness fractions shown as histograms along each axis. The projected histograms in the top panels display the expected behaviour of high completeness at bright magnitudes, falling to lower values for fainter clusters. We estimate the $50\%$ and $90\%$ completeness limits by fitting a logistic function to the decay using a non-linear least squares method; the resulting limits are listed in Table~\ref{tab:completeness_limits}. These magnitude limits are assumed to be constant with age. When translating these magnitude limits into mass limits (see Sect.~\ref{agesmasses}), we find that, overall, we are ${\sim}50\%$ complete to $M \lesssim 10^3~{M}_{\odot}$ for ages $\lesssim100\,\rm{Myr}$, and to $M \lesssim 2\times10^4~{M}_{\odot}$ for ages of ${\sim}10\,\rm{Gyr}$, in both galaxies. We note, however, that the assumption of constant magnitude completeness limits is a simplification, as completeness may vary somewhat with age due to differences in cluster appearance.   

\begin{table}
    \centering
    \caption{Completeness limits.}
    \begin{tabular}{c c c}
    \hline\hline
    Region & $50\%$ limit  & $90\%$ limit\\
    \hline
    NGC\,6822 inner & 19.9 & 18.2 \\
    NGC\,6822 outer & 21.4 & 19.8 \\
    IC\,10 inner & 20.9 & 18.8 \\
    IC\,10 outer & 21.9 & 19.7 \\
    \hline
    \end{tabular}
    \tablefoot{Cluster magnitude completeness limits for the galaxies, separated by inner and outer regions.}
    \label{tab:completeness_limits}
\end{table}

We also examine how the completeness depends on half-light radius, as shown in the vertically-oriented projected histograms in  Fig.\,\ref{fig:completeness_fracs}. Our results indicate that the completeness generally decreases with half-light radius across both regions and galaxies, following a maximum at half-light radii of ${\sim}2$--3\,pc. Recovery rates in the inner regions decline steeply beyond this maximum, whereas those in the outer regions decline more shallowly.

\begin{figure}
    \centering
    \includegraphics[width=\columnwidth]{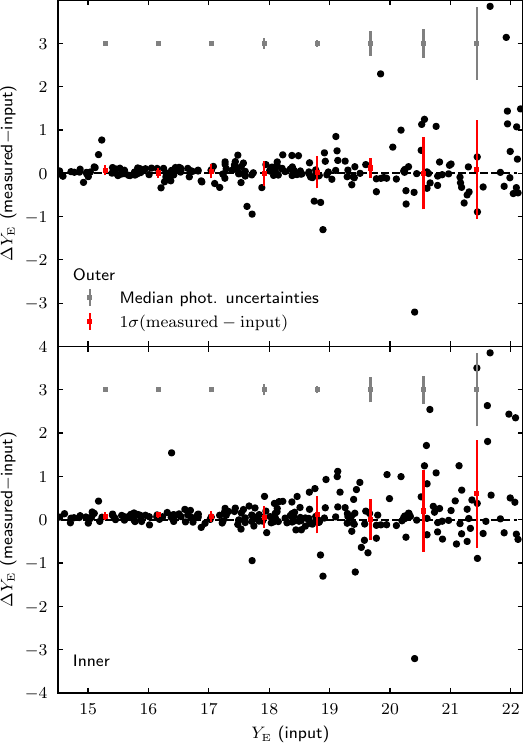}
    \caption{Input versus measured minus input \YE-band magnitudes in the outer (top) and inner (bottom) regions of IC\,10 using synthetic clusters produced with a Moffat profile and \texttt{PARSEC} isochrones. The dashed line indicates an offset of zero. Also plotted, for each magnitude bin, are the mean offsets with associated standard deviations (red) and, for comparison, the median cluster sample photometric uncertainties placed above at a $\Delta$\YE value of 3 (grey).}
    \label{fig:input_mes_mag_comp}
\end{figure}

The synthetic clusters were also used to test the accuracy of our photometry method. Adopting the IC\,10 \YE image as a representative case, we performed photometry in the same way as described in Sect.\,\ref{photometry}, using the same aperture correction relationship and aperture sizes. In Fig.\,\ref{fig:input_mes_mag_comp}, we plot the input vs. measured minus input \YE magnitudes for the separate regions. For sources brighter than \YE $= 19$, the mean offset is very small, 0.01 magnitudes in the outer region and 0.06 in the inner region, indicating excellent agreement, albeit with a slight bias towards fainter recovered magnitudes. For fainter clusters, the mean offset increases to 0.42 and 0.69 magnitudes in the outer and inner regions, respectively. Translating this magnitude offset to stellar mass (see Sect.~\ref{agesmasses}), it corresponds to offsets of $\log_{10}(M/M_\odot){\sim}0.2$ and ${\sim}0.3$, in the outer and inner regions, respectively, across both young and old clusters.

\begin{figure*}
\centering
\includegraphics[width=\columnwidth*2]{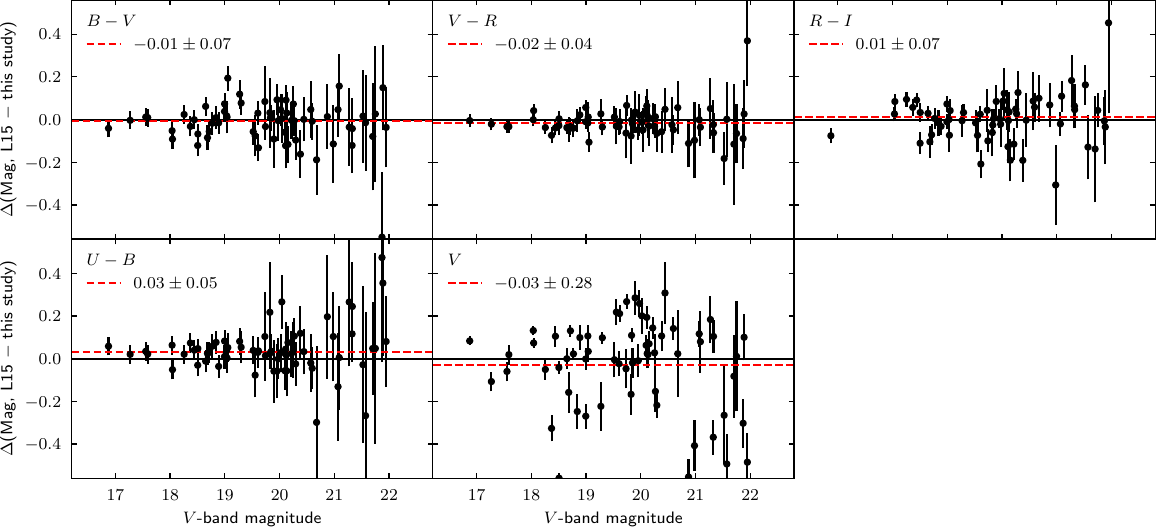}
\caption{Comparison of our $UBVRI$ photometry to that of \citetalias{Lim2015}, using their catalogue of 66 clusters in IC\,10. No extinction corrections have been applied. Error bars reflect the combination of \citetalias{Lim2015} and our photometry uncertainties. The red dashed line is the weighted mean offset (indicated in the top left alongside the associated standard deviation), while the black line indicates an offset of zero. }
\label{fig:Lim_comp}
\end{figure*}

At magnitudes brighter than \YE = 19--20, the standard deviation in each bin remains below ${\sim}0.3$ magnitudes. However, at fainter magnitudes, the scatter increases substantially, with standard deviations reaching up to one magnitude in the faintest bin.  It is evident that the cluster photometric uncertainties are 1--5 times smaller than the observed scatter, indicating that the formal errors underestimate the true uncertainties, as is expected.  To further validate our measurements, we compare our $\it{UBVRI}$ IC\,10 photometry to that of \citetalias{Lim2015} in Fig.\,\ref{fig:Lim_comp}. We find excellent agreement with their values, with the weighted mean offset being $\leq 0.03$ in each colour and 0.03 in the {\it V}-band, with corresponding standard deviations $\leq 0.07$ and 0.28, respectively. Furthermore, there is generally very good agreement among the $V-I$ colours of NGC\,6822 clusters Hubble-VI, Hubble-VII, Hubble-VIII, SC3, SC6 and SC7 (ESCC-NGC6822-01, 02, 03, 05, 07 and 08) with those published in previous studies, as demonstrated in Table~\ref{Table:Comparing_6822_phot}.

\begin{figure}
    \centering
    \includegraphics[width=\columnwidth]{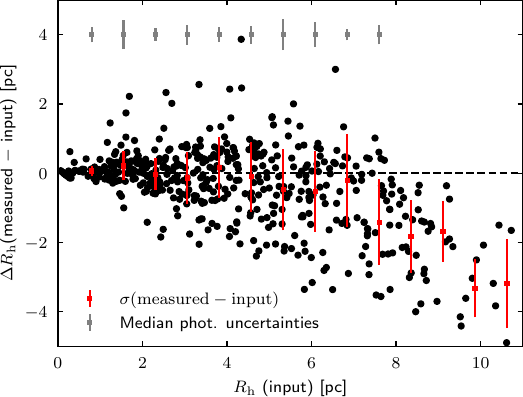}
    \caption{Input versus measured minus input half-light radii (in pc units) derived in the \IE-band for artificial clusters in IC\,10. The dashed line presents an offset of zero. Also plotted, for each magnitude bin, are the mean offsets with associated standard deviations (red) and, for comparison, the median cluster sample photometric uncertainties placed above at a $\Delta R_{\mathrm{h}}$ value of 4 (grey).}
    \label{fig:input_mes_rh_comp}
\end{figure}

\begin{table*}[h!]
\centering
\caption{Comparison of $(V-I)_0$ (extinction-corrected $V-I$ colour) in NGC\,6822 to values from previous studies.}
\begin{tabular}{l c c c c c}
\hline\hline
 & $(V-I)_0$ & $(V-I)_0$ & $(V-I)_0$ & $(V-I)_0$ & $(V-I)_0$\\
ID & This study & \citet{Veljanoski2015} & \citet{Hwang2011} & \citet{Huxor2013} & \citet{Krienke2004} \\
\hline
Hubble-VI  & 0.53 $\pm$ 0.03 & \dots & \dots & \dots & 0.32 \\
Hubble-VII  & 0.94 $\pm$ 0.03  & 0.87 $\pm$ 0.04 & 1.05 & \dots & 0.89 \\
Hubble-VIII & 0.62 $\pm$ 0.04 & \dots & 0.93 & \dots & 0.65 \\
SC3         & 0.87 $\pm$ 0.03 & 0.87 $\pm$ 0.05 & 1.31 & \dots & \dots \\
SC6         & 0.91 $\pm$ 0.03 & 0.87 $\pm$ 0.03 & \dots & 0.84 $\pm$ 0.03 & \dots \\
SC7         & 1.18 $\pm$ 0.03 & 1.02 $\pm$ 0.03 & \dots & 1.05 $\pm$ 0.03 & \dots \\
\hline
\end{tabular}
\label{Table:Comparing_6822_phot}
\end{table*}

\begin{table*}
\centering
\caption{Comparison of $Z/Z_\odot$ in NGC\,6822 to spectroscopic values from previous studies (assuming $Z_{\odot} = 0.02$, consistent with \texttt{BAGPIPES}).}
\begin{tabular}{c c c c c}
\hline\hline
 & $Z/Z_{\odot}$ & $Z/Z_{\odot}$ & $Z/Z_{\odot}$ & $Z/Z_{\odot}$\\
ID & This study & \citet{Larsen2022} & \citet{Chandar2000} & \citet{Hwang2014} \\
\hline
  Hubble-VI & $0.027 \pm 0.010$ & \dots & $0.036 \pm 0.022$ & \dots \\
  Hubble-VII & $0.037 \pm 0.011$ & $0.027 \pm 0.010$ & ${\sim}0.013$ & $0.012 \pm 0.004$ \\
  Hubble-VIII & $0.244 \pm 0.052$ & \dots & \dots & $0.293 \pm 0.129$ \\
  SC3 & $0.191 \pm 0.02$ & \dots & \dots & $0.052 \pm 0.029$ \\
  SC6 & $0.026 \pm 0.005$ & $0.028 \pm  0.001$ & \dots & \dots\\
  SC7 & $0.022 \pm 0.001$ & $0.066 \pm 0.002$ & \dots & \dots\\
\hline
\end{tabular}
\label{Table:Comparing_6822_mets}
\end{table*}

\begin{figure*}
    \centering
    \includegraphics[width=\columnwidth*2]{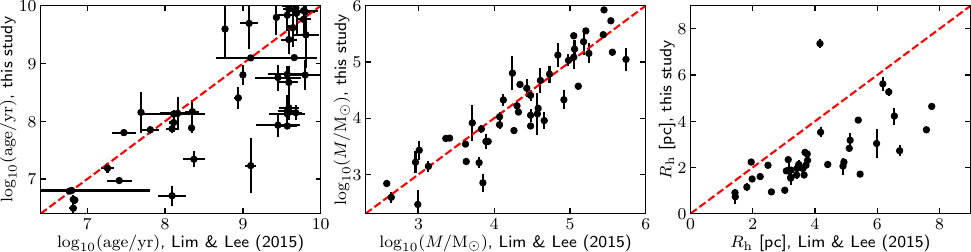}
    \caption{Left and middle: Comparison of our SED–ﬁt masses and ages with those of \citetalias{Lim2015}. The 1:1 line is shown in red. Right: Comparison of our half-light radii estimates with their sample.}
    \label{fig:Lim_comp_SED}
\end{figure*}

Figure\,\ref{fig:input_mes_rh_comp} shows how well we can recover the sizes of the synthetic clusters, using the methodology described in Sect.~\ref{halflight}. Clusters with half-light radii in the 0--5\,pc range are recovered well, but with increasing scatter, whereas for clusters with sizes greater than about 7\,pc, we increasingly underestimate the half-light radius.  This is also true when we compare our results to those of  \citetalias{Lim2015}, shown in the right-hand panel of Fig.\,\ref{fig:Lim_comp_SED}.   While our validation tests indicate an underestimation at larger radii, the discrepancy with \citetalias{Lim2015} appears across all sizes. This may reflect a systematic offset between their profile-fitting methodology and our empirical one. 

Finally, to benchmark the performance of \texttt{BAGPIPES}, we compared the metallicity outputs to spectroscopic literature values, available only for NGC\,6822 \citep{Chandar2000,Hwang2014,Larsen2022}. The outputs from \texttt{BAGPIPES} have the format $Z/Z_{\odot}$, with $Z_{\odot} = 0.02$ \citep{Carnall2018}. For clusters from \citet{Larsen2022}, we transformed their reported $\mathrm{[Fe/H]}$ values into total metallicities ($Z/Z_\odot$) using the corresponding $\alpha$-element abundances and the empirical relation from \citet{Salaris1993}. In contrast, \citet{Hwang2014} and \citet{Chandar2000} reported their metallicities directly in terms of $\mathrm{[Z/H]}$, allowing for a more straightforward comparison. Since \citet{Hwang2014} provided results based on multiple fitting methods, we adopted the mean of the published values for each cluster. The final values used for comparison are listed in Table~\ref{Table:Comparing_6822_mets}. Our metallicity estimates are consistent within errors with the spectroscopic values for Hubble-VI, Hubble-VII, Hubble-VIII, and SC6. However, we find a higher metallicity for SC3 and a lower metallicity for SC7 relative to the literature. Although the comparison sample was limited, the metallicity offsets did not demonstrate a correlation with cluster magnitude, as we might expect if stochastic effects were contributing to these discrepancies (stochastic effects are discussed further in Appendix \ref{stoch_effects}).

A detailed comparison of our derived masses and ages with those from \citetalias{Lim2015} is provided in Fig.\,\ref{fig:Lim_comp_SED}. Both studies utilise the \citet{Bruzual2003} models but differences stem from the version used, variations in SED-fitting techniques, prior selections, and the inclusion of NIR bands in our analysis. Although we observe fairly good agreement in mass, discrepancies are noted in age estimates. Notably, a number of clusters for which \citetalias{Lim2015} finds old ages are found to be considerably younger in our work. We checked if removing the most dust-extincted sources would eliminate this discrepancy, but they persisted, although the overall scatter slightly decreased.

\section{Results and discussion}
In this section, we present the results of our photometric measurements alongside our SED-fitting results, tabulated in Appendix~\ref{results_tables}. Unless otherwise specified, we restrict our analyses to clusters with class 1--4 and with photometric errors less than one magnitude in all bands.  Our catalogues and results should facilitate a variety of future in-depth studies of the cluster populations in NGC\,6822 and IC\,10 and we focus only on a few specific science results in this work. 

\subsection{Photometric properties}
\begin{figure}
    \centering
    \includegraphics[width=\columnwidth]{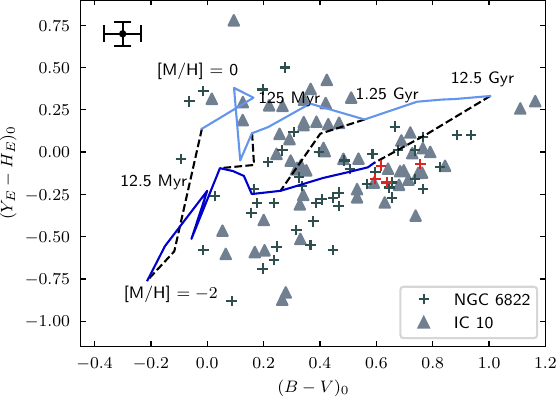}
    \caption{Reddening-corrected $(B-V)_0$ vs. $(\YE-\HE)_0$ colour-colour diagram of our cluster sample. \texttt{PARSEC} simple stellar populations with [M/H] $= -2$ and 0 are overplotted (blue lines) with connecting lines of constant age (dashed lines). A subset of the GCs within NGC\,6822 for which we have data coverage are indicated in red. Median errors are displayed in the top left-hand corner.}
    \label{fig:cc_diagram}
\end{figure}

In Fig.\,\ref{fig:cc_diagram}, we plot the $B-V$ vs. $\YE-\HE$ colours, alongside \texttt{PARSEC} SSP models \citep{BressanA2012,chen2015,pastorelli2020} calculated for two metallicities, $\mathrm{[M/H]} = -2$ and 0. Constant ages are indicated by the connected dashed lines, with increasing age from left to right. It can be seen that while $B-V$ is mostly sensitive to age, $\YE-\HE$ is mostly sensitive to metallicity. We note that the \texttt{PARSEC} SSP models are expected to be consistent with the \texttt{BAGPIPES} output due to their common usage of the \citet{BressanA2012} stellar evolutionary tracks.  

For this initial comparison, we simply correct our photometry for reddening using 
the total $E(B-V)$ towards each galaxy taken from the literature; we use $E(B-V) = 0.81$ for IC\,10 \citep{Massey2007} and $E(B-V) = 0.30$ for NGC\,6822 \citep{Fusco2012}. 

We see that the clusters largely span the expected range of colour-colour space, suggesting the presence of both young and old clusters, in addition to high and low metallicities. That said, we also observe some clusters which have dereddened $\YE-\HE$ colours bluer than what is expected for $\mathrm{[M/H]} = -2$, irrespective of age.  The origin of this offset is not clear, but could potentially be due to one or more of stochastic effects, nebular emission, alpha-enhancement, small-scale reddening variations and photometric uncertainties.   We highlight in red the known GCs within our sample. They fall in the low-metallicity and high-age region, as would be expected and broadly agreeing with their literature values \citep{Hwang2014, Veljanoski2015, Huxor2013, Larsen2022}. 

\begin{figure*}
    \centering
    \includegraphics[width=\columnwidth*2]{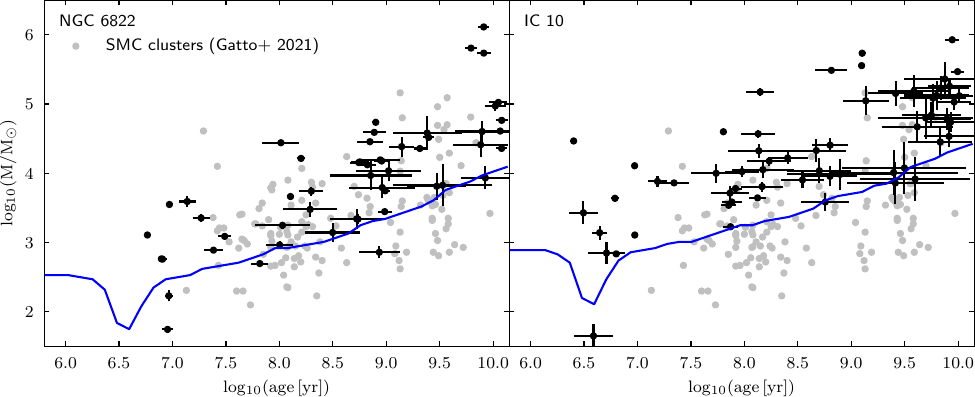}
    \caption{Masses as a function of age as derived by \texttt{BAGPIPES} for the star clusters in NGC\,6822 (left) and IC\,10 (right). The blue line shows the 50\% completeness limit in the inner regions. Also plotted in grey is the sample of SMC clusters from \citet{Gatto2021}.}
    \label{fig:age_vs_mass}
\end{figure*}

\subsection{Ages and masses of the star clusters}  
\label{agesmasses}
We use the outputs of \texttt{BAGPIPES} to separate our cluster sample into young, intermediate, and old sub-populations. We adopt the following age cuts:
\begin{itemize}
  \item $\mathrm{log_{10}(age/yr)} \leq 7.3$ for young clusters;
  \item $7.3 \leq \mathrm{log_{10}(age/yr)}\leq 9.3$ for intermediate-age clusters;
  \item $\mathrm{log_{10}(age/yr)} \geq 9.3$ for old clusters.
\end{itemize}

Based on these classifications, we identify eight and 12 young clusters (classes 1--4) in NGC\,6822 and IC\,10, respectively. We also identify 28 and 32 intermediate-age clusters, and 16 and 27 old clusters in these galaxies.

Figure\,\ref{fig:age_vs_mass} shows the cluster mass (initial mass formed) versus age for both galaxies, along with the 50\% completeness limit calculated for the inner regions of each galaxy. This has been derived by converting the magnitude thresholds in Table~\ref{tab:completeness_limits} into corresponding age and mass limits using the models from \citet{Bruzual2003}. For this, we assume a representative metallicity of $0.1\,\rm Z_{\odot}$ along with reddening values of $E(B-V) = 0.81$ for IC\,10 \citep{Massey2007} and $E(B-V) = 0.3$ for NGC\,6822 \citep{Fusco2012}.  Figure\,\ref{fig:age_vs_mass} also presents masses and ages for a sample of 134 SMC\footnote{As a dIrr with a stellar mass of $\sim 9 \times 10^8~ M_{\odot}$ \citep{Pace2024}, the SMC provides a good analog to the systems we study in this paper, although it is a few times more massive.} clusters, taken from the study by \citet{Gatto2021}. These clusters, primarily situated within the central regions of the SMC (approximately ${\sim}2^{\circ}$–$3^{\circ}$ from the centre), represent only a fraction of the ${\sim} 850$ cluster candidates in the SMC and Magellanic Bridge compiled by \citet{Bica2020}. \citet{Gatto2021} use data from the STEP survey to uniformly derive ages and masses for this sample via colour-magnitude diagram (CMD) fitting, and this homogeneous nature makes it ideal for comparison.  That said,  it is important to keep in mind that the \citet{Gatto2021} sample is biased in terms of its central location (for example, it does not include NGC\,121, which is the only old GC in the SMC). Furthermore, as it is based on clusters that were uncovered in a variety of earlier studies, it is likely to have varying (unquantified) completeness.  Nonetheless, due to the close proximity of the SMC, it is expected that it will reach to lower masses at any given age than our \Euclid search, as indeed can be seen in Fig.\,\ref{fig:age_vs_mass}. Both \texttt{BAGPIPES} and the \texttt{ASteCA} \citep{Perren2015x} version used by \citet{Gatto2021} estimate the initial mass formed. While there may be systematic differences between the initial mass computations, these are expected to be small.

Both NGC\,6822 and IC\,10 host clusters that range in mass from $\sim10^{2}\,M_{\odot}$ to $\sim10^6\,M_{\odot}$, and with ages from $\sim10$\,Myr to $\sim10$\,Gyr, generally consistent with the age range inferred from Fig.~\ref{fig:cc_diagram}. We find that older clusters are typically more massive, which probably reflects the preferential disruption and fading of low-mass clusters over time \citep[e.g.][]{Fall2001}, as well as a possible preference for the formation of more massive star clusters at early times \citep[e.g.][]{Elmegreen1997, Lahen2025}.   On the other hand, clusters younger than a few tens of Myr typically have masses below $10^4\,M_{\odot}$ in both systems. An exception is found in IC\,10, which hosts two young, massive clusters: ESCC-IC10-40 (class 1) and ESCC-IC10-71 (class 4), with $\log_{10}(M/M_{\odot}) = 4.1 \pm 0.01$ and $4.5 \pm 0.01$, and ages of $9.4 \pm 0.3$ and $2.5 \pm 0.2$\,Myr, respectively.\footnote{\citetalias{Lim2015} discuss a potential Super Star Cluster (SSC) candidate in IC\,10, first identified by \citet{Hunter2001} but classified in that study as an OB association.  \citetalias{Lim2015} argue that, based on the stellar density in the HST image, this object could be an SSC. We identified this object in our cluster search (ESCC-IC10-44); however, after classification by our group, it was assigned class 5 so very unlikely to be a cluster, consistent with its original designation.} Observational evidence suggests that the young cluster mass function follows a power law, with a truncation at the high-mass end that appears to correlate with the SFR surface density ($\Sigma_{\rm SFR}$) of the host galaxy \citep{Larsen2009, Adamoa2015, Johnson2017}. Converting the data listed in Table~\ref{table:gal_properties} to SFR surface densities within one half-light radius, we find values of $\log_{10}( \Sigma_{\text{SFR}}/M_{\odot} \rm yr^{-1}kpc^{-2}) = -8.7$ and $-6.5$ for NGC\,6822 and IC\,10, respectively.  While we do not have enough young clusters in our sample to construct meaningful mass functions, we note that the presence of more massive young star clusters in IC\,10 is consistent with this expected dependence of the truncation mass on $\Sigma_{\text{SFR}}$.  

The oldest clusters in the dwarfs reach masses well above $10^5\,M_{\odot}$, with one exceptional GC in NGC\,6822 (SC7=ESCC-NGC6822-08) above even $10^6\,M_{\odot}$.  With a derived initial mass formed of $1.3 \times 10^6\,M_{\odot}$, this cluster is roughly twice as massive as any of the other old clusters in NGC\,6822. Converting this mass to an estimate of the present-day mass (including both living stars and remnants) using FSPS models \citep{Conroy2009x, Conroy2010x}, we obtain a current mass of $7.7 \times 10^5\,M_{\odot}$. This high mass is further supported by the cluster velocity dispersion reported by \citet{Larsen2022}. Using their measurement of 9.9 kms$^{-1}$, we estimate the dynamical mass to be $4.6 \times 10^5\,M_{\odot}$ which is comparable, albeit slightly lower, to our estimated current mass. Our derived metallicity of $Z/Z_{\odot}=0.022 \pm 0.001$ is considerably lower than the one found by \citealt{Larsen2022} (see Table~\ref{Table:Comparing_6822_mets}), but in agreement with the metallicity from fitting a deep HST CMD (McGill et al., in prep). Aside from its high mass, other peculiarities of SC7 include an $\alpha$-abundance close to the solar value \citep{Larsen2022} and a high ellipticity (see Fig.\,\ref{fig:example_clusts}, and \citealt{Huxor2013}).  The existence of this high ellipticity old massive star cluster in the outskirts of NGC\,6822 is particularly intriguing, and it is tempting to speculate that it has been accreted. Indeed, SC7's high mass and ellipticity are reminiscent of 
$\omega$ Centauri, the most massive GC in the MW ($2.5 \times 10^6\,M_{\odot}$; \citealt{vdVen2006}), which is suspected of being the stripped core of a disrupted dwarf galaxy \citep[e.g.][]{2003MNRAS346L11B}.  A similar origin for SC7 may appear less compelling, as NGC\,6822 shows no signs of recent tidal disturbance in its outskirts \citep{Zhang2021}. However, we note that the cluster may originate from a much earlier event, now fully phase-mixed and no longer evident in the galaxy’s present-day structure. 

\begin{figure}
    \centering
    \includegraphics[width=\columnwidth]{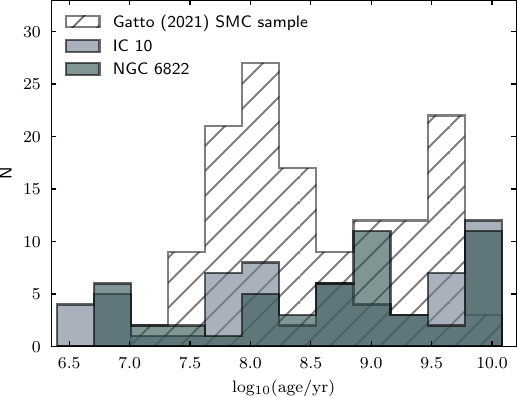}
    \caption{Age distributions of the clusters compared with the SMC clusters taken from \citet{Gatto2021}.}
    \label{fig:Age_dists_smc}
\end{figure}

In Fig.\,\ref{fig:Age_dists_smc}, we examine more closely the relative distribution of cluster ages. All three systems show evidence of continuous cluster formation across time but the central part of the SMC seems to have formed comparatively more 
intermediate-age clusters compared to the relatively flat distributions in both NGC\,6822 and IC\,10.   While there is a hint of an enhanced proportion of ancient clusters [$\mathrm{log_{10}(age/yr)} \geq 9.7$] in NGC\,6822 and IC\,10 compared to the SMC, it should be remembered that the \citet{Gatto2021} sample is biased in terms of areal coverage and that there are potential systematic effects related to the different age determination methods used for the SMC clusters and in this study, as well as their different completeness levels.  

At both young and old ages, IC\,10's cluster population outnumbers that of NGC\,6822. 
Given IC\,10's higher current SFR, we may expect more young clusters [$\mathrm{log_{10}(age/yr)} \leq 7.3$] relative to NGC\,6822, which is indeed the case. 
However, given IC\,10's classification as a starburst dwarf, it is perhaps surprising that it has only 50 percent more young clusters. A possible reason for why this number is not larger is cluster disruption. Some studies suggest that starburst galaxies exhibit particularly short disruption timescales, around 7--10\,Myr, due to rapid gas expulsion. This is supported by observations of diffuse UV light in these galaxies, which is argued to result from the swift dispersal of young clusters \citep{Chandar2005}. Additionally,  the high extinction toward IC\,10 may be obscuring some young clusters, preventing their detection, or they may still be deeply embedded in their dusty natal environments. 

\begin{figure*}
    \centering
    \includegraphics[width=\columnwidth*2]{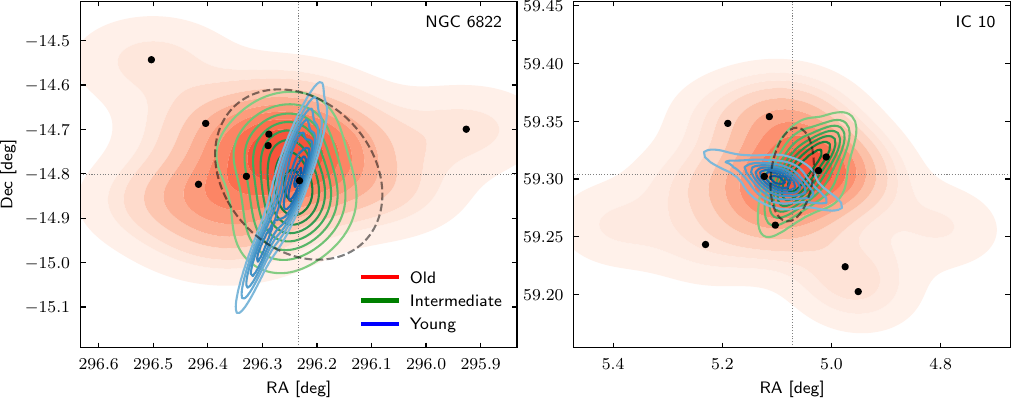}
    \caption{Kernel density estimation contour plots showing the spatial distribution of clusters in NGC\,6822 (left) and IC\,10 (right). Clusters are grouped by age: old (red), intermediate (green) and young (blue). The contours represent the density of clusters at their location within each galaxy. Old clusters are displayed with filled contours to enhance contrast against the other populations, making them easier to distinguish. Also plotted are grid lines marking the nominal centre of each galaxy \citep{Sibbons2012, Gerbrandt2015} and the positions of candidate GCs in black (see Sect.~\ref{gcs}). Note that there are 3 GCs in NGC\,6822 which lie outside the \Euclid FOV and are not shown here. Grey dashed ellipses, representing each galaxy's half-light radius with the appropriate ellipticity and position angle, are overlaid \citep{McConnachie2012, Higgs2021, Jarrett2003}.}
    \label{fig:spatial_plots}
\end{figure*}

\subsubsection{Spatial distributions by age}  

Smoothed density contours of the cluster spatial distributions are displayed in Fig.\,\ref{fig:spatial_plots}. The entire \Euclid FoV is not shown; instead, we have adjusted the RA and Dec ranges to focus on the cluster locations farthest from the galaxy centres, making it easier to highlight the features. The centres of the galaxies are also indicated, with NGC\,6822's centre defined as RA = $19^{\text{h}}44^{\text{m}}56^{\text{s}}$, Dec = $-14^{\text{d}}48^{\text{m}}06^{\text{s}}$ \citep{Sibbons2012} and
for IC\,10, RA = $00^{\text{h}}20^{\text{m}}17.3^{\text{s}}$, Dec = $59^{\text{d}}18^{\text{m}}14^{\text{s}}$, taken from \citet{Gerbrandt2015}. The distributions are centrally concentrated in both systems for all age bins, but especially so for the young clusters. In NGC\,6822, the young clusters are roughly co-located with a couple of previously known star-forming regions; namely, {\it Spitzer} I and Hubble-V. The elongated distribution of young clusters, primarily in a north-south direction, but extending slightly toward the south-east, is also consistent with the young stellar distributions displayed in \citet{Hunt2024}, as well as earlier works \citep{Tantalo2022,HirschauerA2020}, and with the position of the bar. The intermediate-age cluster spatial distribution is more circular, albeit with a slight extension in the NE--SW direction, while the old clusters are more symmetrically distributed yet again.  These strong age-dependent variations in spatial distribution are also seen in IC\,10. The young clusters are primarily elongated in an northeast–southwest direction, while the intermediate-age clusters are oriented nearly perpendicular to this. Small offsets can be seen between the centroids of the cluster distributions but these are not likely to be meaningful given the small numbers of clusters in each bin and the smoothing kernel used. An offset of a few hundred parsecs between the geometric centres of the old and young stellar populations in IC\,10 was previously noted by \citet{Gerbrandt2015}, but the sense of their offset is opposite to what we see for the star cluster populations. Consistent with the nature of dwarf starbursts, clusters of all ages are more centrally concentrated in IC\,10 compared to those in NGC\,6822 (also consistent with the star count maps shown in \citealt{Hunt2024}).

\subsection{Luminosity and colour functions}  
\begin{figure*}
    \centering
    \includegraphics[width=\columnwidth*2]{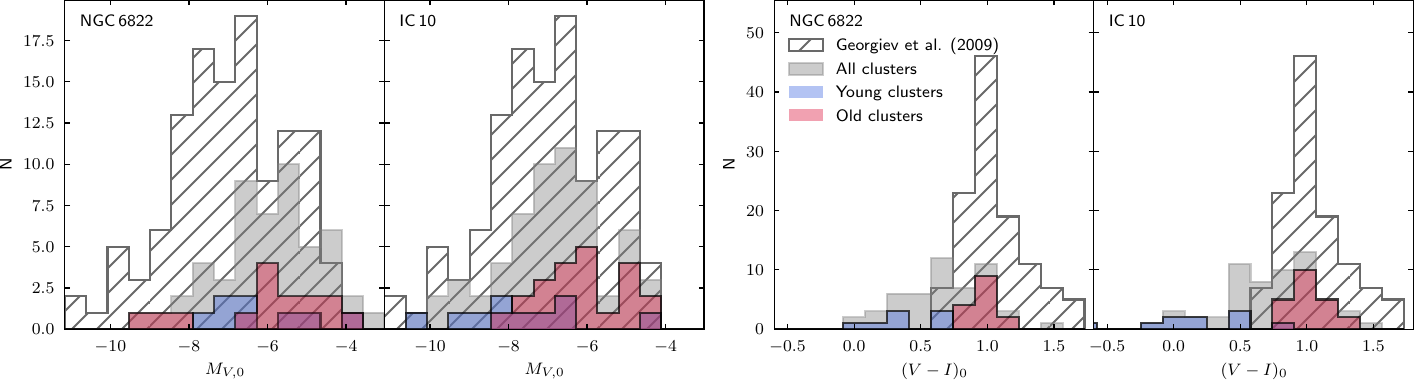}
    \caption{Left panels: Distributions of extinction-corrected absolute {\it V}-band magnitudes for our sample of clusters. The overall distribution is shown as well as the luminosity function split into young (blue) and old (red) age bins. We also show a sample of GCs in more distant dIrr galaxies, taken from \citet{Georgiev2009}. Right panels: The same but for the $(V-I)_0$ colour distributions.}
    \label{fig:lum_color_func_Georgiev_comp}
\end{figure*}

The left panels of Fig.\,\ref{fig:lum_color_func_Georgiev_comp} show the star cluster luminosity functions (LFs), expressed in extinction-corrected absolute {\it V}-band magnitude calculated using the  \texttt{BAGPIPES} outputs for $A_V$ and the distances in Table~\ref{table:gal_properties}. We show the overall LFs for each galaxy, as well as the LFs for the young and old clusters using the definitions introduced in Sect.~\ref{agesmasses}.  For completeness, we also add the NGC\,6822 GCs SC1, SC2, and SC4, which lie outside the \Euclid FoV, using the magnitudes and colours from \citet{Veljanoski2015}.  

The overall LFs for the two dwarfs are broadly similar. They exhibit approximately Gaussian shapes, with the peaks of a Gaussian fit in IC\,10 ($M_{V,0}\sim-6.7$) being slightly brighter than that in NGC\,6822 ($M_{V,0}\sim-5.9$). 
There is tentative evidence for a second peak in the IC\,10 LF at $M_{V,0}\sim-5$, which is reminiscent of the fainter peak in the bimodal LF of M31 halo GCs \citep{Huxor2014,Mackey2019}, but the low numbers of clusters prevent any strong conclusions. The LF of IC\,10 extends to $M_{V,0}\sim-10$ and its most luminous clusters are of young to intermediate age. On the other hand, the most luminous clusters in NGC\,6822 are old clusters, which are roughly one magnitude brighter than any of the old clusters found in IC\,10. 

We also show a comparison sample of 118 candidate GCs in 30 dIrr galaxies taken from \citet{Georgiev2009}.  These dIrrs have $M_V>-16$ and reside in low-density, relatively nearby ($D\lesssim12$~Mpc) environments, similar to NGC\,6822 and IC\,10. Their candidate GCs have been identified through analysis of deep HST images, with 90\% competeness estimated to be $M_V\simeq-4.5$, however most have not been spectroscopically confirmed or age dated. Our old clusters fall within the magnitude range of the GCs in the \citet{Georgiev2009} sample but generally lie toward the fainter end of the distribution. 

The right-hand panels of Fig.\,\ref{fig:lum_color_func_Georgiev_comp} show the $(V-I)_0$ colour distributions. Our old clusters are found within the colour range spanned by the \citet{Georgiev2009} GC sample, while most of our young clusters are found blueward of our old clusters. The colours of the old clusters peak at a similar value compared to the GCs in the \citet{Georgiev2009} sample in both galaxies, with a peak around ${\sim}1.0$. \citet{Georgiev2009} highlight an absence of faint ($M_V \gtrsim -6$) blue GCs in their sample, defined as those with $(V-I)_0\leq 1.0$. We have several old clusters that fit these criteria in both galaxies (with ages between 3 and 12\,Gyr), but we note that the uncertain reddening may be impacting these colours.

\subsection{Half-light radii}
\begin{figure}
    \centering  \includegraphics[width=\columnwidth]{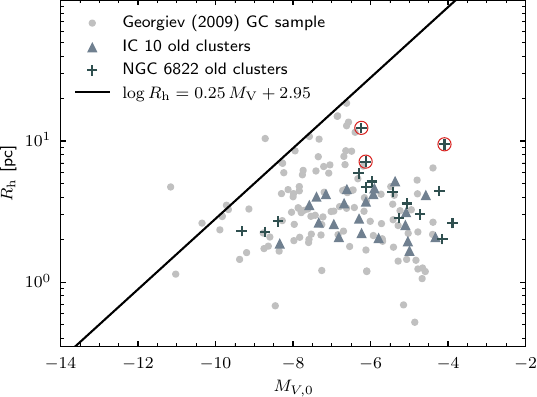}
    \caption{Half-light radii as a function of absolute {\it V}-band magnitude for the old cluster sample, and for the sample of GCs from \citep{Georgiev2009}. The empirical relation  $\logten(R_{\rm h}/{\rm pc}) = 0.25 \,{M_V} + 2.95$ \citep{Mackey2005}, which defines the upper envelope of MW GCs, is overplotted. Small red circles mark the three most extended clusters in NGC\,6822.  }
    \label{fig:Rh_vs_Vmag}
\end{figure}

It is interesting to explore the correlation between the half-light radii of the old clusters and their luminosities. Since $R_{\rm h}$ changes very little over many relaxation times \citep[e.g.][]{Meylan1997}, this can potentially provide insight into the formation process of the most massive, long-lived clusters. Figure\,\ref{fig:Rh_vs_Vmag} shows how the half-light radii of the old clusters depend on absolute {\it V}-band magnitude.  In the MW, GCs show
a clear trend in $R_{\rm h}$ with $M_V$, in the sense that more luminous clusters are more compact. Very few clusters exist above the empirical boundary $\logten(R_{\rm h}/{\rm pc}) = 0.25 \,{M_V} + 2.95$ \citep{Mackey2005}, which is reproduced in Fig.~\ref{fig:Rh_vs_Vmag}. For comparison, the \citet{Georgiev2009} sample is also plotted. Our old clusters lie below the empirical line and fall within the size range of the \citet{Georgiev2009} sample.  Their half-light radii also peak at a similar value to ours (around $\sim2$\,pc), as well as the old [$\mathrm{log_{10}(age/yr)} \geq 9.3$] clusters in the \citet{Gatto2021} sample. The brighter old clusters ($M_{V,0} \leq -6$) have half-light radii within the range ${\sim}1.5$--4.5\,pc, similar to what is seen in MW GCs \citep{Harris1996}. However, toward fainter magnitudes, the spread increases considerably for NGC\,6822, reaching up to 12\,pc, and also extending beyond the half-light radii of the \citet{Georgiev2009} sample at these faint magnitudes. This increased scatter at faint magnitudes has also recently been observed in NGC\,2403 and IC\,342 using \Euclid data  \citep{Larsen25}. On the other hand, the size distribution of old clusters in IC\,10 is relatively flat as a function of magnitude. In both NGC\,6822 and IC\,10, there is a subtle trend where clusters located farther from the galaxy centre tend to have slightly larger half-light radii.  

\begin{figure}
    \centering    \includegraphics[width=\columnwidth]{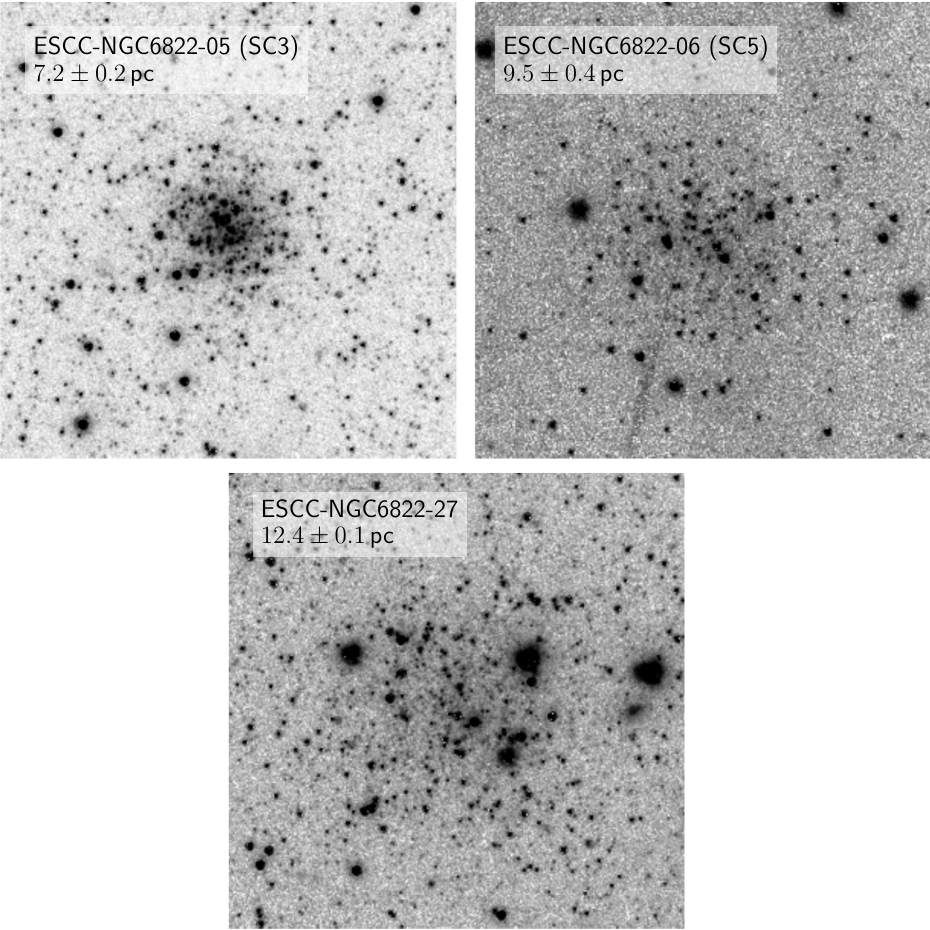}
    \caption{Thumbnails in \Euclid \IE of NGC\,6822's extended clusters. Each cutout is 30\arcsec\,on a side, corresponding to 74~pc at the distance of NGC\,6822.}
    \label{fig:ext_comp}
\end{figure}

Figure\,\ref{fig:Rh_vs_Vmag} shows that NGC\,6822 has three particularly extended old clusters (red circles), with $R_{\rm h}\gtrsim 7$~pc, which have no counterparts in IC\,10.  While a few examples old extended clusters exist in the \citet{Georgiev2009} sample, they are all brighter than the ones highlighted here. Two of
the extended clusters in NGC\,6822, SC3 and SC5, were previously known from ground-based work \citep{Hwang2011, Huxor2013} while the other is newly identified in this work. With 
$R_{\mathrm{h}} = 12.4 \pm 0.1$\,pc, ESCC-NGC6822-27 is the most extended old cluster within our sample, and it has an age of $2.1 \pm 0.3$\,Gyr, a mass of $\mathrm{log_{10}}(M/{M}_{\odot}) = 4.4 \pm 0.04$ and a metallicity of $Z/Z_{\odot}=0.03 \pm 0.01$.
Thumbnails of the extended clusters within our FoV are shown in Fig.\,\ref{fig:ext_comp}, where their very diffuse nature is readily apparent.
Another three extended clusters are known in NGC\,6822 which lie beyond the \Euclid FoV \citep{Hwang2011, Huxor2013}.  A possible explanation for the origin of extended clusters is that they have been born extended and evolved in a weak tidal field \citep{Hurley2010, Bianchini2014bk}, consistent with their existence in the outskirts of a dwarf galaxy such as NGC\,6822. The fact that extended clusters are also seen in the halos of massive galaxies \citep[e.g.][]{Huxor2005,2010MNRAS4041157M,Jang2012} provides tantalising evidence that they have been donated via dwarf galaxy accretion events.

We have also searched for correlations between mass and size, and age and size within our sample.  A modest but statistically significant correlation between cluster mass and size exists in NGC\,6822, but not in IC\,10. In contrast, no clear relationship is found between cluster age and size in either galaxy. While some studies have seen evidence for such trends \citep[e.g.][]{Lee2005, Chandar2016, Mackey2003, Ryon2015, Bastian2012, Scheepmaker2007}, others do not \citep[e.g.][]{Larsen2004, Barmby2009}. Overall, our findings suggest that while mass may influence cluster size to some extent, age does not appear to be a strong driver. 

\subsection{Cluster metallicities, extinctions, and ages} 
\begin{figure}
    \centering
    \includegraphics[width=\columnwidth]{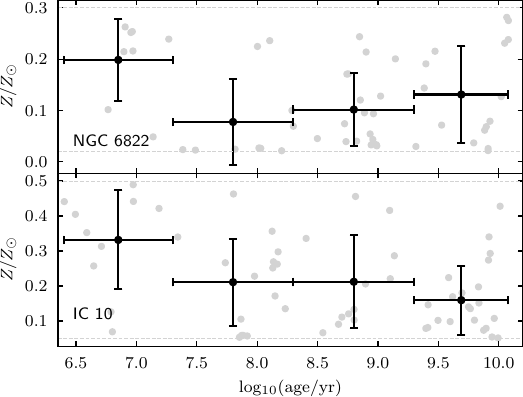}
    \caption{Age–metallicity relation based on \texttt{BAGPIPES} outputs for NGC\,6822 (top) and IC\,10 (bottom). Individual measurements are shown in light grey, with black points representing averages computed in bins of age. Horizontal bars represent bin widths, while vertical error bars show the standard deviation within each bin. Dashed grey lines mark the boundaries of the metallicity priors.  }
    \label{fig:AMR}
\end{figure}

\begin{figure}
    \centering
    \includegraphics[width=\columnwidth]{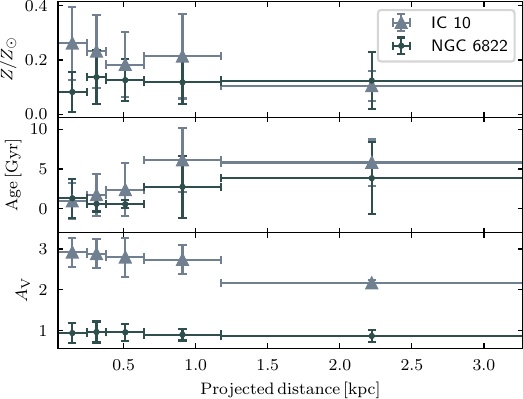}
    \caption{Top: Average cluster metallicities (as output by  \texttt{BAGPIPES}) computed in bins of projected distance from the centre of each galaxy. Horizontal bars represent bin widths, while vertical error bars show the standard deviation within each bin. Middle: The same as the top panel but for age. Bottom: The same but for $A_V$. }
    \label{fig:radial_plots}
\end{figure}

Figure~\ref{fig:AMR} shows how the metallicity of the clusters varies as a function of age, using the \texttt{BAGPIPES} outputs. In addition to the individual cluster values (light grey), we also show the mean metallicity in specific age bins (black points), along with its standard deviation.  The age bins correspond to our previous definitions of young, intermediate, and old populations, with the intermediate-age category further divided into two bins.  Both galaxies show the expected behaviour, with the most recently formed clusters having the highest metallicities and older clusters having lower metallicities. Beyond the first age bin, the age-metallicity relation is flat to within the uncertainties in both systems, with metallicities at a given age being slightly higher in IC\,10 than in NGC\,6822.  Within any age bin, there is significant scatter when examining individual clusters, only some of which is likely to be real. Indeed, the plots also hint at the existence of potential systematics: a preference for low-metallicity fits reaching the floor of the prior in the second age bin, as well as some unexpectedly high-metallicity fits in the oldest age bins. In the latter case, we checked that these clusters 
did not show a correlation with far-IR dust emission (by inspecting SPIRE 500~${\micron}$ and {\it Spitzer} 160~${\micron}$ images), or with strong $\rm H\alpha$ sources; this confirmed that their SEDs were not contaminated by non-stellar emission. One possible explanation for the high metallicities could be due to our extinction priors. These were set by considering the range of reddening values in the published literature; however, these may not capture the full range of reddening across the \Euclid FoV. The \citet{schlegel1998} dust maps, recalibrated by \citet{schlafly2011}, suggest higher $E(B-V)$ values in some places than our adopted limits. If the true extinction is higher than our allowed range, the SED fitting may only be able to reproduce the red colours by favouring older ages and higher metallicities. We explored this by assessing the impact of changing the dust prior in the \texttt{BAGPIPES} fitting process. First, we removed the upper limit on the dust prior and found the old, high-metallicity clusters still broadly occupied the same region. We then replaced the global dust prior for the entire galaxy with a local one specific to each cluster. For this, we exploited the high-resolution reddening map derived from resolved RGB stars (Annibali et al., in prep), who inferred $E(B-V)$ by comparing the CMD positions of stars to a fiducial RGB sequence. Setting the prior for each cluster as a gaussian of width 0.2 centered on the local $E(B-V)$, we redid the fits but found very little change to the results, with the old high-metallicity fits persisting. Alternatively, stochastic effects could be responsible for assigning old ages and high metallicities to these clusters (Appendix~\ref{stoch_effects}).  While Fig.\ref{fig:AMR} illustrates the qualitative trends in the metallicity evolution of the galaxies, a more detailed reconstruction of the age-metallicity relationship of the star clusters would require either CMD fitting or spectroscopy. 

We also searched for radial gradients in age, metallicity, and extinction in the star cluster population of each system, binning the clusters by their projected radial distance from the centre of each galaxy and computing the mean of the quantity of interest.  To facilitate comparison, the radial bins were constructed to have common edges and widths such that there was always a minimum of five clusters per bin. Figure~\ref{fig:radial_plots} shows that there are clear negative radial gradients in metallicity and extinction within IC\,10 but not NGC\,6822. The variation in IC\,10 confirms the presence of significant amounts of internal extinction in that system.  Both systems show positive radial gradients with age, whereby the cluster population becomes increasingly older further out, as expected.  

\subsection{Globular cluster scaling relations}
\label{gcs}
\begin{figure}
    \centering
    \includegraphics[width=\columnwidth]{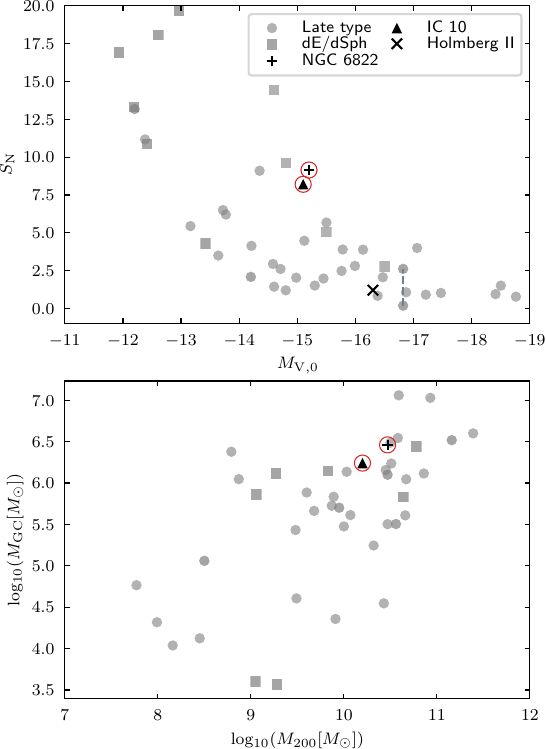}
    \caption{Top: The GC specific frequency $S_{\rm N}$ versus absolute {\it V}-band magnitude $M_{V} $ for nearby dwarf galaxies. NGC\,6822 and IC\,10 are shown by plus and triangle symbols enclosed in red circles. Each galaxy's $N_{\rm GC}$ is calculated from the number of clusters that are class 1 or 2 with ages larger than 5\,Gyr and $Z/Z_{\odot} \leq 0.2$. The grey filled circles and squares represent a sample of LG and late-type galaxies taken from \citet{Forbes2018}.    Bottom: The GC system mass-halo mass relation, where the halo mass is represented by $M_{200}$. The symbols are the same as in the top panel.}
    \label{fig:scaling_relations}
\end{figure}

The GC populations of dwarf galaxies are of particular interest given their likely role in building up the halos of more massive systems. Furthermore, a remarkable diversity in the GC populations of low-mass galaxies has been uncovered in recent years \citep[e.g.][]{Eadie2022,Jones2023, Saifollahi2024} and it is interesting to explore how NGC\,6822 and IC\,10 compare. 

Unfortunately, there is no universally accepted definition of what constitutes a GC \citep[see][]{2019AARv278G}. In the MW and its satellites, GCs are generally identified as ancient halo, bulge and disc star clusters.  On the other hand, accurate age information is rarely available in extragalactic studies, where we also only have knowledge of projected positions. Instead, extragalactic GC candidates are generally identified on the basis of magnitude, colour, and size information, using stellar population models to select clusters older than a few gigayears \citep[e.g.][]{carlsten2022elves-c85, floyd2024phangs-11d,Lim2025}. 
To define a sample of GCs that bridges both the MW and extragalactic conventions, we use the results of our visual classification along with the output of \texttt{BAGPIPES}. Specifically, we adopt the following criteria:

\begin{itemize}
  \item $\mathrm{Class} =$ 1 or 2;  
  \item $\mathrm{Age} \geq 5 \, \mathrm{Gyr}$;
  \item $Z/Z_{\odot} \leq 0.2$.
\end{itemize}

In dwarf galaxies, GCs are typically metal poor \citep[e.g.][]{Larsen2022}, consistent with the mass-metallicity relation. For galaxies with stellar masses like NGC\,6822 and IC\,10, simulations suggest that star clusters formed 5 Gyr ago would be expected to have $\textrm{[Fe/H]} \lesssim -1$ \citep{Kruijssen2019,Ma2016} and so we apply a generous cut to exclude more metal-rich clusters.  There is no well-established age criterion for a GC. In the MW, most GCs are $\geq 10$~Gyr old, but there are also halo star clusters that are several gigayears younger \citep[e.g.][]{Weisz2016}.  Along with the metallicity cut, the age cut of 5~Gyr is designed to capture all long-lived metal-poor star clusters without being overly restrictive. We note that our selection criteria would retain all but one of the six GCs in the Fornax dSph galaxy.  The excluded object would be Fornax 6, which has a spectroscopically derived metallicity of $[\mathrm{Fe/H}] = -0.71 \pm 0.05$ and an estimated age of $\sim 2$ Gyr from Gaia photometry and MIST isochrone fitting \citep{Pace2021}. Given its young age and relatively high metallicity, Fornax 6 is arguably not what is typically considered to be a GC. 
 
Figure~\ref{fig:age_vs_mass} shows that the age cut adopted guarantees that all clusters are more massive than a few times $10^4\,M_{\odot}$. With these criteria, we arrive at 6 GCs in NGC\,6822, and 9 GCs in IC\,10.  Of the star clusters previously identified as NGC\,6822 GCs, Hubble-VII, SC6, and 7 fall within our definition but SC3 does not. The SED-fitting routine assigns a rather young age (2.5\,Gyr) and relatively high metallicity ($Z/Z_{\odot} = 0.19$) to this cluster, which is the faintest of the known GCs in NGC\,6822. Based on a previous spectroscopic analysis of this object \citep[see Table~\ref{Table:Comparing_6822_mets}]{Hwang2014}, as well as analysis of a deep HST CMD (McGill et al., in prep), we believe the \texttt{BAGPIPES} fit is unreliable and that the cluster is older and more metal poor; we therefore include SC3 in the total number of NGC\,6822 GCs. We further add in SC5, for which we could not perform SED-fitting due to lack of $UBVRI$ coverage, as well as SC1, SC2, and SC4, all of which lie outside the \Euclid FOV.  This brings the total number of GCs in NGC\,6822 to 11.  The resulting GC candidates are flagged in the thumbnails shown in Appendix~\ref{appendix_thumbnails} and the spatial distribution of those that fall within the \Euclid FoV are displayed as black points in Fig.\,\ref{fig:spatial_plots}. In both galaxies, the GCs exhibit extended distributions -- in NGC\,6822, 7 of the 11 GCs lie outside the galaxy half-light radius (indicated by dashed grey ellipse) while all 9 of the GCs in IC\,10 lie outside of it.  Interestingly, NGC\,6822's GCs appear to be preferentially located on the eastern side of the galaxy (this remains true even when considering the GCs outwith the \Euclid footprint).  
The most obvious object for comparison is the SMC which, due to its proximity, has many cluster age and metallicity determinations from CMD fitting \citep[e.g.][]{Glatt2008}. The SMC is generally considered to have only one old GC, NGC\,121 \citep[e.g.][]{Forbes2018}\footnote{We note that two faint GC candidates have recently been discovered at large radius in the SMC but their true nature is not yet clear \citep{cerny2021,cerny2023delve-f41}.}. We use the \citet{Bica2020} catalogue to search for high-confidence star clusters (object class ``C") that satisfy our metallicity and age criteria.  We find 14 SMC clusters fall within our definition of GC, which is slightly larger than the numbers found for NGC\,6822 and IC\,10.  

The specific frequency of GCs, defined as the number of GCs per unit of galaxy luminosity, was introduced to measure the richness of GC systems in galaxies \citep{Harris1981} and remains commonly used today.   It is calculated as ${S_{\rm N} = N_{\rm GC} \;10^{0.4(M_{V}  + 15)}}$, where $M_{V} $ is the absolute {\it V}-band magnitude of the host galaxy and $N_{\rm GC}$ is the number of GCs. The $S_{\rm N}$ values are impacted by the galaxy's recent star formation history, affecting its $M_{V} $, and the dynamical evolution history of its GCs, affecting its $N_{\rm GC}$.  Using the galaxy absolute magnitudes from Table~\ref{table:gal_properties}, and  $N_{\rm GC}$ of 11 and 9, we find that $S_{\rm N}$ is $9.1 \pm 3.2$ and $8.2 \pm 3.3$  in NGC\,6822 and IC\,10, respectively. The uncertainty in $S_{\rm N}$ reflects both Poisson noise in the GC counts and the reported uncertainty in $M_{V}$. These values are displayed in the top panel of Fig.\,\ref{fig:scaling_relations}, plotted against $M_{V}$.  This plot also shows $S_{\rm N}$ and $M_{V}$ values for the sample of LG and isolated late-type dwarf galaxies studied by  \citet{Forbes2018}, which includes some of the systems originally presented in \citet{Georgiev2010}. We add further datapoints from the recent studies by \citealt{Karim2024} (IC 2574) and \citealt{Larsen25} (Holmberg\,II), and we include $S_{\rm N}$ values for the SMC calculated assuming both one and 14 GCs (connected via a dashed line).  We see that NGC\,6822 and IC\,10 have high $S_{\rm N}$ values for their luminosity, and lie at the high end of the distribution when compared to the sample of dwarfs as a whole. This is particularly true when considering that this plot only shows dwarf galaxies that have at least one identified GC; many others show none at all. Of the three late-type galaxies with higher $S_{\rm N}$ values, one of them is the LG transition dwarf galaxy Pegasus \citep{Cole2017} while the other two are more distant systems \citep[][KK\,16 and UGC\,685]{Georgiev2010}.  

The bottom panel of Fig.\,\ref{fig:scaling_relations} displays the GC system mass-halo mass relation. For NGC\,6822 and IC\,10, we calculate the total GC system mass by summing the individual cluster masses determined in this work. For clusters SC1, SC2, SC4, and SC5 in NGC\,6822, which lie outside the data coverage for SED fitting, we use luminosities from \citet{Veljanoski2015} and convert them to masses assuming a mass-to-light ratio of 1.88, consistent with the value adopted by \citet{Forbes2018}. We do the same for SC3, as the \texttt{BAGPIPES} fit returned a relatively young age for a globular cluster (2.5\, Gyr). For IC\,10, we adopt the total halo mass from \citet{Oh2015}, while for NGC\,6822 we use the halo mass reported by \citet{Forbes2018}.  For the other galaxies shown, all values are taken directly from \citet{Forbes2018}.  While there is much scatter in this scaling relationship at the low-mass end, the positions of NGC\,6822 and IC\,10 are consistent with expectations for galaxies of their halo mass.  

\section{Conclusions}
We presented a detailed and homogeneous analysis of the star cluster systems in two LG dwarf galaxies, NGC\,6822 and IC\,10. We used images from the \Euclid ERO programme to conduct a blind search for star clusters across the main bodies and into the remote halos of both systems. We classified star clusters into various confidence classes according to their appearance in \IE images, the highest spatial resolution imagery provided by \Euclid. We undertook an extensive literature search to compile all previously identified star cluster candidates in these galaxies, and assigned classifications to them, too, based on \IE appearance. Each newly discovered or literature cluster was assigned to class 1, 2, 3, or 4, going from most likely to least likely. Two additional categories are used for literature candidates:  class 5 indicates a source that does not resemble a cluster on \Euclid \IE images, while class 6 denotes objects that are definitely not clusters in \Euclid \IE images. Using archival LGGS $UBVRI$ imagery in combination with \Euclid \IE, \YE, \JE, and \HE data, we derive homogeneous 9-band integrated photometry for our cluster sample, as well as size measurements. Furthermore, we used the \texttt{BAGPIPES} SED-fitting code to determine the age, mass, metallicity, and line-of-sight extinction to each cluster.   Through injecting a range of artificial star clusters into the \Euclid \IE images, we conclude that our sample is ${\sim}50\%$ complete to $M \lesssim 10^3~{M}_{\odot}$ for ages $\lesssim100\,\rm{Myr}$, and to $M \lesssim 2\times10^4~{M}_{\odot}$ for ages of ${\sim}10\,\rm{Gyr}$, in both galaxies. 

Our star cluster study is unique in terms of its fidelity (with clusters being largely resolved and hence of high confidence), its probing nearly the full spatial extents of the two dwarfs, it being sensitive to both low- and high-mass clusters with a large variety of ages, and its systematic approach to deriving quantities of interest. It demonstrates how the high spatial resolution and NIR capabilities of \Euclid can be effectively combined with ground-based optical photometry to enable comprehensive local volume star cluster studies  -- studies of this type will be increasingly common in the near future when Euclid Wide Survey data can be combined with the data from the Legacy Survey of Space and Time \citep{DDP2022, Usher2023}.  The catalogues we release are intended to support a wide range of future investigations, including comparisons with resolved star CMD analyses. In this work, we focus on just a few specific topics.

Our primary results are listed below. 
\begin{enumerate}
\item We identified 30 new clusters (classes one to four) in NGC\,6822 and 16 in IC\,10, taking the total candidates in these systems within our FoV to 52 and 71, respectively, when including those previously reported. An additional three previously known clusters in NGC\,6822 lie outside our field of view. Additionally, we reclassified 21 and 32 previously identified candidates as either class 5 or 6.
\item The clusters identified have masses spanning from $10^{1.7}$ to $10^{6.1}\,M_\odot$, and span ages from 10~Myr to 10~Gyr. IC\,10 has more young clusters than NGC\,6822, and its young clusters also reach higher masses ($\gtrsim10^4\,M_{\odot}$), consistent with this galaxy having a higher truncation mass on account of its higher SFR surface density.  We found several examples of old massive ($\gtrsim10^5\,M_{\odot}$) clusters in both dwarfs that could be the descendants of the young star clusters recently uncovered in low-mass galaxies at high redshift \citep[e.g.][]{Mowla2024, Adamo2024}.  We highlighted the existence of a particularly exceptional object in the outskirts of NGC\,6822, SC7, for which we made the first mass determination of $10^{6.1}\,M_\odot$, rivalling the most massive GC in the MW. Combined with its high ellipticity and moderately high metallicity, this object bears many hallmarks of being the accreted core of a now-defunct smaller system.  
\item We examined the spatial distribution of the star clusters as a function of age in each system. In NGC\,6822,  the young clusters trace out an elongated N-S structure aligned with the bar, whereas the intermediate-age and old clusters have increasingly circular distributions.  These strongly age-dependent spatial distributions are also seen in IC\,10, with its young clusters primarily elongated in a northeast–southwest direction.  
\item The size-magnitude distribution of the old star clusters is broadly consistent with that of the MW as well as that of GCs in nearby late-type galaxies.  In NGC\,6822, we observed an increase in the spread of half-light radii at faint magnitudes that is not seen in IC\,10. Previous work has demonstrated that NGC\,6822 hosts a population of extended GCs, and we identified a new example in this work, ESCC-NGC6822-27, which has 
$R_{\mathrm{h}} = 12.4 \pm 0.11\,\mathrm{pc}$, an age of $2.1 \pm 0.3$\,Gyr, a mass of $\mathrm{log_{10}}(M/{M}_{\odot}) = 4.4 \pm 0.04$, and a metallicity of $Z/Z_{\odot}=0.03 \pm 0.01$.
\item We constructed age-metallicity relationships for both systems and observed the expected behaviour of the most recently formed clusters having the highest metallicities and the older clusters having lower metallicities. Beyond $\sim10^8$~yr, the age-metallicity relation is essentially flat to within the uncertainties in both systems, with metallicities at a given age being higher in IC\,10 than in NGC\,6822.   The age-metallicity relationships revealed a few hints of systematics in the SED-fitting outputs, and a more detailed examination will require either CMD fitting or spectroscopy.  The IC\,10 cluster population exhibits clear negative radial gradients in metallicity and extinction but NGC\,6822's does not. Both systems show positive radial gradients with age, whereby the cluster population becomes increasingly older further out, as expected.  
\item Lastly, we used well-defined criteria to select a subsample of clusters as candidate GCs; this yields 11 objects in NGC\,6822 (folding in objects from previous studies) and 9 in IC\,10. The GCs are more broadly distributed than other star clusters, with 7 of the 11 GCs lying outside the galaxy half-light radius in NGC\,6822 and all 9 of the GCs in IC\,10 lying outside of it.  We placed the systems on the well-known GC scaling relations and found that both galaxies have high $S_{\rm N}$ values for their luminosity, and lie at the high end of the distribution when compared to the population of dwarfs as a whole.  We note that although the new star clusters uncovered in this work have increased the GC count in both galaxies, more GCs may still await detection.  Indeed, the \emph{Euclid} ERO images do not cover the entire halos of these systems, where more GCs may reside (including 3 previously known GCs in NGC\,6822). This highlights the need for systematic, high-sensitivity, wide-field surveys of dwarf galaxies to obtain an accurate census of their GC populations.

\end{enumerate}

\section*{Data Availability}
Tables~\ref{table:ngc6822_search_results},~\ref{table:ic10_search_results},~\ref{Table:6822_phot},~\ref{Table:ic10_phot},~\ref{Table:6822_SED} and~\ref{Table:ic10_SED} are available in electronic form at the CDS via anonymous ftp to cdsarc.u-strasbg.fr (130.79.128.5) or via \url{http://cdsweb.u-strasbg.fr/cgi-bin/qcat?J/A+A/}.

\begin{acknowledgements}
JMH acknowledges
funding from the Bell Burnell Graduate Scholarship Fund (BB0015). AMNF is supported by UK Research and Innovation (UKRI) under the UK government’s Horizon Europe funding guarantee [grant number EP/Z534353/1] and by the Science and Technology Facilities
Council [grant number ST/Y001281/1]. We thank Philip Massey for helpful clarifications on the LGGS data, and Adam Carnall for sharing insights and discussion on SED fitting with \texttt{BAGPIPES}. The Euclid Consortium acknowledges the European Space
Agency and a number of agencies and institutes that have supported the development of \Euclid, in particular the Agenzia Spaziale Italiana, the Austrian Forschungsförderungsgesellschaft funded through BMK, the Belgian Science Policy, the Canadian Euclid Consortium, the Deutsches Zentrum für Luft1245 und Raumfahrt, the DTU Space and the Niels Bohr Institute in Denmark, the
French Centre National d’Etudes Spatiales, the Fundação para a Ciência e a
Tecnologia, the Hungarian Academy of Sciences, the Ministerio de Ciencia,
Innovación y Universidades, the National Aeronautics and Space Administration, the National Astronomical Observatory of Japan, the Netherlandse On1250 derzoekschool Voor Astronomie, the Norwegian Space Agency, the Research
Council of Finland, the Romanian Space Agency, the State Secretariat for Education, Research, and Innovation (SERI) at the Swiss Space Office (SSO), and the United Kingdom Space Agency. A complete and detailed list is available
on the Euclid web site (\url{www.euclid-ec.org}). This work has made use of the Early Release Observations (ERO) data from the Euclid mission of the European Space Agency (ESA),
2024, \url{https://doi.org/10.57780/esa-qmocze3}, and the $\rm H\alpha$ and UBVRI images obtained as part of the Survey of Local Group Galaxies Currently Forming Stars.  This research made use of the following software, packages and \texttt{python} libraries: SVO Filter Profile Service \enquote{Carlos Rodrigo}, \texttt{TOPCAT} (\url{https://www.star.bristol.ac.uk/mbt/topcat}), \texttt{SAOImage DS9} (\url{https://ds9.si.edu}), \texttt{NUMPY} (\citealp{numpy}), \texttt{SCIPY} (\citealp{scipy}), \texttt{ASTROPY} (\citealp{astropy}).
\end{acknowledgements}
\bibliography{bibtex,Euclid}

@ARTICLE{EuclidSkyOverview,
author = {{Euclid Collaboration: Mellier}, Y. and {Abdurro'uf} and {Acevedo~Barroso}, J.A. and others},
	title = {Euclid - I. Overview of the Euclid mission},
	DOI= "10.1051/0004-6361/202450810",
	url= "https://doi.org/10.1051/0004-6361/202450810",
	journal = {A\&A},
	year = 2025,
	volume = 697,
	pages = "A1",
}

@ARTICLE{EuclidSkyVIS,
author = {{Euclid Collaboration: Cropper}, M. and {Al-Bahlawan}, A. and {Amiaux}, J. and others},
	title = {Euclid - II. The VIS instrument},
	DOI= "10.1051/0004-6361/202450996",
	url= "https://doi.org/10.1051/0004-6361/202450996",
	journal = {A\&A},
	year = 2025,
	volume = 697,
	pages = "A2",
}

@ARTICLE{EuclidSkyNISP,
author = {{Euclid Collaboration: Jahnke}, K. and {Gillard}, W. and {Schirmer}, M. and others},
	title = {Euclid - III. The NISP Instrument},
	DOI= "10.1051/0004-6361/202450786",
	url= "https://doi.org/10.1051/0004-6361/202450786",
	journal = {A\&A},
	year = 2025,
	volume = 697,
	pages = "A3",
}

@misc{EROcite,
author = "{Euclid Early Release Observations}",
howpublished = "\url{https://doi.org/10.57780/esa-qmocze3}",
year = 2024
}

@ARTICLE{Scaramella-EP1,
       author = {{Euclid Collaboration: Scaramella}, R. and {Amiaux}, J. and {Mellier}, Y. and others},
        title = "{Euclid preparation. I. The Euclid Wide Survey}",
      journal = {\aap},
         year = 2022,
        month = jun,
       volume = {662},
          eid = {A112},
        pages = {A112},
          doi = {10.1051/0004-6361/202141938},
archivePrefix = {arXiv},
       eprint = {2108.01201},
 primaryClass = {astro-ph.CO},
       adsurl = {https://ui.adsabs.harvard.edu/abs/2022A&A...662A.112E}
}

@ARTICLE{Larsen25,
       author = {{Larsen}, S.~S. and {Ferguson}, A.~M.~N. and {Howell}, J.~M. and others},
       title = "{Euclid: Star clusters in IC 342, NGC 2403, and Holmberg II}",
      journal = {A\&A, submitted},
         year = 2025,
        month = mar,
          eid = {arXiv:2503.16637},
        pages = {arXiv:2503.16637},
          doi = {10.48550/arXiv.2503.16637},
archivePrefix = {arXiv},
       eprint = {2503.16637},
 primaryClass = {astro-ph.GA},
       adsurl = {https://ui.adsabs.harvard.edu/abs/2025arXiv250316637L}
}

@ARTICLE{Conroy2009x,
       author = {{Conroy}, Charlie and {Gunn}, James E. and {White}, Martin},
        title = "{The Propagation of Uncertainties in Stellar Population Synthesis Modeling. I. The Relevance of Uncertain Aspects of Stellar Evolution and the Initial Mass Function to the Derived Physical Properties of Galaxies}",
      journal = {\apj},
         year = 2009,
        month = jul,
       volume = {699},
       number = {1},
        pages = {486-506},
          doi = {10.1088/0004-637X/699/1/486},
archivePrefix = {arXiv},
       eprint = {0809.4261},
 primaryClass = {astro-ph},
       adsurl = {https://ui.adsabs.harvard.edu/abs/2009ApJ...699..486C}
}

@ARTICLE{Conroy2010x,
       author = {{Conroy}, Charlie and {Gunn}, James E.},
        title = "{The Propagation of Uncertainties in Stellar Population Synthesis Modeling. III. Model Calibration, Comparison, and Evaluation}",
      journal = {\apj},
         year = 2010,
        month = apr,
       volume = {712},
       number = {2},
        pages = {833-857},
          doi = {10.1088/0004-637X/712/2/833},
archivePrefix = {arXiv},
       eprint = {0911.3151},
 primaryClass = {astro-ph.CO},
       adsurl = {https://ui.adsabs.harvard.edu/abs/2010ApJ...712..833C}
}

@ARTICLE{Perren2015x,
       author = {{Perren}, G.~I. and {V{\'a}zquez}, R.~A. and {Piatti}, A.~E.},
        title = "{ASteCA: Automated Stellar Cluster Analysis}",
      journal = {\aap},
         year = 2015,
        month = apr,
       volume = {576},
          eid = {A6},
        pages = {A6},
          doi = {10.1051/0004-6361/201424946},
archivePrefix = {arXiv},
       eprint = {1412.2366},
 primaryClass = {astro-ph.GA},
       adsurl = {https://ui.adsabs.harvard.edu/abs/2015A&A...576A...6P}
}

@ARTICLE{Brodie2006,
       author = {{Brodie}, Jean P. and {Strader}, Jay},
        title = "{Extragalactic Globular Clusters and Galaxy Formation}",
      journal = {\araa},
         year = 2006,
        month = sep,
       volume = {44},
       number = {1},
        pages = {193-267},
          doi = {10.1146/annurev.astro.44.051905.092441},
archivePrefix = {arXiv},
       eprint = {astro-ph/0602601},
 primaryClass = {astro-ph},
       adsurl = {https://ui.adsabs.harvard.edu/abs/2006ARA&A..44..193B}
}

@ARTICLE{Renzini1988,
       author = {{Renzini}, Alvio and {Fusi Pecci}, Flavio},
        title = "{Tests of evolutionary sequences using color-magnitude diagrams of globular clusters.}",
      journal = {\araa},
         year = 1988,
        month = jan,
       volume = {26},
        pages = {199-244},
          doi = {10.1146/annurev.aa.26.090188.001215},
       adsurl = {https://ui.adsabs.harvard.edu/abs/1988ARA&A..26..199R}
}

@ARTICLE{Portegies2010,
       author = {{Portegies Zwart}, Simon F. and {McMillan}, Stephen L.~W. and {Gieles}, Mark},
        title = "{Young Massive Star Clusters}",
      journal = {\araa},
         year = 2010,
        month = sep,
       volume = {48},
        pages = {431-493},
          doi = {10.1146/annurev-astro-081309-130834},
archivePrefix = {arXiv},
       eprint = {1002.1961},
 primaryClass = {astro-ph.GA},
       adsurl = {https://ui.adsabs.harvard.edu/abs/2010ARA&A..48..431P}
}

@ARTICLE{Hunter2001,
       author = {{Hunter}, Deidre A.},
        title = "{The Stellar Population and Star Clusters in the Unusual Local Group Galaxy IC 10}",
      journal = {\apj},
         year = 2001,
        month = sep,
       volume = {559},
       number = {1},
        pages = {225-242},
          doi = {10.1086/322399},
archivePrefix = {arXiv},
       eprint = {astro-ph/0105456},
 primaryClass = {astro-ph},
       adsurl = {https://ui.adsabs.harvard.edu/abs/2001ApJ...559..225H}
}

@ARTICLE{Kirby2013,
       author = {{Kirby}, Evan N. and {Cohen}, Judith G. and {Guhathakurta}, Puragra and {Cheng}, Lucy and {Bullock}, James S. and {Gallazzi}, Anna},
        title = "{The Universal Stellar Mass-Stellar Metallicity Relation for Dwarf Galaxies}",
      journal = {\apj},
         year = 2013,
        month = dec,
       volume = {779},
       number = {2},
          eid = {102},
        pages = {102},
          doi = {10.1088/0004-637X/779/2/102},
archivePrefix = {arXiv},
       eprint = {1310.0814},
 primaryClass = {astro-ph.GA},
       adsurl = {https://ui.adsabs.harvard.edu/abs/2013ApJ...779..102K}
}

@ARTICLE{DellAgli2018,
       author = {{Dell'Agli}, F. and {Di Criscienzo}, M. and {Ventura}, P. and {Limongi}, M. and {Garc{\'\i}a-Hern{\'a}ndez}, D.~A. and {Marini}, E. and {Rossi}, C.},
        title = "{Evolved stars in the Local Group galaxies - II. AGB, RSG stars, and dust production in IC10}",
      journal = {\mnras},
         year = 2018,
        month = oct,
       volume = {479},
       number = {4},
        pages = {5035-5048},
          doi = {10.1093/mnras/sty1614},
archivePrefix = {arXiv},
       eprint = {1806.04160},
 primaryClass = {astro-ph.SR},
       adsurl = {https://ui.adsabs.harvard.edu/abs/2018MNRAS.479.5035D}
}

@ARTICLE{Kim2009,
       author = {{Kim}, Minsun and {Kim}, Eunhyeuk and {Hwang}, Narae and {Lee}, Myung Gyoon and {Im}, Myungshin and {Karoji}, Hiroshi and {Noumaru}, Junichi and {Tanaka}, Ichi},
        title = "{Reddening and Distance of the Local Group Starburst Galaxy IC 10}",
      journal = {\apj},
         year = 2009,
        month = sep,
       volume = {703},
       number = {1},
        pages = {816-828},
          doi = {10.1088/0004-637X/703/1/816},
archivePrefix = {arXiv},
       eprint = {0907.4844},
 primaryClass = {astro-ph.CO},
       adsurl = {https://ui.adsabs.harvard.edu/abs/2009ApJ...703..816K}
}

@ARTICLE{Richer2001,
       author = {{Richer}, M.~G. and {Bullejos}, A. and {Borissova}, J. and {McCall}, M.~L. and {Lee}, H. and {Kurtev}, R. and {Georgiev}, L. and {Kingsburgh}, R.~L. and {Ross}, R. and {Rosado}, M.},
        title = "{IC 10: More evidence that it is a blue compact dwarf}",
      journal = {\aap},
         year = 2001,
        month = apr,
       volume = {370},
        pages = {34-42},
          doi = {10.1051/0004-6361:20010206},
archivePrefix = {arXiv},
       eprint = {astro-ph/0103066},
 primaryClass = {astro-ph},
       adsurl = {https://ui.adsabs.harvard.edu/abs/2001A&A...370...34R}
}

@ARTICLE{Demers2004,
       author = {{Demers}, S. and {Battinelli}, P. and {Letarte}, B.},
        title = "{A Carbon star approach to IC 10: Distance and correct size}",
      journal = {\aap},
         year = 2004,
        month = sep,
       volume = {424},
        pages = {125-132},
          doi = {10.1051/0004-6361:20040552},
       adsurl = {https://ui.adsabs.harvard.edu/abs/2004A&A...424..125D}
}

@ARTICLE{Sanna2010,
       author = {{Sanna}, N. and {Bono}, G. and {Stetson}, P.~B. and {Ferraro}, I. and {Monelli}, M. and {Nonino}, M. and {Prada Moroni}, P.~G. and {Bresolin}, R. and {Buonanno}, R. and {Caputo}, F. and {Cignoni}, M. and {Degl'Innocenti}, S. and {Iannicola}, G. and {Matsunaga}, N. and {Pietrinferni}, A. and {Romaniello}, M. and {Storm}, J. and {Walker}, A.~R.},
        title = "{On the Radial Extent of the Dwarf Irregular Galaxy IC10}",
      journal = {\apjl},
         year = 2010,
        month = oct,
       volume = {722},
       number = {2},
        pages = {L244-L249},
          doi = {10.1088/2041-8205/722/2/L244},
archivePrefix = {arXiv},
       eprint = {1009.3917},
 primaryClass = {astro-ph.SR},
       adsurl = {https://ui.adsabs.harvard.edu/abs/2010ApJ...722L.244S}
}

@ARTICLE{Gerbrandt2015,
       author = {{Gerbrandt}, Stephanie A.~N. and {McConnachie}, Alan W. and {Irwin}, Mike},
        title = "{The red extended structure of IC 10, the nearest blue compact galaxy}",
      journal = {\mnras},
         year = 2015,
        month = nov,
       volume = {454},
       number = {1},
        pages = {1000-1011},
          doi = {10.1093/mnras/stv2029},
archivePrefix = {arXiv},
       eprint = {1509.05436},
 primaryClass = {astro-ph.GA},
       adsurl = {https://ui.adsabs.harvard.edu/abs/2015MNRAS.454.1000G}
}

@ARTICLE{Gieren2006,
       author = {{Gieren}, Wolfgang and {Pietrzy{\'n}ski}, Grzegorz and {Nalewajko}, Krzysztof and {Soszy{\'n}ski}, Igor and {Bresolin}, Fabio and {Kudritzki}, Rolf-Peter and {Minniti}, Dante and {Romanowsky}, Aaron},
        title = "{The Araucaria Project: An Accurate Distance to the Local Group Galaxy NGC 6822 from Near-Infrared Photometry of Cepheid Variables}",
      journal = {\apj},
         year = 2006,
        month = aug,
       volume = {647},
       number = {2},
        pages = {1056-1064},
          doi = {10.1086/505574},
archivePrefix = {arXiv},
       eprint = {astro-ph/0605231},
 primaryClass = {astro-ph},
       adsurl = {https://ui.adsabs.harvard.edu/abs/2006ApJ...647.1056G}
}

@ARTICLE{Efremova2011,
       author = {{Efremova}, Boryana V. and {Bianchi}, Luciana and {Thilker}, David A. and {Neill}, James D. and {Burgarella}, Denis and {Wyder}, Ted K. and {Madore}, Barry F. and {Rey}, Soo-Chang and {Barlow}, Tom A. and {Conrow}, Tim and {Forster}, Karl and {Friedman}, Peter G. and {Martin}, D. Christopher and {Morrissey}, Patrick and {Neff}, Susan G. and {Schiminovich}, David and {Seibert}, Mark and {Small}, Todd},
        title = "{The Recent Star Formation in NGC 6822: An Ultraviolet Study}",
      journal = {\apj},
         year = 2011,
        month = apr,
       volume = {730},
       number = {2},
          eid = {88},
        pages = {88},
          doi = {10.1088/0004-637X/730/2/88},
archivePrefix = {arXiv},
       eprint = {1101.6051},
 primaryClass = {astro-ph.GA},
       adsurl = {https://ui.adsabs.harvard.edu/abs/2011ApJ...730...88E}
}

@ARTICLE{Tantalo2022,
       author = {{Tantalo}, Maria and {Dall'Ora}, Massimo and {Bono}, Giuseppe and {Stetson}, Peter B. and {Fabrizio}, Michele and {Ferraro}, Ivan and {Nonino}, Mario and {Braga}, Vittorio F. and {da Silva}, Ronaldo and {Fiorentino}, Giuliana and {Iannicola}, Giacinto and {Marengo}, Massimo and {Monelli}, Matteo and {Mullen}, Joseph P. and {Pietrinferni}, Adriano and {Salaris}, Maurizio},
        title = "{On the Dwarf Irregular Galaxy NGC 6822. I. Young, Intermediate, and Old Stellar Populations}",
      journal = {\apj},
         year = 2022,
        month = jul,
       volume = {933},
       number = {2},
          eid = {197},
        pages = {197},
          doi = {10.3847/1538-4357/ac7468},
archivePrefix = {arXiv},
       eprint = {2205.15143},
 primaryClass = {astro-ph.GA},
       adsurl = {https://ui.adsabs.harvard.edu/abs/2022ApJ...933..197T}
}

@ARTICLE{Battinelli2006,
       author = {{Battinelli}, P. and {Demers}, S. and {Kunkel}, W.~E.},
        title = "{Photometric survey of the polar ring galaxy NGC 6822}",
      journal = {\aap},
         year = 2006,
        month = may,
       volume = {451},
       number = {1},
        pages = {99-108},
          doi = {10.1051/0004-6361:20054718},
archivePrefix = {arXiv},
       eprint = {astro-ph/0603558},
 primaryClass = {astro-ph},
       adsurl = {https://ui.adsabs.harvard.edu/abs/2006A&A...451...99B}
}

@ARTICLE{Massey2007,
       author = {{Massey}, Philip and {Olsen}, K.~A.~G. and {Hodge}, Paul W. and {Jacoby}, George H. and {McNeill}, Reagin T. and {Smith}, R.~C. and {Strong}, Shay B.},
        title = "{A Survey of Local Group Galaxies Currently Forming Stars. II. UBVRI Photometry of Stars in Seven Dwarfs and a Comparison of the Entire Sample}",
      journal = {\aj},
         year = 2007,
        month = may,
       volume = {133},
       number = {5},
        pages = {2393-2417},
          doi = {10.1086/513319},
archivePrefix = {arXiv},
       eprint = {astro-ph/0702236},
 primaryClass = {astro-ph},
       adsurl = {https://ui.adsabs.harvard.edu/abs/2007AJ....133.2393M}
}

@ARTICLE{Huxor2014,
       author = {{Huxor}, A.~P. and {Mackey}, A.~D. and {Ferguson}, A.~M.~N. and {Irwin}, M.~J. and {Martin}, N.~F. and {Tanvir}, N.~R. and {Veljanoski}, J. and {McConnachie}, A. and {Fishlock}, C.~K. and {Ibata}, R. and {Lewis}, G.~F.},
        title = "{The outer halo globular cluster system of M31 - I. The final PAndAS catalogue}",
      journal = {\mnras},
         year = 2014,
        month = aug,
       volume = {442},
       number = {3},
        pages = {2165-2187},
          doi = {10.1093/mnras/stu771},
archivePrefix = {arXiv},
       eprint = {1404.5807},
 primaryClass = {astro-ph.GA},
       adsurl = {https://ui.adsabs.harvard.edu/abs/2014MNRAS.442.2165H}
}

@ARTICLE{Cookd2019,
       author = {{Cook}, D.~O. and {Lee}, J.~C. and {Adamo}, A. and {Kim}, H. and {Chandar}, R. and {Whitmore}, B.~C. and {Mok}, A. and {Ryon}, J.~E. and {Dale}, D.~A. and {Calzetti}, D. and {Andrews}, J.~E. and {Aloisi}, A. and {Ashworth}, G. and {Bright}, S.~N. and {Brown}, T.~M. and {Christian}, C. and {Cignoni}, M. and {Clayton}, G.~C. and {da Silva}, R. and {de Mink}, S.~E. and {Dobbs}, C.~L. and {Elmegreen}, B.~G. and {Elmegreen}, D.~M. and {Evans}, A.~S. and {Fumagalli}, M. and {Gallagher}, J.~S. and {Gouliermis}, D.~A. and {Grasha}, K. and {Grebel}, E.~K. and {Herrero}, A. and {Hunter}, D.~A. and {Jensen}, E.~I. and {Johnson}, K.~E. and {Kahre}, L. and {Kennicutt}, R.~C. and {Krumholz}, M.~R. and {Lee}, N.~J. and {Lennon}, D. and {Linden}, S. and {Martin}, C. and {Messa}, M. and {Nair}, P. and {Nota}, A. and {{\"O}stlin}, G. and {Parziale}, R.~C. and {Pellerin}, A. and {Regan}, M.~W. and {Sabbi}, E. and {Sacchi}, E. and {Schaerer}, D. and {Schiminovich}, D. and {Shabani}, F. and {Slane}, F.~A. and {Small}, J. and {Smith}, C.~L. and {Smith}, L.~J. and {Taibi}, S. and {Thilker}, D.~A. and {de la Torre}, I.~C. and {Tosi}, M. and {Turner}, J.~A. and {Ubeda}, L. and {Van Dyk}, S.~D. and {Walterbos}, R. AM and {Wofford}, A.},
        title = "{Star cluster catalogues for the LEGUS dwarf galaxies}",
      journal = {\mnras},
         year = 2019,
        month = apr,
       volume = {484},
       number = {4},
        pages = {4897-4919},
          doi = {10.1093/mnras/stz331},
archivePrefix = {arXiv},
       eprint = {1902.00082},
 primaryClass = {astro-ph.GA},
       adsurl = {https://ui.adsabs.harvard.edu/abs/2019MNRAS.484.4897C}
}

@ARTICLE{Elson1987,
       author = {{Elson}, Rebecca A.~W. and {Fall}, S. Michael and {Freeman}, Kenneth C.},
        title = "{The Structure of Young Star Clusters in the Large Magellanic Cloud}",
      journal = {\apj},
         year = 1987,
        month = dec,
       volume = {323},
        pages = {54},
          doi = {10.1086/165807},
       adsurl = {https://ui.adsabs.harvard.edu/abs/1987ApJ...323...54E}
}

@ARTICLE{King1966,
       author = {{King}, Ivan R.},
        title = "{The structure of star clusters. III. Some simple dynamical models}",
      journal = {\aj},
         year = 1966,
        month = feb,
       volume = {71},
        pages = {64},
          doi = {10.1086/109857},
       adsurl = {https://ui.adsabs.harvard.edu/abs/1966AJ.....71...64K}
}

@misc{Bradley2023,
author       = {Larry Bradley and
                Brigitta Sip{\H o}cz and
                Thomas Robitaille and
                Erik Tollerud and
                Z\`e Vin{\'{\i}}cius and
                Christoph Deil and
                Kyle Barbary and
                Tom J Wilson and
                Ivo Busko and
                Axel Donath and
                Hans Moritz G{\"u}nther and
                Mihai Cara and
                P. L. Lim and
                Sebastian Me{\ss}linger and
                Simon Conseil and
                Azalee Bostroem and
                Michael Droettboom and
                E. M. Bray and
                Lars Andersen Bratholm and
                Geert Barentsen and
                Matt Craig and
                Shivangee Rathi and
                Sergio Pascual and
                Gabriel Perren and
                Iskren Y. Georgiev and
                Miguel de Val-Borro and
                Wolfgang Kerzendorf and
                Yoonsoo P. Bach and
                Bruno Quint and
                Harrison Souchereau},
title        = {astropy/photutils: 1.8.0},
month        = may,
year         = 2023,
publisher    = {Zenodo},
version      = {1.8.0},
doi          = {10.5281/zenodo.7946442},
url          = {https://doi.org/10.5281/zenodo.7946442}
}

@ARTICLE{Lang2010,
       author = {{Lang}, Dustin and {Hogg}, David W. and {Mierle}, Keir and {Blanton}, Michael and {Roweis}, Sam},
        title = "{Astrometry.net: Blind Astrometric Calibration of Arbitrary Astronomical Images}",
      journal = {\aj},
         year = 2010,
        month = may,
       volume = {139},
       number = {5},
        pages = {1782-1800},
          doi = {10.1088/0004-6256/139/5/1782},
archivePrefix = {arXiv},
       eprint = {0910.2233},
 primaryClass = {astro-ph.IM},
       adsurl = {https://ui.adsabs.harvard.edu/abs/2010AJ....139.1782L}
}

@ARTICLE{Lim2015,
       author = {{Lim}, Sungsoon and {Lee}, Myung Gyoon},
        title = "{The Star Cluster System in the Local Group Starburst Galaxy IC 10}",
      journal = {\apj},
         year = 2015,
        month = may,
       volume = {804},
       number = {2},
          eid = {123},
        pages = {123},
          doi = {10.1088/0004-637X/804/2/123},
archivePrefix = {arXiv},
       eprint = {1503.03372},
 primaryClass = {astro-ph.GA},
       adsurl = {https://ui.adsabs.harvard.edu/abs/2015ApJ...804..123L}
}

@ARTICLE{Chandar2010,
       author = {{Chandar}, Rupali and {Whitmore}, Bradley C. and {Kim}, Hwihyun and {Kaleida}, Catherine and {Mutchler}, Max and {Calzetti}, Daniela and {Saha}, Abhijit and {O'Connell}, Robert and {Balick}, Bruce and {Bond}, Howard and {Carollo}, Marcella and {Disney}, Michael and {Dopita}, Michael A. and {Frogel}, Jay A. and {Hall}, Donald and {Holtzman}, Jon A. and {Kimble}, Randy A. and {McCarthy}, Patrick and {Paresce}, Francesco and {Silk}, Joe and {Trauger}, John and {Walker}, Alistair R. and {Windhorst}, Rogier A. and {Young}, Erick},
        title = "{The Luminosity, Mass, and Age Distributions of Compact Star Clusters in M83 Based on Hubble Space Telescope/Wide Field Camera 3 Observations}",
      journal = {\apj},
         year = 2010,
        month = aug,
       volume = {719},
       number = {1},
        pages = {966-978},
          doi = {10.1088/0004-637X/719/1/966},
archivePrefix = {arXiv},
       eprint = {1007.5237},
 primaryClass = {astro-ph.GA},
       adsurl = {https://ui.adsabs.harvard.edu/abs/2010ApJ...719..966C}
}

@ARTICLE{Hodge1977,
       author = {{Hodge}, P.~W.},
        title = "{The structure and content of NGC 6822.}",
      journal = {\apjs},
         year = 1977,
        month = jan,
       volume = {33},
        pages = {69},
          doi = {10.1086/190419},
       adsurl = {https://ui.adsabs.harvard.edu/abs/1977ApJS...33...69H}
}

@ARTICLE{Green2018,
       author = {{Green}, {Gregory M.}},
        title = "{dustmaps: A Python interface for maps of interstellar dust}",
      journal = {The Journal of Open Source Software},
         year = "2018",
        month = "Jun",
       volume = {3},
       number = {26},
        pages = {695},
          doi = {10.21105/joss.00695},
       adsurl = {https://ui.adsabs.harvard.edu/abs/2018JOSS....3..695G}
}

@ARTICLE{Schlegel1998,
       author = {{Schlegel}, David J. and {Finkbeiner}, Douglas P. and {Davis}, Marc},
        title = "{Maps of Dust Infrared Emission for Use in Estimation of Reddening and Cosmic Microwave Background Radiation Foregrounds}",
      journal = {\apj},
         year = 1998,
        month = jun,
       volume = {500},
       number = {2},
        pages = {525-553},
          doi = {10.1086/305772},
archivePrefix = {arXiv},
       eprint = {astro-ph/9710327},
 primaryClass = {astro-ph},
       adsurl = {https://ui.adsabs.harvard.edu/abs/1998ApJ...500..525S}
}

@ARTICLE{Schlafly2011,
       author = {{Schlafly}, Edward F. and {Finkbeiner}, Douglas P.},
        title = "{Measuring Reddening with Sloan Digital Sky Survey Stellar Spectra and Recalibrating SFD}",
      journal = {\apj},
         year = 2011,
        month = aug,
       volume = {737},
       number = {2},
          eid = {103},
        pages = {103},
          doi = {10.1088/0004-637X/737/2/103},
archivePrefix = {arXiv},
       eprint = {1012.4804},
 primaryClass = {astro-ph.GA},
       adsurl = {https://ui.adsabs.harvard.edu/abs/2011ApJ...737..103S}
}

@ARTICLE{Kroupa2001,
       author = {{Kroupa}, Pavel},
        title = "{On the variation of the initial mass function}",
      journal = {\mnras},
         year = 2001,
        month = apr,
       volume = {322},
       number = {2},
        pages = {231-246},
          doi = {10.1046/j.1365-8711.2001.04022.x},
archivePrefix = {arXiv},
       eprint = {astro-ph/0009005},
 primaryClass = {astro-ph},
       adsurl = {https://ui.adsabs.harvard.edu/abs/2001MNRAS.322..231K}
}

@ARTICLE{Skillman1989,
       author = {{Skillman}, Evan D. and {Kennicutt}, R.~C. and {Hodge}, P.~W.},
        title = "{Oxygen Abundances in Nearby Dwarf Irregular Galaxies}",
      journal = {\apj},
         year = 1989,
        month = dec,
       volume = {347},
        pages = {875},
          doi = {10.1086/168178},
       adsurl = {https://ui.adsabs.harvard.edu/abs/1989ApJ...347..875S}
}

@ARTICLE{Garnett1990,
       author = {{Garnett}, Donald R.},
        title = "{Nitrogen in Irregular Galaxies}",
      journal = {\apj},
         year = 1990,
        month = nov,
       volume = {363},
        pages = {142},
          doi = {10.1086/169324},
       adsurl = {https://ui.adsabs.harvard.edu/abs/1990ApJ...363..142G}
}

@ARTICLE{Lee2006,
       author = {{Lee}, Henry and {Skillman}, Evan D. and {Venn}, Kim A.},
        title = "{The Spatial Homogeneity of Nebular and Stellar Oxygen Abundances in the Local Group Dwarf Irregular Galaxy NGC 6822}",
      journal = {\apj},
         year = 2006,
        month = may,
       volume = {642},
       number = {2},
        pages = {813-833},
          doi = {10.1086/500568},
archivePrefix = {arXiv},
       eprint = {astro-ph/0512428},
 primaryClass = {astro-ph},
       adsurl = {https://ui.adsabs.harvard.edu/abs/2006ApJ...642..813L}
}

@ARTICLE{Karachentsev1993,
       author = {{Karachentsev}, I.~D. and {Tikhonov}, N.~A.},
        title = "{Photometric distances to the nearby galaxies IC 10, IC 342, and UGCA 86, visible through the Milky Way.}",
      journal = {\aaps},
         year = 1993,
        month = aug,
       volume = {100},
        pages = {227-235},
       adsurl = {https://ui.adsabs.harvard.edu/abs/1993A&AS..100..227K}
}

@ARTICLE{Pastorelli2020,
       author = {{Pastorelli}, Giada and {Marigo}, Paola and {Girardi}, L{\'e}o and {Aringer}, Bernhard and {Chen}, Yang and {Rubele}, Stefano and {Trabucchi}, Michele and {Bladh}, Sara and {Boyer}, Martha L. and {Bressan}, Alessandro and {Dalcanton}, Julianne J. and {Groenewegen}, Martin A.~T. and {Lebzelter}, Thomas and {Mowlavi}, Nami and {Chubb}, Katy L. and {Cioni}, Maria-Rosa L. and {de Grijs}, Richard and {Ivanov}, Valentin D. and {Nanni}, Ambra and {van Loon}, Jacco Th and {Zaggia}, Simone},
        title = "{Constraining the thermally pulsing asymptotic giant branch phase with resolved stellar populations in the Large Magellanic Cloud}",
      journal = {\mnras},
         year = 2020,
        month = nov,
       volume = {498},
       number = {3},
        pages = {3283-3301},
          doi = {10.1093/mnras/staa2565},
archivePrefix = {arXiv},
       eprint = {2008.08595},
 primaryClass = {astro-ph.SR},
       adsurl = {https://ui.adsabs.harvard.edu/abs/2020MNRAS.498.3283P}
}

@ARTICLE{Chen2015,
       author = {{Chen}, Yang and {Bressan}, Alessandro and {Girardi}, L{\'e}o and {Marigo}, Paola and {Kong}, Xu and {Lanza}, Antonio},
        title = "{PARSEC evolutionary tracks of massive stars up to 350 M$_{{\ensuremath{\odot}}}$ at metallicities 0.0001 {\ensuremath{\leq}} Z {\ensuremath{\leq}} 0.04}",
      journal = {\mnras},
         year = 2015,
        month = sep,
       volume = {452},
       number = {1},
        pages = {1068-1080},
          doi = {10.1093/mnras/stv1281},
archivePrefix = {arXiv},
       eprint = {1506.01681},
 primaryClass = {astro-ph.SR},
       adsurl = {https://ui.adsabs.harvard.edu/abs/2015MNRAS.452.1068C}
}

@ARTICLE{BressanA2012,
       author = {{Bressan}, Alessandro and {Marigo}, Paola and {Girardi}, L{\'e}o. and {Salasnich}, Bernardo and {Dal Cero}, Claudia and {Rubele}, Stefano and {Nanni}, Ambra},
        title = "{PARSEC: stellar tracks and isochrones with the PAdova and TRieste Stellar Evolution Code}",
      journal = {\mnras},
         year = 2012,
        month = nov,
       volume = {427},
       number = {1},
        pages = {127-145},
          doi = {10.1111/j.1365-2966.2012.21948.x},
archivePrefix = {arXiv},
       eprint = {1208.4498},
 primaryClass = {astro-ph.SR},
       adsurl = {https://ui.adsabs.harvard.edu/abs/2012MNRAS.427..127B}
}

@ARTICLE{Georgiev2009,
       author = {{Georgiev}, Iskren Y. and {Puzia}, Thomas H. and {Hilker}, Michael and {Goudfrooij}, Paul},
        title = "{Globular cluster systems in nearby dwarf galaxies - I. HST/ACS observations and dynamical properties of globular clusters at low environmental density}",
      journal = {\mnras},
         year = 2009,
        month = jan,
       volume = {392},
       number = {2},
        pages = {879-893},
          doi = {10.1111/j.1365-2966.2008.14104.x},
archivePrefix = {arXiv},
       eprint = {0810.3660},
 primaryClass = {astro-ph},
       adsurl = {https://ui.adsabs.harvard.edu/abs/2009MNRAS.392..879G}
}

@ARTICLE{Veljanoski2015,
       author = {{Veljanoski}, J. and {Ferguson}, A.~M.~N. and {Mackey}, A.~D. and {Huxor}, A.~P. and {Hurley}, J.~R. and {Bernard}, E.~J. and {C{\^o}t{\'e}}, P. and {Irwin}, M.~J. and {Martin}, N.~F. and {Burgett}, W.~S. and {Chambers}, K.~C. and {Flewelling}, H. and {Kudritzki}, R. and {Waters}, C.},
        title = "{The globular cluster system of NGC 6822}",
      journal = {\mnras},
         year = 2015,
        month = sep,
       volume = {452},
       number = {1},
        pages = {320-332},
          doi = {10.1093/mnras/stv1259},
archivePrefix = {arXiv},
       eprint = {1506.00951},
 primaryClass = {astro-ph.GA},
       adsurl = {https://ui.adsabs.harvard.edu/abs/2015MNRAS.452..320V}
}

@ARTICLE{Hwang2014,
       author = {{Hwang}, Narae and {Park}, Hong Soo and {Lee}, Myung Gyoon and {Lim}, Sungsoon and {Hodge}, Paul W. and {Kim}, Sang Chul and {Miller}, Bryan and {Weisz}, Daniel},
        title = "{Spectroscopic Study of Extended Star Clusters in Dwarf Galaxy NGC 6822}",
      journal = {\apj},
         year = 2014,
        month = mar,
       volume = {783},
       number = {1},
          eid = {49},
        pages = {49},
          doi = {10.1088/0004-637X/783/1/49},
archivePrefix = {arXiv},
       eprint = {1401.2492},
 primaryClass = {astro-ph.GA},
       adsurl = {https://ui.adsabs.harvard.edu/abs/2014ApJ...783...49H}
}

@ARTICLE{Larsen2022,
       author = {{Larsen}, S.~S. and {Eitner}, P. and {Magg}, E. and {Bergemann}, M. and {Moltzer}, C.~A.~S. and {Brodie}, J.~P. and {Romanowsky}, A.~J. and {Strader}, J.},
        title = "{The chemical composition of globular clusters in the Local Group}",
      journal = {\aap},
         year = 2022,
        month = apr,
       volume = {660},
          eid = {A88},
        pages = {A88},
          doi = {10.1051/0004-6361/202142243},
archivePrefix = {arXiv},
       eprint = {2112.00081},
 primaryClass = {astro-ph.GA},
       adsurl = {https://ui.adsabs.harvard.edu/abs/2022A&A...660A..88L}
}

@ARTICLE{Huxor2013,
       author = {{Huxor}, A.~P. and {Ferguson}, A.~M.~N. and {Veljanoski}, J. and {Mackey}, A.~D. and {Tanvir}, N.~R.},
        title = "{Three newly discovered globular clusters in NGC 6822}",
      journal = {\mnras},
         year = 2013,
        month = feb,
       volume = {429},
       number = {2},
        pages = {1039-1044},
          doi = {10.1093/mnras/sts387},
archivePrefix = {arXiv},
       eprint = {1211.3896},
 primaryClass = {astro-ph.CO},
       adsurl = {https://ui.adsabs.harvard.edu/abs/2013MNRAS.429.1039H}
}

@ARTICLE{Sibbons2012,
       author = {{Sibbons}, L.~F. and {Ryan}, S.~G. and {Cioni}, M. -R.~L. and {Irwin}, M. and {Napiwotzki}, R.},
        title = "{The AGB population of NGC 6822: distribution and the C/M ratio from JHK photometry}",
      journal = {\aap},
         year = 2012,
        month = apr,
       volume = {540},
          eid = {A135},
        pages = {A135},
          doi = {10.1051/0004-6361/201118365},
archivePrefix = {arXiv},
       eprint = {1202.3285},
 primaryClass = {astro-ph.CO},
       adsurl = {https://ui.adsabs.harvard.edu/abs/2012A&A...540A.135S}
}

@ARTICLE{Tikhonov2009,
       author = {{Tikhonov}, N.~A. and {Galazutdinova}, O.~A.},
        title = "{Stellar population of the irregular galaxy IC 10}",
      journal = {Astronomy Letters},
         year = 2009,
        month = nov,
       volume = {35},
       number = {11},
        pages = {748-763},
          doi = {10.1134/S1063773709110036},
       adsurl = {https://ui.adsabs.harvard.edu/abs/2009AstL...35..748T}
}

@ARTICLE{Krienke2004,
       author = {{Krienke}, Karl and {Hodge}, Paul},
        title = "{Newly Identified Star Clusters in NGC 6822, and the Age Distribution of Its Cluster System}",
      journal = {\pasp},
         year = 2004,
        month = jun,
       volume = {116},
       number = {820},
        pages = {497-505},
          doi = {10.1086/420873},
       adsurl = {https://ui.adsabs.harvard.edu/abs/2004PASP..116..497K}
}

@ARTICLE{Hwang2011,
       author = {{Hwang}, Narae and {Lee}, Myung Gyoon and {Lee}, Jong Chul and {Park}, Won-Kee and {Park}, Hong Soo and {Kim}, Sang Chul and {Park}, Jang-Hyun},
        title = "{Extended Star Clusters in the Remote Halo of the Intriguing Dwarf Galaxy NGC 6822}",
      journal = {\apj},
         year = 2011,
        month = sep,
       volume = {738},
       number = {1},
          eid = {58},
        pages = {58},
          doi = {10.1088/0004-637X/738/1/58},
archivePrefix = {arXiv},
       eprint = {1106.2878},
 primaryClass = {astro-ph.CO},
       adsurl = {https://ui.adsabs.harvard.edu/abs/2011ApJ...738...58H}
}

@ARTICLE{Mackey2005,
       author = {{Mackey}, A.~D. and {van den Bergh}, Sidney},
        title = "{The properties of Galactic globular cluster subsystems}",
      journal = {\mnras},
         year = 2005,
        month = jun,
       volume = {360},
       number = {2},
        pages = {631-645},
          doi = {10.1111/j.1365-2966.2005.09080.x},
archivePrefix = {arXiv},
       eprint = {astro-ph/0504142},
 primaryClass = {astro-ph},
       adsurl = {https://ui.adsabs.harvard.edu/abs/2005MNRAS.360..631M}
}

@ARTICLE{Bica2020,
       author = {{Bica}, Eduardo and {Westera}, Pieter and {Kerber}, Leandro de O. and {Dias}, Bruno and {Maia}, Francisco and {Santos}, Jo{\~a}o F.~C., Jr. and {Barbuy}, Beatriz and {Oliveira}, Raphael A.~P.},
        title = "{An Updated Small Magellanic Cloud and Magellanic Bridge Catalog of Star Clusters, Associations, and Related Objects}",
      journal = {\aj},
         year = 2020,
        month = mar,
       volume = {159},
       number = {3},
          eid = {82},
        pages = {82},
          doi = {10.3847/1538-3881/ab6595},
archivePrefix = {arXiv},
       eprint = {1907.08642},
 primaryClass = {astro-ph.GA},
       adsurl = {https://ui.adsabs.harvard.edu/abs/2020AJ....159...82B}
}

@ARTICLE{Gatto2021,
       author = {{Gatto}, M. and {Ripepi}, V. and {Bellazzini}, M. and {Tosi}, M. and {Cignoni}, M. and {Tortora}, C. and {Leccia}, S. and {Clementini}, G. and {Grebel}, E.~K. and {Longo}, G. and {Marconi}, M. and {Musella}, I.},
        title = "{STEP survey - II. Structural analysis of 170 star clusters in the SMC}",
      journal = {\mnras},
         year = 2021,
        month = nov,
       volume = {507},
       number = {3},
        pages = {3312-3330},
          doi = {10.1093/mnras/stab2297},
archivePrefix = {arXiv},
       eprint = {2108.02791},
 primaryClass = {astro-ph.GA},
       adsurl = {https://ui.adsabs.harvard.edu/abs/2021MNRAS.507.3312G}
}

@ARTICLE{Massey1995,
       author = {{Massey}, Philip and {Armandroff}, Taft E. and {Pyke}, Randall and {Patel}, Kanan and {Wilson}, Christine D.},
        title = "{Hot, Luminous Stars in Selected Regions of NGC 6822, M31, and M33}",
      journal = {\aj},
         year = 1995,
        month = dec,
       volume = {110},
        pages = {2715},
          doi = {10.1086/117725},
       adsurl = {https://ui.adsabs.harvard.edu/abs/1995AJ....110.2715M}
}

@ARTICLE{Gallart1996,
       author = {{Gallart}, C. and {Aparicio}, A. and {Vilchez}, J.~M.},
        title = "{The Local Group Dwarf Irregular Galaxy NGC 6822.I.The Stellar Content}",
      journal = {\aj},
         year = 1996,
        month = nov,
       volume = {112},
        pages = {1928},
          doi = {10.1086/118153},
       adsurl = {https://ui.adsabs.harvard.edu/abs/1996AJ....112.1928G}
}

@ARTICLE{Dotter2007,
       author = {{Dotter}, Aaron and {Chaboyer}, Brian and {Jevremovi{\'c}}, Darko and {Baron}, E. and {Ferguson}, Jason W. and {Sarajedini}, Ata and {Anderson}, Jay},
        title = "{The ACS Survey of Galactic Globular Clusters. II. Stellar Evolution Tracks, Isochrones, Luminosity Functions, and Synthetic Horizontal-Branch Models}",
      journal = {\aj},
         year = 2007,
        month = jul,
       volume = {134},
       number = {1},
        pages = {376-390},
          doi = {10.1086/517915},
archivePrefix = {arXiv},
       eprint = {0706.0847},
 primaryClass = {astro-ph},
       adsurl = {https://ui.adsabs.harvard.edu/abs/2007AJ....134..376D}
}

@ARTICLE{Fusco2012,
       author = {{Fusco}, F. and {Buonanno}, R. and {Bono}, G. and {Cassisi}, S. and {Monelli}, M. and {Pietrinferni}, A.},
        title = "{Distance and reddening of the Local Group dwarf irregular galaxy NGC 6822}",
      journal = {\aap},
         year = 2012,
        month = dec,
       volume = {548},
          eid = {A129},
        pages = {A129},
          doi = {10.1051/0004-6361/201220554},
archivePrefix = {arXiv},
       eprint = {1210.6753},
 primaryClass = {astro-ph.CO},
       adsurl = {https://ui.adsabs.harvard.edu/abs/2012A&A...548A.129F}
}

@ARTICLE{Rich2014,
       author = {{Rich}, Jeffrey A. and {Persson}, S.~E. and {Freedman}, Wendy L. and {Madore}, Barry F. and {Monson}, Andrew J. and {Scowcroft}, Victoria and {Seibert}, Mark},
        title = "{A New Cepheid Distance Measurement and Method for NGC 6822}",
      journal = {\apj},
         year = 2014,
        month = oct,
       volume = {794},
       number = {2},
          eid = {107},
        pages = {107},
          doi = {10.1088/0004-637X/794/2/107},
archivePrefix = {arXiv},
       eprint = {1409.6830},
 primaryClass = {astro-ph.GA},
       adsurl = {https://ui.adsabs.harvard.edu/abs/2014ApJ...794..107R}
}

@ARTICLE{Sanna2008,
       author = {{Sanna}, N. and {Bono}, G. and {Stetson}, P.~B. and {Monelli}, M. and {Pietrinferni}, A. and {Drozdovsky}, I. and {Caputo}, F. and {Cassisi}, S. and {Gennaro}, M. and {Prada Moroni}, P.~G. and {Buonanno}, R. and {Corsi}, C.~E. and {Degl'Innocenti}, S. and {Ferraro}, I. and {Iannicola}, G. and {Nonino}, M. and {Pulone}, L. and {Romaniello}, M. and {Walker}, A.~R.},
        title = "{On the Distance and Reddening of the Starburst Galaxy IC 10}",
      journal = {\apjl},
         year = 2008,
        month = dec,
       volume = {688},
       number = {2},
        pages = {L69},
          doi = {10.1086/595551},
archivePrefix = {arXiv},
       eprint = {0810.1210},
 primaryClass = {astro-ph},
       adsurl = {https://ui.adsabs.harvard.edu/abs/2008ApJ...688L..69S}
}

@ARTICLE{Bruzual2003,
       author = {{Bruzual}, G. and {Charlot}, S.},
        title = "{Stellar population synthesis at the resolution of 2003}",
      journal = {\mnras},
         year = 2003,
        month = oct,
       volume = {344},
       number = {4},
        pages = {1000-1028},
          doi = {10.1046/j.1365-8711.2003.06897.x},
archivePrefix = {arXiv},
       eprint = {astro-ph/0309134},
 primaryClass = {astro-ph},
       adsurl = {https://ui.adsabs.harvard.edu/abs/2003MNRAS.344.1000B}
}

@ARTICLE{Carnall2018,
       author = {{Carnall}, A.~C. and {McLure}, R.~J. and {Dunlop}, J.~S. and {Dav{\'e}}, R.},
        title = "{Inferring the star formation histories of massive quiescent galaxies with BAGPIPES: evidence for multiple quenching mechanisms}",
      journal = {\mnras},
         year = 2018,
        month = nov,
       volume = {480},
       number = {4},
        pages = {4379-4401},
          doi = {10.1093/mnras/sty2169},
archivePrefix = {arXiv},
       eprint = {1712.04452},
 primaryClass = {astro-ph.GA},
       adsurl = {https://ui.adsabs.harvard.edu/abs/2018MNRAS.480.4379C}
}

@ARTICLE{Chandar2000,
       author = {{Chandar}, Rupali and {Bianchi}, Luciana and {Ford}, Holland C.},
        title = "{Spectroscopy of Star Cluster Candidates and H II Regions in NGC 6822}",
      journal = {\aj},
         year = 2000,
        month = dec,
       volume = {120},
       number = {6},
        pages = {3088-3097},
          doi = {10.1086/316859},
       adsurl = {https://ui.adsabs.harvard.edu/abs/2000AJ....120.3088C}
}

@ARTICLE{Marigo2013,
       author = {{Marigo}, Paola and {Bressan}, Alessandro and {Nanni}, Ambra and {Girardi}, L{\'e}o and {Pumo}, Maria Letizia},
        title = "{Evolution of thermally pulsing asymptotic giant branch stars - I. The COLIBRI code}",
      journal = {\mnras},
         year = 2013,
        month = sep,
       volume = {434},
       number = {1},
        pages = {488-526},
          doi = {10.1093/mnras/stt1034},
archivePrefix = {arXiv},
       eprint = {1305.4485},
 primaryClass = {astro-ph.SR},
       adsurl = {https://ui.adsabs.harvard.edu/abs/2013MNRAS.434..488M}
}

@article{Skilling2006,
author = {John Skilling},
title = {{Nested sampling for general Bayesian computation}},
volume = {1},
journal = {Bayesian Analysis},
number = {4},
publisher = {International Society for Bayesian Analysis},
pages = {833 -- 859},
year = {2006},
doi = {10.1214/06-BA127},
URL = {https://doi.org/10.1214/06-BA127}
}

@ARTICLE{Feroz2008,
       author = {{Feroz}, F. and {Hobson}, M.~P.},
        title = "{Multimodal nested sampling: an efficient and robust alternative to Markov Chain Monte Carlo methods for astronomical data analyses}",
      journal = {\mnras},
         year = 2008,
        month = feb,
       volume = {384},
       number = {2},
        pages = {449-463},
          doi = {10.1111/j.1365-2966.2007.12353.x},
archivePrefix = {arXiv},
       eprint = {0704.3704},
 primaryClass = {astro-ph},
       adsurl = {https://ui.adsabs.harvard.edu/abs/2008MNRAS.384..449F}
}

@ARTICLE{Feroz2009,
       author = {{Feroz}, F. and {Hobson}, M.~P. and {Bridges}, M.},
        title = "{MULTINEST: an efficient and robust Bayesian inference tool for cosmology and particle physics}",
      journal = {\mnras},
         year = 2009,
        month = oct,
       volume = {398},
       number = {4},
        pages = {1601-1614},
          doi = {10.1111/j.1365-2966.2009.14548.x},
archivePrefix = {arXiv},
       eprint = {0809.3437},
 primaryClass = {astro-ph},
       adsurl = {https://ui.adsabs.harvard.edu/abs/2009MNRAS.398.1601F}
}

@ARTICLE{Buchner2014,
       author = {{Buchner}, J. and {Georgakakis}, A. and {Nandra}, K. and {Hsu}, L. and {Rangel}, C. and {Brightman}, M. and {Merloni}, A. and {Salvato}, M. and {Donley}, J. and {Kocevski}, D.},
        title = "{X-ray spectral modelling of the AGN obscuring region in the CDFS: Bayesian model selection and catalogue}",
      journal = {\aap},
         year = 2014,
        month = apr,
       volume = {564},
          eid = {A125},
        pages = {A125},
          doi = {10.1051/0004-6361/201322971},
archivePrefix = {arXiv},
       eprint = {1402.0004},
 primaryClass = {astro-ph.HE},
       adsurl = {https://ui.adsabs.harvard.edu/abs/2014A&A...564A.125B}
}

@ARTICLE{Sharina2010,
       author = {{Sharina}, Margarita E. and {Chandar}, Rupali and {Puzia}, Thomas H. and {Goudfrooij}, Paul and {Davoust}, Emmanuel},
        title = "{SAO RAS 6-m telescope spectroscopic observations of globular clusters in nearby galaxies}",
      journal = {\mnras},
         year = 2010,
        month = jun,
       volume = {405},
       number = {2},
        pages = {839-856},
          doi = {10.1111/j.1365-2966.2010.16510.x},
archivePrefix = {arXiv},
       eprint = {1002.2144},
 primaryClass = {astro-ph.CO},
       adsurl = {https://ui.adsabs.harvard.edu/abs/2010MNRAS.405..839S}
}

@ARTICLE{Hubble1925,
       author = {{Hubble}, E.~P.},
        title = "{NGC 6822, a remote stellar system.}",
      journal = {\apj},
         year = 1925,
        month = dec,
       volume = {62},
        pages = {409-433},
          doi = {10.1086/142943},
       adsurl = {https://ui.adsabs.harvard.edu/abs/1925ApJ....62..409H}
}

@ARTICLE{Wyder2000,
       author = {{Wyder}, Ted K. and {Hodge}, Paul W. and {Zucker}, Daniel B.},
        title = "{Hubble Space Telescope Observations of the Hubble Clusters in NGC 6822: Ages and Structure}",
      journal = {\pasp},
         year = 2000,
        month = sep,
       volume = {112},
       number = {775},
        pages = {1162-1176},
          doi = {10.1086/316614},
       adsurl = {https://ui.adsabs.harvard.edu/abs/2000PASP..112.1162W}
}

@ARTICLE{Patrick2015,
       author = {{Patrick}, L.~R. and {Evans}, C.~J. and {Davies}, B. and {Kudritzki}, R. -P. and {Gazak}, J.~Z. and {Bergemann}, M. and {Plez}, B. and {Ferguson}, A.~M.~N.},
        title = "{Red Supergiant Stars as Cosmic Abundance Probes: KMOS Observations in NGC 6822}",
      journal = {\apj},
         year = 2015,
        month = apr,
       volume = {803},
       number = {1},
          eid = {14},
        pages = {14},
          doi = {10.1088/0004-637X/803/1/14},
archivePrefix = {arXiv},
       eprint = {1501.07601},
 primaryClass = {astro-ph.SR},
       adsurl = {https://ui.adsabs.harvard.edu/abs/2015ApJ...803...14P}
}

@ARTICLE{Venn2001,
       author = {{Venn}, K.~A. and {Lennon}, D.~J. and {Kaufer}, A. and {McCarthy}, J.~K. and {Przybilla}, N. and {Kudritzki}, R.~P. and {Lemke}, M. and {Skillman}, E.~D. and {Smartt}, S.~J.},
        title = "{First Stellar Abundances in NGC 6822 from VLT-UVES and Keck-HIRES Spectroscopy}",
      journal = {\apj},
         year = 2001,
        month = feb,
       volume = {547},
       number = {2},
        pages = {765-776},
          doi = {10.1086/318424},
archivePrefix = {arXiv},
       eprint = {astro-ph/0009213},
 primaryClass = {astro-ph},
       adsurl = {https://ui.adsabs.harvard.edu/abs/2001ApJ...547..765V}
}

@ARTICLE{Lee2003,
       author = {{Lee}, Henry and {McCall}, Marshall L. and {Kingsburgh}, Robin L. and {Ross}, Robert and {Stevenson}, Chris C.},
        title = "{Uncovering Additional Clues to Galaxy Evolution. I. Dwarf Irregular Galaxies in the Field}",
      journal = {\aj},
         year = 2003,
        month = jan,
       volume = {125},
       number = {1},
        pages = {146-165},
          doi = {10.1086/345384},
archivePrefix = {arXiv},
       eprint = {astro-ph/0210098},
 primaryClass = {astro-ph},
       adsurl = {https://ui.adsabs.harvard.edu/abs/2003AJ....125..146L}
}

@ARTICLE{HirschauerA2020,
       author = {{Hirschauer}, Alec S. and {Gray}, Laurin and {Meixner}, Margaret and {Jones}, Olivia C. and {Srinivasan}, Sundar and {Boyer}, Martha L. and {Sargent}, B.~A.},
        title = "{Dusty Stellar Birth and Death in the Metal-poor Galaxy NGC 6822}",
      journal = {\apj},
         year = 2020,
        month = apr,
       volume = {892},
       number = {2},
          eid = {91},
        pages = {91},
          doi = {10.3847/1538-4357/ab7b60},
archivePrefix = {arXiv},
       eprint = {2002.12437},
 primaryClass = {astro-ph.GA},
       adsurl = {https://ui.adsabs.harvard.edu/abs/2020ApJ...892...91H}
}

@ARTICLE{Krumholz2019,
       author = {{Krumholz}, Mark R. and {McKee}, Christopher F. and {Bland-Hawthorn}, Joss},
        title = "{Star Clusters Across Cosmic Time}",
      journal = {\mnras},
         year = 2019,
        month = aug,
       volume = {57},
        pages = {227-303},
          doi = {10.1146/annurev-astro-091918-104430},
archivePrefix = {arXiv},
       eprint = {1812.01615},
 primaryClass = {astro-ph.GA},
       adsurl = {https://ui.adsabs.harvard.edu/abs/2019ARA&A..57..227K}
}

@ARTICLE{Scheepmaker2007,
       author = {{Scheepmaker}, R.~A. and {Haas}, M.~R. and {Gieles}, M. and {Bastian}, N. and {Larsen}, S.~S. and {Lamers}, H.~J.~G.~L.~M.},
        title = "{ACS imaging of star clusters in M 51. I. Identification and radius distribution}",
      journal = {\aap},
         year = 2007,
        month = jul,
       volume = {469},
       number = {3},
        pages = {925-940},
          doi = {10.1051/0004-6361:20077511},
archivePrefix = {arXiv},
       eprint = {0704.3604},
 primaryClass = {astro-ph},
       adsurl = {https://ui.adsabs.harvard.edu/abs/2007A&A...469..925S}
}

@ARTICLE{Larsen2004,
       author = {{Larsen}, S.~S.},
        title = "{The structure and environment of young stellar clusters in spiral galaxies}",
      journal = {\aap},
         year = 2004,
        month = mar,
       volume = {416},
        pages = {537-553},
          doi = {10.1051/0004-6361:20034533},
archivePrefix = {arXiv},
       eprint = {astro-ph/0312338},
 primaryClass = {astro-ph},
       adsurl = {https://ui.adsabs.harvard.edu/abs/2004A&A...416..537L}
}

@ARTICLE{Mackey2003,
       author = {{Mackey}, A.~D. and {Gilmore}, G.~F.},
        title = "{Surface brightness profiles and structural parameters for 53 rich stellar clusters in the Large Magellanic Cloud}",
      journal = {\mnras},
         year = 2003,
        month = jan,
       volume = {338},
       number = {1},
        pages = {85-119},
          doi = {10.1046/j.1365-8711.2003.06021.x},
archivePrefix = {arXiv},
       eprint = {astro-ph/0209031},
 primaryClass = {astro-ph},
       adsurl = {https://ui.adsabs.harvard.edu/abs/2003MNRAS.338...85M}
}

@ARTICLE{Ryon2015,
       author = {{Ryon}, J.~E. and {Bastian}, N. and {Adamo}, A. and {Konstantopoulos}, I.~S. and {Gallagher}, J.~S. and {Larsen}, S. and {Hollyhead}, K. and {Silva-Villa}, E. and {Smith}, L.~J.},
        title = "{Sizes and shapes of young star cluster light profiles in M83}",
      journal = {\mnras},
         year = 2015,
        month = sep,
       volume = {452},
       number = {1},
        pages = {525-539},
          doi = {10.1093/mnras/stv1282},
archivePrefix = {arXiv},
       eprint = {1506.02042},
 primaryClass = {astro-ph.GA},
       adsurl = {https://ui.adsabs.harvard.edu/abs/2015MNRAS.452..525R}
}

@ARTICLE{Chandar2016,
       author = {{Chandar}, Rupali and {Whitmore}, Bradley C. and {Dinino}, Daiana and {Kennicutt}, Robert C. and {Chien}, L. -H. and {Schinnerer}, Eva and {Meidt}, Sharon},
        title = "{The Age, Mass, and Size Distributions of Star Clusters in M51}",
      journal = {\apj},
         year = 2016,
        month = jun,
       volume = {824},
       number = {2},
          eid = {71},
        pages = {71},
          doi = {10.3847/0004-637X/824/2/71},
       adsurl = {https://ui.adsabs.harvard.edu/abs/2016ApJ...824...71C}
}

@ARTICLE{Lee2005,
       author = {{Lee}, Myung Gyoon and {Chandar}, Rupali and {Whitmore}, Bradley C.},
        title = "{Properties of Resolved Star Clusters in M51}",
      journal = {\aj},
         year = 2005,
        month = nov,
       volume = {130},
       number = {5},
        pages = {2128-2139},
          doi = {10.1086/491786},
archivePrefix = {arXiv},
       eprint = {astro-ph/0510144},
 primaryClass = {astro-ph},
       adsurl = {https://ui.adsabs.harvard.edu/abs/2005AJ....130.2128L}
}

@ARTICLE{Bastian2012,
       author = {{Bastian}, N. and {Adamo}, A. and {Gieles}, M. and {Silva-Villa}, E. and {Lamers}, H.~J.~G.~L.~M. and {Larsen}, S.~S. and {Smith}, L.~J. and {Konstantopoulos}, I.~S. and {Zackrisson}, E.},
        title = "{Stellar clusters in M83: formation, evolution, disruption and the influence of the environment}",
      journal = {\mnras},
         year = 2012,
        month = jan,
       volume = {419},
       number = {3},
        pages = {2606-2622},
          doi = {10.1111/j.1365-2966.2011.19909.x},
archivePrefix = {arXiv},
       eprint = {1109.6015},
 primaryClass = {astro-ph.CO},
       adsurl = {https://ui.adsabs.harvard.edu/abs/2012MNRAS.419.2606B}
}

@ARTICLE{Barmby2009,
       author = {{Barmby}, P. and {Perina}, S. and {Bellazzini}, M. and {Cohen}, J.~G. and {Hodge}, P.~W. and {Huchra}, J.~P. and {Kissler-Patig}, M. and {Puzia}, T.~H. and {Strader}, J.},
        title = "{A Hubble Space Telescope/WFPC2 Survey of Bright Young Clusters in M31. III. Structural Parameters}",
      journal = {\aj},
         year = 2009,
        month = dec,
       volume = {138},
       number = {6},
        pages = {1667-1680},
          doi = {10.1088/0004-6256/138/6/1667},
archivePrefix = {arXiv},
       eprint = {0909.4053},
 primaryClass = {astro-ph.GA},
       adsurl = {https://ui.adsabs.harvard.edu/abs/2009AJ....138.1667B}
}

@ARTICLE{Fusco2014,
       author = {{Fusco}, F. and {Buonanno}, R. and {Hidalgo}, S.~L. and {Aparicio}, A. and {Pietrinferni}, A. and {Bono}, G. and {Monelli}, M. and {Cassisi}, S.},
        title = "{A state-of-the-art analysis of the dwarf irregular galaxy NGC 6822}",
      journal = {\aap},
         year = 2014,
        month = dec,
       volume = {572},
          eid = {A26},
        pages = {A26},
          doi = {10.1051/0004-6361/201323075},
archivePrefix = {arXiv},
       eprint = {1409.5247},
 primaryClass = {astro-ph.GA},
       adsurl = {https://ui.adsabs.harvard.edu/abs/2014A&A...572A..26F}
}

@ARTICLE{Hodge1980,
       author = {{Hodge}, P.~W.},
        title = "{The recent evolutionary history of the galaxies NGC 6822 and IC 1613.}",
      journal = {\apj},
         year = 1980,
        month = oct,
       volume = {241},
        pages = {125-131},
          doi = {10.1086/158323},
       adsurl = {https://ui.adsabs.harvard.edu/abs/1980ApJ...241..125H}
}

@ARTICLE{McConnachie2012,
       author = {{McConnachie}, Alan W.},
        title = "{The Observed Properties of Dwarf Galaxies in and around the Local Group}",
      journal = {\aj},
         year = 2012,
        month = jul,
       volume = {144},
       number = {1},
          eid = {4},
        pages = {4},
          doi = {10.1088/0004-6256/144/1/4},
archivePrefix = {arXiv},
       eprint = {1204.1562},
 primaryClass = {astro-ph.CO},
       adsurl = {https://ui.adsabs.harvard.edu/abs/2012AJ....144....4M}
}

@ARTICLE{Forbes2018,
       author = {{Forbes}, Duncan A. and {Read}, Justin I. and {Gieles}, Mark and {Collins}, Michelle L.~M.},
        title = "{Extending the globular cluster system-halo mass relation to the lowest galaxy masses}",
      journal = {\mnras},
         year = 2018,
        month = dec,
       volume = {481},
       number = {4},
        pages = {5592-5605},
          doi = {10.1093/mnras/sty2584},
archivePrefix = {arXiv},
       eprint = {1809.07831},
 primaryClass = {astro-ph.GA},
       adsurl = {https://ui.adsabs.harvard.edu/abs/2018MNRAS.481.5592F}
}

@ARTICLE{Karim2024,
       author = {{Karim}, Noushin and {Collins}, Michelle L.~M. and {Forbes}, Duncan A. and {Read}, Justin I.},
        title = "{Discovery of Globular Cluster Candidates in the Dwarf Irregular Galaxy IC 2574 Using HST/ACS Imaging}",
      journal = {\mnras},
         year = 2024,
        month = jun,
       volume = {530},
       number = {4},
        pages = {4936-4949},
          doi = {10.1093/mnras/stae611},
archivePrefix = {arXiv},
       eprint = {2402.16955},
 primaryClass = {astro-ph.GA},
       adsurl = {https://ui.adsabs.harvard.edu/abs/2024MNRAS.530.4936K}
}

@ARTICLE{Oh2015, author = {{Oh}, Se-Heon and {Hunter}, Deidre A. and {Brinks}, Elias and {Elmegreen}, Bruce G. and {Schruba}, Andreas and {Walter}, Fabian and {Rupen}, Michael P. and {Young}, Lisa M. and {Simpson}, Caroline E. and {Johnson}, Megan C. and {Herrmann}, Kimberly A. and {Ficut-Vicas}, Dana and {Cigan}, Phil and {Heesen}, Volker and {Ashley}, Trisha and {Zhang}, Hong-Xin}, title = "{High-resolution Mass Models of Dwarf Galaxies from LITTLE THINGS}", journal = {\aj}, year = 2015, month = jun, volume = {149}, number = {6}, eid = {180}, pages = {180}, doi = {10.1088/0004-6256/149/6/180}, archivePrefix = {arXiv}, eprint = {1502.01281}, primaryClass = {astro-ph.GA}, adsurl = {https://ui.adsabs.harvard.edu/abs/2015AJ....149..180O} }

@ARTICLE{Popescu2010,
       author = {{Popescu}, Bogdan and {Hanson}, M.~M.},
        title = "{MASSCLEANage{\textemdash}Stellar Cluster Ages from Integrated Colors}",
      journal = {\apj},
         year = 2010,
        month = nov,
       volume = {724},
       number = {1},
        pages = {296-305},
          doi = {10.1088/0004-637X/724/1/296},
archivePrefix = {arXiv},
       eprint = {1007.3718},
 primaryClass = {astro-ph.GA},
       adsurl = {https://ui.adsabs.harvard.edu/abs/2010ApJ...724..296P}
}

@ARTICLE{Cardelli1989,
       author = {{Cardelli}, Jason A. and {Clayton}, Geoffrey C. and {Mathis}, John S.},
        title = "{The Relationship between Infrared, Optical, and Ultraviolet Extinction}",
      journal = {\apj},
         year = 1989,
        month = oct,
       volume = {345},
        pages = {245},
          doi = {10.1086/167900},
       adsurl = {https://ui.adsabs.harvard.edu/abs/1989ApJ...345..245C}
}

@ARTICLE{Harris1996,
       author = {{Harris}, William E.},
        title = "{A Catalog of Parameters for Globular Clusters in the Milky Way}",
      journal = {\aj},
         year = 1996,
        month = oct,
       volume = {112},
        pages = {1487},
          doi = {10.1086/118116},
       adsurl = {https://ui.adsabs.harvard.edu/abs/1996AJ....112.1487H}
}

@ARTICLE{Hurley2010,
       author = {{Hurley}, Jarrod R. and {Mackey}, A. Dougal},
        title = "{N-body models of extended star clusters}",
      journal = {\mnras},
         year = 2010,
        month = nov,
       volume = {408},
       number = {4},
        pages = {2353-2363},
          doi = {10.1111/j.1365-2966.2010.17285.x},
archivePrefix = {arXiv},
       eprint = {1008.4991},
 primaryClass = {astro-ph.GA},
       adsurl = {https://ui.adsabs.harvard.edu/abs/2010MNRAS.408.2353H}
}

@ARTICLE{Harris1981,
       author = {{Harris}, W.~E. and {van den Bergh}, S.},
        title = "{Globular clusters in galaxies beyond the local group. I. New cluster systems in selected northern ellipticals.}",
      journal = {\aj},
         year = 1981,
        month = nov,
       volume = {86},
        pages = {1627-1642},
          doi = {10.1086/113047},
       adsurl = {https://ui.adsabs.harvard.edu/abs/1981AJ.....86.1627H}
}

@ARTICLE{Johnson2017,
       author = {{Johnson}, L. Clifton and {Seth}, Anil C. and {Dalcanton}, Julianne J. and {Beerman}, Lori C. and {Fouesneau}, Morgan and {Weisz}, Daniel R. and {Bell}, Timothy A. and {Dolphin}, Andrew E. and {Sandstrom}, Karin and {Williams}, Benjamin F.},
        title = "{Panchromatic Hubble Andromeda Treasury. XVIII. The High-mass Truncation of the Star Cluster Mass Function}",
      journal = {\apj},
         year = 2017,
        month = apr,
       volume = {839},
       number = {2},
          eid = {78},
        pages = {78},
          doi = {10.3847/1538-4357/aa6a1f},
archivePrefix = {arXiv},
       eprint = {1703.10312},
 primaryClass = {astro-ph.GA},
       adsurl = {https://ui.adsabs.harvard.edu/abs/2017ApJ...839...78J}
}

@ARTICLE{Khatamsaz2024,
       author = {{Khatamsaz}, Fatemeh and {Abdollahi}, Mahdi and {Abdollahi}, Hedieh and {Javadi}, Atefeh and {van Loon}, Jacco Th.},
        title = "{Star Formation History of the Local Group Dwarf Irregular Galaxy, NGC 6822}",
      journal = {arXiv e-prints},
         year = 2024,
        month = dec,
          eid = {arXiv:2412.05646},
        pages = {arXiv:2412.05646},
          doi = {10.48550/arXiv.2412.05646},
archivePrefix = {arXiv},
       eprint = {2412.05646},
 primaryClass = {astro-ph.GA},
       adsurl = {https://ui.adsabs.harvard.edu/abs/2024arXiv241205646K}
}

@ARTICLE{Chandar2005,
       author = {{Chandar}, Rupali and {Leitherer}, Claus and {Tremonti}, Christy A. and {Calzetti}, Daniela and {Aloisi}, Alessandra and {Meurer}, Gerhardt R. and {de Mello}, Duilia},
        title = "{The Stellar Content of Nearby Star-forming Galaxies. III. Unravelling the Nature of the Diffuse Ultraviolet Light}",
      journal = {\apj},
         year = 2005,
        month = jul,
       volume = {628},
       number = {1},
        pages = {210-230},
          doi = {10.1086/430592},
archivePrefix = {arXiv},
       eprint = {astro-ph/0505024},
 primaryClass = {astro-ph},
       adsurl = {https://ui.adsabs.harvard.edu/abs/2005ApJ...628..210C}
}

@ARTICLE{Cuillandre2024,
       author = {{Cuillandre}, J. -C. and {Bertin}, E. and {Bolzonella}, M. and {Bouy}, H. and {Gwyn}, S. and {Isani}, S. and {Kluge}, M. and {Lai}, O. and {Lan{\c{c}}on}, A. and {Lang}, D.~A. and {Laureijs}, R. and {Saifollahi}, T. and {Schirmer}, M. and {Stone}, C. and {Abdurro'uf} and {Aghanim}, N. and {Altieri}, B. and {Annibali}, F. and {Atek}, H. and {Awad}, P. and {Baes}, M. and {Ba{\~n}ados}, E. and {Barrado}, D. and {Belladitta}, S. and {Belokurov}, V. and {Boselli}, A. and {Bournaud}, F. and {Bovy}, J. and {Bowler}, R.~A.~A. and {Buenadicha}, G. and {Buitrago}, F. and {Cantiello}, M. and {Carollo}, D. and {Codis}, S. and {Collins}, M.~L.~M. and {Congedo}, G. and {Dalessandro}, E. and {de Lapparent}, V. and {De Paolis}, F. and {Diego}, J.~M. and {Dimauro}, P. and {Dinis}, J. and {Dole}, H. and {Duc}, P. -A. and {Erkal}, D. and {Ezziati}, M. and {Ferguson}, A.~M.~N. and {Ferr{\'e}-Mateu}, A. and {Franco}, A. and {Gavazzi}, R. and {George}, K. and {Gillard}, W. and {Golden-Marx}, J.~B. and {Goldman}, B. and {Gonzalez}, A.~H. and {Habas}, R. and {Hartley}, W.~G. and {Hatch}, N.~A. and {Kohley}, R. and {Hoar}, J. and {Howell}, J.~M. and {Hunt}, L.~K. and {Jablonka}, P. and {Jauzac}, M. and {Kang}, Y. and {Knapen}, J.~H. and {Kneib}, J. -P. and {Kuzma}, P.~B. and {Larsen}, S.~S. and {Marchal}, O. and {Mart{\'\i}n-Fleitas}, J. and {Marcos-Arenal}, P. and {Marleau}, F.~R. and {Mart{\'\i}n}, E.~L. and {Massari}, D. and {McConnachie}, A.~W. and {Meneghetti}, M. and {Miluzio}, M. and {Miro Carretero}, J. and {Miyatake}, H. and {Mondelin}, M. and {Montes}, M. and {Mora}, A. and {M{\"u}ller}, O. and {Nally}, C. and {Noeske}, K. and {Nucita}, A.~A. and {Oesch}, P.~A. and {Oguri}, M. and {Peletier}, R.~F. and {Poulain}, M. and {Quilley}, L. and {Racca}, G.~D. and {Rejkuba}, M. and {Rhodes}, J. and {Rocci}, P. -F. and {Rom{\'a}n}, J. and {Sacquegna}, S. and {Saremi}, E. and {Scaramella}, R. and {Schinnerer}, E. and {Serjeant}, S. and {Sola}, E. and {Sorce}, J.~G. and {Tarsitano}, F. and {Tereno}, I. and {Toft}, S. and {Tortora}, C. and {Urbano}, M. and {Venhola}, A. and {Voggel}, K. and {Weaver}, J.~R. and {Xu}, X. and {{\v{Z}}erjal}, M. and {Z{\"o}ller}, R. and {Andreon}, S. and {Auricchio}, N. and {Baccigalupi}, C. and {Baldi}, M. and {Balestra}, A. and {Bardelli}, S. and {Basset}, A. and {Bender}, R. and {Bodendorf}, C. and {Branchini}, E. and {Brau-Nogue}, S. and {Brescia}, M. and {Brinchmann}, J. and {Camera}, S. and {Capobianco}, V. and {Carbone}, C. and {Carretero}, J. and {Casas}, S. and {Castander}, F.~J. and {Castellano}, M. and {Cavuoti}, S. and {Cimatti}, A. and {Conselice}, C.~J. and {Conversi}, L. and {Copin}, Y. and {Courbin}, F. and {Courtois}, H.~M. and {Cropper}, M. and {Cuby}, J. -G. and {Da Silva}, A. and {Degaudenzi}, H. and {Di Giorgio}, A.~M. and {Douspis}, M. and {Duncan}, C.~A.~J. and {Dupac}, X. and {Dusini}, S. and {Fabricius}, M. and {Farina}, M. and {Farrens}, S. and {Ferriol}, S. and {Fosalba}, P. and {Fotopoulou}, S. and {Frailis}, M. and {Franceschi}, E. and {Galeotta}, S. and {Garilli}, B. and {Gillis}, B. and {Giocoli}, C. and {G{\'o}mez-Alvarez}, P. and {Grazian}, A. and {Grupp}, F. and {Guzzo}, L. and {Haugan}, S.~V.~H. and {Hoekstra}, H. and {Holmes}, W. and {Hook}, I. and {Hormuth}, F. and {Hornstrup}, A. and {Hudelot}, P. and {Jahnke}, K. and {Jhabvala}, M. and {Keih{\"a}nen}, E. and {Kermiche}, S. and {Kiessling}, A. and {Kilbinger}, M. and {Kitching}, T. and {Kubik}, B. and {Kuijken}, K. and {K{\"u}mmel}, M. and {Kunz}, M. and {Kurki-Suonio}, H. and {Lahav}, O. and {Liebing}, P. and {Ligori}, S. and {Lilje}, P.~B. and {Lindholm}, V. and {Lloro}, I. and {Maino}, D. and {Maiorano}, E. and {Mansutti}, O. and {Marggraf}, O. and {Markovic}, K. and {Martinet}, N. and {Marulli}, F. and {Massey}, R.},
        title = "{Euclid: Early Release Observations {\textendash} Programme overview and pipeline for compact- and diffuse-emission photometry}",
      journal = {\aap},
         year = 2025,
        month = may,
       volume = {697},
          eid = {A6},
        pages = {A6},
          doi = {10.1051/0004-6361/202450803},
archivePrefix = {arXiv},
       eprint = {2405.13496},
 primaryClass = {astro-ph.IM},
       adsurl = {https://ui.adsabs.harvard.edu/abs/2025A&A...697A...6C}
}

@ARTICLE{Gouliermis2010,
       author = {{Gouliermis}, Dimitrios A. and {Schmeja}, Stefan and {Klessen}, Ralf S. and {de Blok}, W.~J.~G. and {Walter}, Fabian},
        title = "{Hierarchical Stellar Structures in the Local Group Dwarf Galaxy NGC 6822}",
      journal = {\apj},
         year = 2010,
        month = dec,
       volume = {725},
       number = {2},
        pages = {1717-1734},
          doi = {10.1088/0004-637X/725/2/1717},
archivePrefix = {arXiv},
       eprint = {1010.1940},
 primaryClass = {astro-ph.GA},
       adsurl = {https://ui.adsabs.harvard.edu/abs/2010ApJ...725.1717G}
}

@ARTICLE{Hunt2024,
       author = {{Hunt}, L.~K. and {Annibali}, F. and {Cuillandre}, J. -C. and {Ferguson}, A.~M.~N. and {Jablonka}, P. and {Larsen}, S.~S. and {Marleau}, F.~R. and {Schinnerer}, E. and {Schirmer}, M. and {Stone}, C. and {Tortora}, C. and {Saifollahi}, T. and {Lan{\c{c}}on}, A. and {Bolzonella}, M. and {Gwyn}, S. and {Kluge}, M. and {Laureijs}, R. and {Carollo}, D. and {Collins}, M.~L.~M. and {Dimauro}, P. and {Duc}, P. -A. and {Erkal}, D. and {Howell}, J.~M. and {Nally}, C. and {Saremi}, E. and {Scaramella}, R. and {Belokurov}, V. and {Conselice}, C.~J. and {Knapen}, J.~H. and {McConnachie}, A.~W. and {McDonald}, I. and {Miro Carretero}, J. and {Roman}, J. and {Sauvage}, M. and {Sola}, E. and {Aghanim}, N. and {Altieri}, B. and {Andreon}, S. and {Auricchio}, N. and {Awan}, S. and {Azzollini}, R. and {Baldi}, M. and {Balestra}, A. and {Bardelli}, S. and {Basset}, A. and {Bender}, R. and {Bonino}, D. and {Branchini}, E. and {Brescia}, M. and {Brinchmann}, J. and {Camera}, S. and {Candini}, G.~P. and {Capobianco}, V. and {Carbone}, C. and {Carretero}, J. and {Casas}, S. and {Castellano}, M. and {Cavuoti}, S. and {Cimatti}, A. and {Congedo}, G. and {Conversi}, L. and {Copin}, Y. and {Corcione}, L. and {Courbin}, F. and {Courtois}, H.~M. and {Cropper}, M. and {Da Silva}, A. and {Degaudenzi}, H. and {De Lucia}, G. and {Di Giorgio}, A.~M. and {Dinis}, J. and {Dubath}, F. and {Dupac}, X. and {Dusini}, S. and {Farina}, M. and {Farrens}, S. and {Ferriol}, S. and {Fosalba}, P. and {Frailis}, M. and {Franceschi}, E. and {Fumana}, M. and {Galeotta}, S. and {Garilli}, B. and {George}, K. and {Gillard}, W. and {Gillis}, B. and {Giocoli}, C. and {G{\'o}mez-Alvarez}, P. and {Granett}, B.~R. and {Grazian}, A. and {Grupp}, F. and {Guzzo}, L. and {Haugan}, S.~V.~H. and {Hoar}, J. and {Hoekstra}, H. and {Holliman}, M.~S. and {Holmes}, W. and {Hook}, I. and {Hormuth}, F. and {Hornstrup}, A. and {Hudelot}, P. and {Jahnke}, K. and {Keih{\"a}nen}, E. and {Kermiche}, S. and {Kiessling}, A. and {Kilbinger}, M. and {Kitching}, T. and {Kohley}, R. and {Kubik}, B. and {Kuijken}, K. and {K{\"u}mmel}, M. and {Kunz}, M. and {Kurki-Suonio}, H. and {Lahav}, O. and {Le Mignant}, D. and {Lilje}, P.~B. and {Lindholm}, V. and {Lloro}, I. and {Maiorano}, E. and {Mansutti}, O. and {Marggraf}, O. and {Markovic}, K. and {Martinet}, N. and {Marulli}, F. and {Massey}, R. and {Maurogordato}, S. and {McCracken}, H.~J. and {Medinaceli}, E. and {Mei}, S. and {Mellier}, Y. and {Meneghetti}, M. and {Merlin}, E. and {Meylan}, G. and {Moresco}, M. and {Moscardini}, L. and {Munari}, E. and {Nakajima}, R. and {Nichol}, R.~C. and {Niemi}, S. -M. and {Nightingale}, J.~W. and {Padilla}, C. and {Paltani}, S. and {Pasian}, F. and {Pedersen}, K. and {Percival}, W.~J. and {Pettorino}, V. and {Pires}, S. and {Polenta}, G. and {Poncet}, M. and {Popa}, L.~A. and {Pozzetti}, L. and {Racca}, G.~D. and {Raison}, F. and {Rebolo}, R. and {Refregier}, A. and {Renzi}, A. and {Rhodes}, J. and {Riccio}, G. and {Romelli}, E. and {Roncarelli}, M. and {Rossetti}, E. and {Saglia}, R. and {Sapone}, D. and {Sartoris}, B. and {Schneider}, P. and {Schrabback}, T. and {Scodeggio}, M. and {Secroun}, A. and {Seidel}, G. and {Serrano}, S. and {Sirignano}, C. and {Sirri}, G. and {Skottfelt}, J. and {Stanco}, L. and {Tallada-Cresp{\'\i}}, P. and {Tavagnacco}, D. and {Taylor}, A.~N. and {Teplitz}, H.~I. and {Tereno}, I. and {Toledo-Moreo}, R. and {Torradeflot}, F. and {Tutusaus}, I. and {Valentijn}, E.~A. and {Valenziano}, L. and {Vassallo}, T. and {Verdoes Kleijn}, G. and {Veropalumbo}, A. and {Wang}, Y. and {Weller}, J. and {Williams}, O.~R. and {Zamorani}, G. and {Zucca}, E. and {Burigana}, C. and {Scottez}, V. and {Miluzio}, M. and {Simon}, P. and {Mora}, A. and {Mart{\'\i}n-Fleitas}, J. and {Scott}, D.},
        title = "{Euclid: Early Release Observations {\textendash} Deep anatomy of nearby galaxies}",
      journal = {\aap},
         year = 2025,
        month = may,
       volume = {697},
          eid = {A9},
        pages = {A9},
          doi = {10.1051/0004-6361/202450781},
archivePrefix = {arXiv},
       eprint = {2405.13499},
 primaryClass = {astro-ph.GA},
       adsurl = {https://ui.adsabs.harvard.edu/abs/2025A&A...697A...9H}
}

@ARTICLE{Adamoa2015,
       author = {{Adamo}, A. and {Kruijssen}, J.~M.~D. and {Bastian}, N. and {Silva-Villa}, E. and {Ryon}, J.},
        title = "{Probing the role of the galactic environment in the formation of stellar clusters, using M83 as a test bench}",
      journal = {\mnras},
         year = 2015,
        month = sep,
       volume = {452},
       number = {1},
        pages = {246-260},
          doi = {10.1093/mnras/stv1203},
archivePrefix = {arXiv},
       eprint = {1505.07475},
 primaryClass = {astro-ph.GA},
       adsurl = {https://ui.adsabs.harvard.edu/abs/2015MNRAS.452..246A}
}

@ARTICLE{Larsen2009,
       author = {{Larsen}, S.~S.},
        title = "{The mass function of young star clusters in spiral galaxies}",
      journal = {\aap},
         year = 2009,
        month = feb,
       volume = {494},
       number = {2},
        pages = {539-551},
          doi = {10.1051/0004-6361:200811212},
archivePrefix = {arXiv},
       eprint = {0812.1400},
 primaryClass = {astro-ph},
       adsurl = {https://ui.adsabs.harvard.edu/abs/2009A&A...494..539L}
}

@ARTICLE{Nidever2013,
       author = {{Nidever}, David L. and {Ashley}, Trisha and {Slater}, Colin T. and {Ott}, J{\"u}rgen and {Johnson}, Megan and {Bell}, Eric F. and {Stanimirovi{\'c}}, Sne{\v{z}}ana and {Putman}, Mary and {Majewski}, Steven R. and {Simpson}, Caroline E. and {J{\"u}tte}, Eva and {Oosterloo}, Tom A. and {Butler Burton}, W.},
        title = "{Evidence for an Interaction in the Nearest Starbursting Dwarf Irregular Galaxy IC 10}",
      journal = {\apjl},
         year = 2013,
        month = dec,
       volume = {779},
       number = {2},
          eid = {L15},
        pages = {L15},
          doi = {10.1088/2041-8205/779/2/L15},
archivePrefix = {arXiv},
       eprint = {1310.7573},
 primaryClass = {astro-ph.GA},
       adsurl = {https://ui.adsabs.harvard.edu/abs/2013ApJ...779L..15N}
}

@ARTICLE{Zhang2021,
       author = {{Zhang}, Shumeng and {Mackey}, Dougal and {Da Costa}, Gary S.},
        title = "{A panoramic view of the Local Group dwarf galaxy NGC 6822}",
      journal = {\mnras},
         year = 2021,
        month = dec,
       volume = {508},
       number = {2},
        pages = {2098-2113},
          doi = {10.1093/mnras/stab2642},
archivePrefix = {arXiv},
       eprint = {2108.04431},
 primaryClass = {astro-ph.GA},
       adsurl = {https://ui.adsabs.harvard.edu/abs/2021MNRAS.508.2098Z}
}

@ARTICLE{Pace2024,
       author = {{Pace}, Andrew B.},
        title = "{The Local Volume Database: a library of the observed properties of nearby dwarf galaxies and star clusters}",
      journal = {arXiv e-prints},
         year = 2024,
        month = nov,
          eid = {arXiv:2411.07424},
        pages = {arXiv:2411.07424},
          doi = {10.48550/arXiv.2411.07424},
archivePrefix = {arXiv},
       eprint = {2411.07424},
 primaryClass = {astro-ph.GA},
       adsurl = {https://ui.adsabs.harvard.edu/abs/2024arXiv241107424P}
}

@ARTICLE{Higgs2021,
       author = {{Higgs}, C.~R. and {McConnachie}, A.~W. and {Annau}, N. and {Irwin}, M. and {Battaglia}, G. and {C{\^o}t{\'e}}, P. and {Lewis}, G.~F. and {Venn}, K.},
        title = "{Solo dwarfs II: the stellar structure of isolated Local Group dwarf galaxies}",
      journal = {\mnras},
         year = 2021,
        month = may,
       volume = {503},
       number = {1},
        pages = {176-199},
          doi = {10.1093/mnras/stab002},
archivePrefix = {arXiv},
       eprint = {2101.03189},
 primaryClass = {astro-ph.GA},
       adsurl = {https://ui.adsabs.harvard.edu/abs/2021MNRAS.503..176H}
}

@Article{         numpy,
 title         = {Array programming with {NumPy}},
 author        = {Charles R. Harris and K. Jarrod Millman and St{\'{e}}fan J.
                 van der Walt and Ralf Gommers and Pauli Virtanen and David
                 Cournapeau and Eric Wieser and Julian Taylor and Sebastian
                 Berg and Nathaniel J. Smith and Robert Kern and Matti Picus
                 and Stephan Hoyer and Marten H. van Kerkwijk and Matthew
                 Brett and Allan Haldane and Jaime Fern{\'{a}}ndez del
                 R{\'{i}}o and Mark Wiebe and Pearu Peterson and Pierre
                 G{\'{e}}rard-Marchant and Kevin Sheppard and Tyler Reddy and
                 Warren Weckesser and Hameer Abbasi and Christoph Gohlke and
                 Travis E. Oliphant},
 year          = {2020},
 month         = sep,
 journal       = {\nat},
 volume        = {585},
 number        = {7825},
 pages         = {357--362},
 doi           = {10.1038/s41586-020-2649-2},
 publisher     = {Springer Science and Business Media {LLC}},
 url           = {https://doi.org/10.1038/s41586-020-2649-2}
}

@ARTICLE{scipy,
  author  = {Virtanen, Pauli and Gommers, Ralf and Oliphant, Travis E. and
            Haberland, Matt and Reddy, Tyler and Cournapeau, David and
            Burovski, Evgeni and Peterson, Pearu and Weckesser, Warren and
            Bright, Jonathan and {van der Walt}, St{\'e}fan J. and
            Brett, Matthew and Wilson, Joshua and Millman, K. Jarrod and
            Mayorov, Nikolay and Nelson, Andrew R. J. and Jones, Eric and
            Kern, Robert and Larson, Eric and Carey, C J and
            Polat, {\.I}lhan and Feng, Yu and Moore, Eric W. and
            {VanderPlas}, Jake and Laxalde, Denis and Perktold, Josef and
            Cimrman, Robert and Henriksen, Ian and Quintero, E. A. and
            Harris, Charles R. and Archibald, Anne M. and
            Ribeiro, Ant{\^o}nio H. and Pedregosa, Fabian and
            {van Mulbregt}, Paul and {SciPy 1.0 Contributors}},
  title   = {{{SciPy} 1.0: Fundamental Algorithms for Scientific
            Computing in Python}},
  journal = {Nature Methods},
  year    = {2020},
  volume  = {17},
  pages   = {261--272},
  adsurl  = {https://rdcu.be/b08Wh},
  doi     = {10.1038/s41592-019-0686-2},
}

@ARTICLE{astropy,
       author = {{Astropy Collaboration} and {Price-Whelan}, Adrian M. and {Lim}, Pey Lian and {Earl}, Nicholas and {Starkman}, Nathaniel and {Bradley}, Larry and {Shupe}, David L. and {Patil}, Aarya A. and {Corrales}, Lia and {Brasseur}, C.~E. and {N{\"o}the}, Maximilian and {Donath}, Axel and {Tollerud}, Erik and {Morris}, Brett M. and {Ginsburg}, Adam and {Vaher}, Eero and {Weaver}, Benjamin A. and {Tocknell}, James and {Jamieson}, William and {van Kerkwijk}, Marten H. and {Robitaille}, Thomas P. and {Merry}, Bruce and {Bachetti}, Matteo and {G{\"u}nther}, H. Moritz and {Aldcroft}, Thomas L. and {Alvarado-Montes}, Jaime A. and {Archibald}, Anne M. and {B{\'o}di}, Attila and {Bapat}, Shreyas and {Barentsen}, Geert and {Baz{\'a}n}, Juanjo and {Biswas}, Manish and {Boquien}, M{\'e}d{\'e}ric and {Burke}, D.~J. and {Cara}, Daria and {Cara}, Mihai and {Conroy}, Kyle E. and {Conseil}, Simon and {Craig}, Matthew W. and {Cross}, Robert M. and {Cruz}, Kelle L. and {D'Eugenio}, Francesco and {Dencheva}, Nadia and {Devillepoix}, Hadrien A.~R. and {Dietrich}, J{\"o}rg P. and {Eigenbrot}, Arthur Davis and {Erben}, Thomas and {Ferreira}, Leonardo and {Foreman-Mackey}, Daniel and {Fox}, Ryan and {Freij}, Nabil and {Garg}, Suyog and {Geda}, Robel and {Glattly}, Lauren and {Gondhalekar}, Yash and {Gordon}, Karl D. and {Grant}, David and {Greenfield}, Perry and {Groener}, Austen M. and {Guest}, Steve and {Gurovich}, Sebastian and {Handberg}, Rasmus and {Hart}, Akeem and {Hatfield-Dodds}, Zac and {Homeier}, Derek and {Hosseinzadeh}, Griffin and {Jenness}, Tim and {Jones}, Craig K. and {Joseph}, Prajwel and {Kalmbach}, J. Bryce and {Karamehmetoglu}, Emir and {Ka{\l}uszy{\'n}ski}, Miko{\l}aj and {Kelley}, Michael S.~P. and {Kern}, Nicholas and {Kerzendorf}, Wolfgang E. and {Koch}, Eric W. and {Kulumani}, Shankar and {Lee}, Antony and {Ly}, Chun and {Ma}, Zhiyuan and {MacBride}, Conor and {Maljaars}, Jakob M. and {Muna}, Demitri and {Murphy}, N.~A. and {Norman}, Henrik and {O'Steen}, Richard and {Oman}, Kyle A. and {Pacifici}, Camilla and {Pascual}, Sergio and {Pascual-Granado}, J. and {Patil}, Rohit R. and {Perren}, Gabriel I. and {Pickering}, Timothy E. and {Rastogi}, Tanuj and {Roulston}, Benjamin R. and {Ryan}, Daniel F. and {Rykoff}, Eli S. and {Sabater}, Jose and {Sakurikar}, Parikshit and {Salgado}, Jes{\'u}s and {Sanghi}, Aniket and {Saunders}, Nicholas and {Savchenko}, Volodymyr and {Schwardt}, Ludwig and {Seifert-Eckert}, Michael and {Shih}, Albert Y. and {Jain}, Anany Shrey and {Shukla}, Gyanendra and {Sick}, Jonathan and {Simpson}, Chris and {Singanamalla}, Sudheesh and {Singer}, Leo P. and {Singhal}, Jaladh and {Sinha}, Manodeep and {Sip{\H{o}}cz}, Brigitta M. and {Spitler}, Lee R. and {Stansby}, David and {Streicher}, Ole and {{\v{S}}umak}, Jani and {Swinbank}, John D. and {Taranu}, Dan S. and {Tewary}, Nikita and {Tremblay}, Grant R. and {de Val-Borro}, Miguel and {Van Kooten}, Samuel J. and {Vasovi{\'c}}, Zlatan and {Verma}, Shresth and {de Miranda Cardoso}, Jos{\'e} Vin{\'\i}cius and {Williams}, Peter K.~G. and {Wilson}, Tom J. and {Winkel}, Benjamin and {Wood-Vasey}, W.~M. and {Xue}, Rui and {Yoachim}, Peter and {Zhang}, Chen and {Zonca}, Andrea and {Astropy Project Contributors}},
        title = "{The Astropy Project: Sustaining and Growing a Community-oriented Open-source Project and the Latest Major Release (v5.0) of the Core Package}",
      journal = {\apj},
         year = 2022,
        month = aug,
       volume = {935},
       number = {2},
          eid = {167},
        pages = {167},
          doi = {10.3847/1538-4357/ac7c74},
archivePrefix = {arXiv},
       eprint = {2206.14220},
 primaryClass = {astro-ph.IM},
       adsurl = {https://ui.adsabs.harvard.edu/abs/2022ApJ...935..167A}
}

@ARTICLE{Saifollahi2024,
       author = {{Saifollahi}, T. and {Voggel}, K. and {Lan{\c{c}}on}, A. and {Cantiello}, Michele and {Raj}, M.~A. and {Cuillandre}, J. -C. and {Larsen}, S.~S. and {Marleau}, F.~R. and {Venhola}, A. and {Schirmer}, M. and {Carollo}, D. and {Duc}, P. -A. and {Ferguson}, A.~M.~N. and {Hunt}, L.~K. and {K{\"u}mmel}, M. and {Laureijs}, R. and {Marchal}, O. and {Nucita}, A.~A. and {Peletier}, R.~F. and {Poulain}, M. and {Rejkuba}, M. and {S{\'a}nchez-Janssen}, R. and {Urbano}, M. and {Abdurro'uf} and {Altieri}, B. and {Baes}, M. and {Bolzonella}, M. and {Conselice}, C.~J. and {Cote}, P. and {Dimauro}, P. and {Gonzalez}, A.~H. and {Habas}, R. and {Hudelot}, P. and {Kluge}, M. and {Lonare}, P. and {Massari}, D. and {Romelli}, E. and {Scaramella}, R. and {Sola}, E. and {Stone}, C. and {Tortora}, C. and {van Mierlo}, S.~E. and {Knapen}, J.~H. and {Mart{\'\i}n-Fleitas}, J. and {Mora}, A. and {Rom{\'a}n}, J. and {Aghanim}, N. and {Amara}, A. and {Andreon}, S. and {Auricchio}, N. and {Baldi}, M. and {Balestra}, A. and {Bardelli}, S. and {Basset}, A. and {Bender}, R. and {Bonino}, D. and {Branchini}, E. and {Brescia}, M. and {Brinchmann}, J. and {Camera}, S. and {Capobianco}, V. and {Carbone}, C. and {Carretero}, J. and {Casas}, S. and {Castellano}, M. and {Cavuoti}, S. and {Cimatti}, A. and {Congedo}, G. and {Conversi}, L. and {Copin}, Y. and {Courbin}, F. and {Courtois}, H.~M. and {Cropper}, M. and {Da Silva}, A. and {Degaudenzi}, H. and {Di Giorgio}, A.~M. and {Dinis}, J. and {Dubath}, F. and {Dupac}, X. and {Dusini}, S. and {Fabricius}, M. and {Farina}, M. and {Farrens}, S. and {Ferriol}, S. and {Fosalba}, P. and {Frailis}, M. and {Franceschi}, E. and {Fumana}, M. and {Galeotta}, S. and {Garilli}, B. and {Gillard}, W. and {Gillis}, B. and {Giocoli}, C. and {G{\'o}mez-Alvarez}, P. and {Granett}, B.~R. and {Grazian}, A. and {Grupp}, F. and {Guzzo}, L. and {Haugan}, S.~V.~H. and {Hoar}, J. and {Hoekstra}, H. and {Holmes}, W. and {Hook}, I. and {Hormuth}, F. and {Hornstrup}, A. and {Jahnke}, K. and {Jhabvala}, M. and {Keih{\"a}nen}, E. and {Kermiche}, S. and {Kiessling}, A. and {Kitching}, T. and {Kohley}, R. and {Kubik}, B. and {Kuijken}, K. and {Kunz}, M. and {Kurki-Suonio}, H. and {Lahav}, O. and {Le Mignant}, D. and {Ligori}, S. and {Lilje}, P.~B. and {Lindholm}, V. and {Lloro}, I. and {Maino}, D. and {Maiorano}, E. and {Mansutti}, O. and {Marggraf}, O. and {Markovic}, K. and {Martinet}, N. and {Marulli}, F. and {Massey}, R. and {Maurogordato}, S. and {McCracken}, H.~J. and {Medinaceli}, E. and {Mei}, S. and {Melchior}, M. and {Mellier}, Y. and {Meneghetti}, M. and {Meylan}, G. and {Moresco}, M. and {Moscardini}, L. and {Munari}, E. and {Nakajima}, R. and {Nichol}, R.~C. and {Niemi}, S. -M. and {Padilla}, C. and {Paltani}, S. and {Pasian}, F. and {Pedersen}, K. and {Percival}, W.~J. and {Pettorino}, V. and {Pires}, S. and {Polenta}, G. and {Poncet}, M. and {Popa}, L.~A. and {Pozzetti}, L. and {Racca}, G.~D. and {Raison}, F. and {Rebolo}, R. and {Refregier}, A. and {Renzi}, A. and {Rhodes}, J. and {Riccio}, G. and {Roncarelli}, M. and {Rossetti}, E. and {Saglia}, R. and {Sapone}, D. and {Sartoris}, B. and {Schneider}, P. and {Schrabback}, T. and {Secroun}, A. and {Seidel}, G. and {Serrano}, S. and {Sirignano}, C. and {Sirri}, G. and {Stanco}, L. and {Tallada-Cresp{\'\i}}, P. and {Taylor}, A.~N. and {Teplitz}, H.~I. and {Tereno}, I. and {Toledo-Moreo}, R. and {Torradeflot}, F. and {Tsyganov}, A. and {Tutusaus}, I. and {Valentijn}, E.~A. and {Valenziano}, L. and {Vassallo}, T. and {Verdoes Kleijn}, G. and {Veropalumbo}, A. and {Wang}, Y. and {Weller}, J. and {Williams}, O.~R. and {Zamorani}, G. and {Zucca}, E. and {Biviano}, A. and {Burigana}, C. and {Scottez}, V. and {Simon}, P. and {Balogh}, M. and {Scott}, D.},
        title = "{Euclid: Early Release Observations {\textendash} Globular clusters in the Fornax galaxy cluster, from dwarf galaxies to the intracluster field}",
      journal = {\aap},
         year = 2025,
        month = may,
       volume = {697},
          eid = {A10},
        pages = {A10},
          doi = {10.1051/0004-6361/202450784},
archivePrefix = {arXiv},
       eprint = {2405.13500},
 primaryClass = {astro-ph.GA},
       adsurl = {https://ui.adsabs.harvard.edu/abs/2025A&A...697A..10S}
}

@ARTICLE{Fouesneau2010,
       author = {{Fouesneau}, M. and {Lan{\c{c}}on}, A.},
        title = "{Accounting for stochastic fluctuations when analysing the integrated light of star clusters. I. First systematics}",
      journal = {\aap},
         year = 2010,
        month = oct,
       volume = {521},
          eid = {A22},
        pages = {A22},
          doi = {10.1051/0004-6361/201014084},
archivePrefix = {arXiv},
       eprint = {1003.2334},
 primaryClass = {astro-ph.SR},
       adsurl = {https://ui.adsabs.harvard.edu/abs/2010A&A...521A..22F}
}

@ARTICLE{Kruijssen2019,
       author = {{Kruijssen}, J.~M. Diederik},
        title = "{The minimum metallicity of globular clusters and its physical origin - implications for the galaxy mass-metallicity relation and observations of proto-globular clusters at high redshift}",
      journal = {\mnras},
         year = 2019,
        month = jun,
       volume = {486},
       number = {1},
        pages = {L20-L25},
          doi = {10.1093/mnrasl/slz052},
archivePrefix = {arXiv},
       eprint = {1904.09987},
 primaryClass = {astro-ph.GA},
       adsurl = {https://ui.adsabs.harvard.edu/abs/2019MNRAS.486L..20K}
}

@ARTICLE{Ma2016,
       author = {{Ma}, Xiangcheng and {Hopkins}, Philip F. and {Faucher-Gigu{\`e}re}, Claude-Andr{\'e} and {Zolman}, Nick and {Muratov}, Alexander L. and {Kere{\v{s}}}, Du{\v{s}}an and {Quataert}, Eliot},
        title = "{The origin and evolution of the galaxy mass-metallicity relation}",
      journal = {\mnras},
         year = 2016,
        month = feb,
       volume = {456},
       number = {2},
        pages = {2140-2156},
          doi = {10.1093/mnras/stv2659},
archivePrefix = {arXiv},
       eprint = {1504.02097},
 primaryClass = {astro-ph.GA},
       adsurl = {https://ui.adsabs.harvard.edu/abs/2016MNRAS.456.2140M}
}

@ARTICLE{Adamo2024,
       author = {{Adamo}, Angela and {Bradley}, Larry D. and {Vanzella}, Eros and {Claeyssens}, Ad{\'e}la{\"\i}de and {Welch}, Brian and {Diego}, Jose M. and {Mahler}, Guillaume and {Oguri}, Masamune and {Sharon}, Keren and {Abdurro'uf} and {Hsiao}, Tiger Yu-Yang and {Xu}, Xinfeng and {Messa}, Matteo and {Lassen}, Augusto E. and {Zackrisson}, Erik and {Brammer}, Gabriel and {Coe}, Dan and {Kokorev}, Vasily and {Ricotti}, Massimo and {Zitrin}, Adi and {Fujimoto}, Seiji and {Inoue}, Akio K. and {Resseguier}, Tom and {Rigby}, Jane R. and {Jim{\'e}nez-Teja}, Yolanda and {Windhorst}, Rogier A. and {Hashimoto}, Takuya and {Tamura}, Yoichi},
        title = "{Bound star clusters observed in a lensed galaxy 460 Myr after the Big Bang}",
      journal = {\nat},
         year = 2024,
        month = aug,
       volume = {632},
       number = {8025},
        pages = {513-516},
          doi = {10.1038/s41586-024-07703-7},
archivePrefix = {arXiv},
       eprint = {2401.03224},
 primaryClass = {astro-ph.GA},
       adsurl = {https://ui.adsabs.harvard.edu/abs/2024Natur.632..513A}
}

@ARTICLE{Crnojevic2016,
       author = {{Crnojevi{\'c}}, D. and {Sand}, D.~J. and {Zaritsky}, D. and {Spekkens}, K. and {Willman}, B. and {Hargis}, J.~R.},
        title = "{Deep Imaging of Eridanus II and Its Lone Star Cluster}",
      journal = {\apjl},
         year = 2016,
        month = jun,
       volume = {824},
       number = {1},
          eid = {L14},
        pages = {L14},
          doi = {10.3847/2041-8205/824/1/L14},
archivePrefix = {arXiv},
       eprint = {1604.08590},
 primaryClass = {astro-ph.GA},
       adsurl = {https://ui.adsabs.harvard.edu/abs/2016ApJ...824L..14C}
}

@ARTICLE{Jang2012,
       author = {{Jang}, In Sung and {Lim}, Sungsoon and {Park}, Hong Soo and {Lee}, Myung Gyoon},
        title = "{Discovery of the Most Isolated Globular Cluster in the Local Universe}",
      journal = {\apjl},
         year = 2012,
        month = may,
       volume = {751},
       number = {1},
          eid = {L19},
        pages = {L19},
          doi = {10.1088/2041-8205/751/1/L19},
archivePrefix = {arXiv},
       eprint = {1205.3559},
 primaryClass = {astro-ph.CO},
       adsurl = {https://ui.adsabs.harvard.edu/abs/2012ApJ...751L..19J}
}

@ARTICLE{Demers2006,
       author = {{Demers}, Serge and {Battinelli}, Paolo and {Kunkel}, William E.},
        title = "{A Local Group Polar Ring Galaxy: NGC 6822}",
      journal = {\apjl},
         year = 2006,
        month = jan,
       volume = {636},
       number = {2},
        pages = {L85-L88},
          doi = {10.1086/500207},
archivePrefix = {arXiv},
       eprint = {astro-ph/0510778},
 primaryClass = {astro-ph},
       adsurl = {https://ui.adsabs.harvard.edu/abs/2006ApJ...636L..85D}
}

@ARTICLE{Lahen2025,
       author = {{Lah{\'e}n}, Natalia and {Rantala}, Antti and {Naab}, Thorsten and {Partmann}, Christian and {Johansson}, Peter H. and {Hislop}, Jessica May},
        title = "{The formation, evolution and disruption of star clusters with improved gravitational dynamics in simulated dwarf galaxies}",
      journal = {\mnras},
         year = 2025,
        month = feb,
          doi = {10.1093/mnras/staf350},
 primaryClass = {astro-ph.GA},
       adsurl = {https://ui.adsabs.harvard.edu/abs/2025MNRAS.tmp..325L}
}

@ARTICLE{Johnson2015,
       author = {{Johnson}, L. Clifton and {Seth}, Anil C. and {Dalcanton}, Julianne J. and {Wallace}, Matthew L. and {Simpson}, Robert J. and {Lintott}, Chris J. and {Kapadia}, Amit and {Skillman}, Evan D. and {Caldwell}, Nelson and {Fouesneau}, Morgan and {Weisz}, Daniel R. and {Williams}, Benjamin F. and {Beerman}, Lori C. and {Gouliermis}, Dimitrios A. and {Sarajedini}, Ata},
        title = "{PHAT Stellar Cluster Survey. II. Andromeda Project Cluster Catalog}",
      journal = {\apj},
         year = 2015,
        month = apr,
       volume = {802},
       number = {2},
          eid = {127},
        pages = {127},
          doi = {10.1088/0004-637X/802/2/127},
archivePrefix = {arXiv},
       eprint = {1501.04966},
 primaryClass = {astro-ph.GA},
       adsurl = {https://ui.adsabs.harvard.edu/abs/2015ApJ...802..127J}
}

@ARTICLE{Johnson2022,
       author = {{Johnson}, L. Clifton and {Wainer}, Tobin M. and {Torresvillanueva}, Estephani E. and {Seth}, Anil C. and {Williams}, Benjamin F. and {Durbin}, Meredith J. and {Dalcanton}, Julianne J. and {Weisz}, Daniel R. and {Bell}, Eric F. and {Guhathakurta}, Puragra and {Skillman}, Evan and {Smercina}, Adam and {Phatter Collaboration}},
        title = "{The Panchromatic Hubble Andromeda Treasury: Triangulum Extended Region (PHATTER). IV. Star Cluster Catalog}",
      journal = {\apj},
         year = 2022,
        month = oct,
       volume = {938},
       number = {1},
          eid = {81},
        pages = {81},
          doi = {10.3847/1538-4357/ac8def},
archivePrefix = {arXiv},
       eprint = {2208.11760},
 primaryClass = {astro-ph.GA},
       adsurl = {https://ui.adsabs.harvard.edu/abs/2022ApJ...938...81J}
}

@ARTICLE{Grasha2019,
       author = {{Grasha}, K. and {Calzetti}, D. and {Adamo}, A. and {Kennicutt}, R.~C. and {Elmegreen}, B.~G. and {Messa}, M. and {Dale}, D.~A. and {Fedorenko}, K. and {Mahadevan}, S. and {Grebel}, E.~K. and {Fumagalli}, M. and {Kim}, H. and {Dobbs}, C.~L. and {Gouliermis}, D.~A. and {Ashworth}, G. and {Gallagher}, J.~S. and {Smith}, L.~J. and {Tosi}, M. and {Whitmore}, B.~C. and {Schinnerer}, E. and {Colombo}, D. and {Hughes}, A. and {Leroy}, A.~K. and {Meidt}, S.~E.},
        title = "{The spatial relation between young star clusters and molecular clouds in M51 with LEGUS}",
      journal = {\mnras},
         year = 2019,
        month = mar,
       volume = {483},
       number = {4},
        pages = {4707-4723},
          doi = {10.1093/mnras/sty3424},
archivePrefix = {arXiv},
       eprint = {1812.06109},
 primaryClass = {astro-ph.GA},
       adsurl = {https://ui.adsabs.harvard.edu/abs/2019MNRAS.483.4707G}
}

@ARTICLE{Rhode2004,
       author = {{Rhode}, Katherine L. and {Zepf}, Stephen E.},
        title = "{The Globular Cluster Systems of the Early-Type Galaxies NGC 3379, NGC 4406, and NGC 4594 and Implications for Galaxy Formation}",
      journal = {\aj},
         year = 2004,
        month = jan,
       volume = {127},
       number = {1},
        pages = {302-317},
          doi = {10.1086/380616},
archivePrefix = {arXiv},
       eprint = {astro-ph/0310277},
 primaryClass = {astro-ph},
       adsurl = {https://ui.adsabs.harvard.edu/abs/2004AJ....127..302R}
}

@ARTICLE{Massari2019,
       author = {{Massari}, D. and {Koppelman}, H.~H. and {Helmi}, A.},
        title = "{Origin of the system of globular clusters in the Milky Way}",
      journal = {\aap},
         year = 2019,
        month = oct,
       volume = {630},
          eid = {L4},
        pages = {L4},
          doi = {10.1051/0004-6361/201936135},
archivePrefix = {arXiv},
       eprint = {1906.08271},
 primaryClass = {astro-ph.GA},
       adsurl = {https://ui.adsabs.harvard.edu/abs/2019A&A...630L...4M}
}

@ARTICLE{Monty2024,
       author = {{Monty}, Stephanie and {Belokurov}, Vasily and {Sanders}, Jason L. and {Hansen}, Terese T. and {Sakari}, Charli M. and {McKenzie}, Madeleine and {Myeong}, GyuChul and {Davies}, Elliot Y. and {Ardern-Arentsen}, Anke and {Massari}, Davide},
        title = "{The ratio of [Eu/{\ensuremath{\alpha}}] differentiates accreted/in situ Milky Way stars across metallicities, as indicated by both field stars and globular clusters}",
      journal = {\mnras},
         year = 2024,
        month = sep,
       volume = {533},
       number = {2},
        pages = {2420-2440},
          doi = {10.1093/mnras/stae1895},
archivePrefix = {arXiv},
       eprint = {2405.08963},
 primaryClass = {astro-ph.GA},
       adsurl = {https://ui.adsabs.harvard.edu/abs/2024MNRAS.533.2420M}
}

@ARTICLE{Mackey2019,
       author = {{Mackey}, A.~D. and {Ferguson}, A.~M.~N. and {Huxor}, A.~P. and {Veljanoski}, J. and {Lewis}, G.~F. and {McConnachie}, A.~W. and {Martin}, N.~F. and {Ibata}, R.~A. and {Irwin}, M.~J. and {C{\^o}t{\'e}}, P. and {Collins}, M.~L.~M. and {Tanvir}, N.~R. and {Bate}, N.~F.},
        title = "{The outer halo globular cluster system of M31 - III. Relationship to the stellar halo}",
      journal = {\mnras},
         year = 2019,
        month = apr,
       volume = {484},
       number = {2},
        pages = {1756-1789},
          doi = {10.1093/mnras/stz072},
archivePrefix = {arXiv},
       eprint = {1810.10719},
 primaryClass = {astro-ph.GA},
       adsurl = {https://ui.adsabs.harvard.edu/abs/2019MNRAS.484.1756M}
}

@ARTICLE{Foster2014,
       author = {{Foster}, C. and {Lux}, H. and {Romanowsky}, A.~J. and {Mart{\'\i}nez-Delgado}, D. and {Zibetti}, S. and {Arnold}, J.~A. and {Brodie}, J.~P. and {Ciardullo}, R. and {GaBany}, R.~J. and {Merrifield}, M.~R. and {Singh}, N. and {Strader}, J.},
        title = "{Kinematics and simulations of the stellar stream in the halo of the Umbrella Galaxy}",
      journal = {\mnras},
         year = 2014,
        month = aug,
       volume = {442},
       number = {4},
        pages = {3544-3564},
          doi = {10.1093/mnras/stu1074},
archivePrefix = {arXiv},
       eprint = {1406.5511},
 primaryClass = {astro-ph.GA},
       adsurl = {https://ui.adsabs.harvard.edu/abs/2014MNRAS.442.3544F}
}

@ARTICLE{Mowla2024,
       author = {{Mowla}, Lamiya and {Iyer}, Kartheik and {Asada}, Yoshihisa and {Desprez}, Guillaume and {Tan}, Vivian Yun Yan and {Martis}, Nicholas and {Sarrouh}, Ghassan and {Strait}, Victoria and {Abraham}, Roberto and {Brada{\v{c}}}, Maru{\v{s}}a and {Brammer}, Gabriel and {Muzzin}, Adam and {Pacifici}, Camilla and {Ravindranath}, Swara and {Sawicki}, Marcin and {Willott}, Chris and {Estrada-Carpenter}, Vince and {Jahan}, Nusrath and {Noirot}, Ga{\"e}l and {Matharu}, Jasleen and {Rihtar{\v{s}}i{\v{c}}}, Gregor and {Zabl}, Johannes},
        title = "{Formation of a low-mass galaxy from star clusters in a 600-million-year-old Universe}",
      journal = {\nat},
         year = 2024,
        month = dec,
       volume = {636},
       number = {8042},
        pages = {332-336},
          doi = {10.1038/s41586-024-08293-0},
       adsurl = {https://ui.adsabs.harvard.edu/abs/2024Natur.636..332M}
}

@ARTICLE{Adamo2017,
       author = {{Adamo}, A. and {Ryon}, J.~E. and {Messa}, M. and {Kim}, H. and {Grasha}, K. and {Cook}, D.~O. and {Calzetti}, D. and {Lee}, J.~C. and {Whitmore}, B.~C. and {Elmegreen}, B.~G. and {Ubeda}, L. and {Smith}, L.~J. and {Bright}, S.~N. and {Runnholm}, A. and {Andrews}, J.~E. and {Fumagalli}, M. and {Gouliermis}, D.~A. and {Kahre}, L. and {Nair}, P. and {Thilker}, D. and {Walterbos}, R. and {Wofford}, A. and {Aloisi}, A. and {Ashworth}, G. and {Brown}, T.~M. and {Chandar}, R. and {Christian}, C. and {Cignoni}, M. and {Clayton}, G.~C. and {Dale}, D.~A. and {de Mink}, S.~E. and {Dobbs}, C. and {Elmegreen}, D.~M. and {Evans}, A.~S. and {Gallagher}, III, J.~S. and {Grebel}, E.~K. and {Herrero}, A. and {Hunter}, D.~A. and {Johnson}, K.~E. and {Kennicutt}, R.~C. and {Krumholz}, M.~R. and {Lennon}, D. and {Levay}, K. and {Martin}, C. and {Nota}, A. and {{\"O}stlin}, G. and {Pellerin}, A. and {Prieto}, J. and {Regan}, M.~W. and {Sabbi}, E. and {Sacchi}, E. and {Schaerer}, D. and {Schiminovich}, D. and {Shabani}, F. and {Tosi}, M. and {Van Dyk}, S.~D. and {Zackrisson}, E.},
        title = "{Legacy ExtraGalactic UV Survey with The Hubble Space Telescope: Stellar Cluster Catalogs and First Insights Into Cluster Formation and Evolution in NGC 628}",
      journal = {\apj},
         year = 2017,
        month = jun,
       volume = {841},
       number = {2},
          eid = {131},
        pages = {131},
          doi = {10.3847/1538-4357/aa7132},
archivePrefix = {arXiv},
       eprint = {1705.01588},
 primaryClass = {astro-ph.GA},
       adsurl = {https://ui.adsabs.harvard.edu/abs/2017ApJ...841..131A}
}

@ARTICLE{Cote1998,
       author = {{C{\^o}t{\'e}}, Patrick and {Marzke}, Ronald O. and {West}, Michael J.},
        title = "{The Formation of Giant Elliptical Galaxies and Their Globular Cluster Systems}",
      journal = {\apj},
         year = 1998,
        month = jul,
       volume = {501},
       number = {2},
        pages = {554-570},
          doi = {10.1086/305838},
archivePrefix = {arXiv},
       eprint = {astro-ph/9804319},
 primaryClass = {astro-ph},
       adsurl = {https://ui.adsabs.harvard.edu/abs/1998ApJ...501..554C}
}

@ARTICLE{Nersesian2019,
       author = {{Nersesian}, A. and {Xilouris}, E.~M. and {Bianchi}, S. and {Galliano}, F. and {Jones}, A.~P. and {Baes}, M. and {Casasola}, V. and {Cassar{\`a}}, L.~P. and {Clark}, C.~J.~R. and {Davies}, J.~I. and {Decleir}, M. and {Dobbels}, W. and {De Looze}, I. and {De Vis}, P. and {Fritz}, J. and {Galametz}, M. and {Madden}, S.~C. and {Mosenkov}, A.~V. and {Tr{\v{c}}ka}, A. and {Verstocken}, S. and {Viaene}, S. and {Lianou}, S.},
        title = "{Old and young stellar populations in DustPedia galaxies and their role in dust heating}",
      journal = {\aap},
         year = 2019,
        month = apr,
       volume = {624},
          eid = {A80},
        pages = {A80},
          doi = {10.1051/0004-6361/201935118},
archivePrefix = {arXiv},
       eprint = {1903.05933},
 primaryClass = {astro-ph.GA},
       adsurl = {https://ui.adsabs.harvard.edu/abs/2019A&A...624A..80N}
}

@ARTICLE{Hodge1988,
       author = {{Hodge}, Paul and {Kennicutt}, Jr., Robert C. and {Lee}, Myung Gyoon},
        title = "{The H II Regions of NGC 6822. I. an Atlas of 157 H II Regions}",
      journal = {\pasp},
         year = 1988,
        month = aug,
       volume = {100},
        pages = {917},
          doi = {10.1086/132254},
       adsurl = {https://ui.adsabs.harvard.edu/abs/1988PASP..100..917H}
}

@ARTICLE{deBlok2000,
       author = {{de Blok}, W.~J.~G. and {Walter}, F.},
        title = "{Evidence for Tidal Interaction and a Supergiant H I Shell in the Local Group Dwarf Galaxy NGC 6822}",
      journal = {\apjl},
         year = 2000,
        month = jul,
       volume = {537},
       number = {2},
        pages = {L95-L98},
          doi = {10.1086/312777},
archivePrefix = {arXiv},
       eprint = {astro-ph/0005473},
 primaryClass = {astro-ph},
       adsurl = {https://ui.adsabs.harvard.edu/abs/2000ApJ...537L..95D}
}

@ARTICLE{Letarte2002,
       author = {{Letarte}, Bruno and {Demers}, Serge and {Battinelli}, Paolo and {Kunkel}, W.~E.},
        title = "{The Extent of NGC 6822 Revealed by Its C Star Population}",
      journal = {\aj},
         year = 2002,
        month = feb,
       volume = {123},
       number = {2},
        pages = {832-839},
          doi = {10.1086/338319},
archivePrefix = {arXiv},
       eprint = {astro-ph/0110026},
 primaryClass = {astro-ph},
       adsurl = {https://ui.adsabs.harvard.edu/abs/2002AJ....123..832L}
}

@ARTICLE{Battaglia2022,
       author = {{Battaglia}, G. and {Taibi}, S. and {Thomas}, G.~F. and {Fritz}, T.~K.},
        title = "{Gaia early DR3 systemic motions of Local Group dwarf galaxies and orbital properties with a massive Large Magellanic Cloud}",
      journal = {\aap},
         year = 2022,
        month = jan,
       volume = {657},
          eid = {A54},
        pages = {A54},
          doi = {10.1051/0004-6361/202141528},
archivePrefix = {arXiv},
       eprint = {2106.08819},
 primaryClass = {astro-ph.GA},
       adsurl = {https://ui.adsabs.harvard.edu/abs/2022A&A...657A..54B}
}

@ARTICLE{Bennet2024,
       author = {{Bennet}, Paul and {Patel}, Ekta and {Sohn}, Sangmo Tony and {del Pino Molina}, Andr{\'e}s and {van der Marel}, Roeland P. and {Libralato}, Mattia and {Watkins}, Laura L. and {Aparicio}, Antonio and {Besla}, Gurtina and {Gallart}, Carme and {Fardal}, Mark A. and {Monelli}, Matteo and {Sacchi}, Elena and {Tollerud}, Erik and {Weisz}, Daniel R.},
        title = "{Proper Motions and Orbits of Distant Local Group Dwarf Galaxies from a Combination of Gaia and Hubble Data}",
      journal = {\apj},
         year = 2024,
        month = aug,
       volume = {971},
       number = {1},
          eid = {98},
        pages = {98},
          doi = {10.3847/1538-4357/ad5349},
archivePrefix = {arXiv},
       eprint = {2312.09276},
 primaryClass = {astro-ph.GA},
       adsurl = {https://ui.adsabs.harvard.edu/abs/2024ApJ...971...98B}
}

@ARTICLE{Huxor2008,
       author = {{Huxor}, A.~P. and {Tanvir}, N.~R. and {Ferguson}, A.~M.~N. and {Irwin}, M.~J. and {Ibata}, R. and {Bridges}, T. and {Lewis}, G.~F.},
        title = "{Globular clusters in the outer halo of M31: the survey}",
      journal = {\mnras},
         year = 2008,
        month = apr,
       volume = {385},
       number = {4},
        pages = {1989-1997},
          doi = {10.1111/j.1365-2966.2008.12882.x},
archivePrefix = {arXiv},
       eprint = {0801.0002},
 primaryClass = {astro-ph},
       adsurl = {https://ui.adsabs.harvard.edu/abs/2008MNRAS.385.1989H}
}

@ARTICLE{Barnard1884,
       author = {{Barnard}, E.~E.},
        title = "{New Nebula near General Catalogue No 4510}",
      journal = {Astronomische Nachrichten},
         year = 1884,
        month = oct,
       volume = {110},
        pages = {125},
          doi = {10.1002/asna.18841100805},
       adsurl = {https://ui.adsabs.harvard.edu/abs/1884AN....110..125B}
}

@ARTICLE{Weldrake2003,
       author = {{Weldrake}, D.~T.~F. and {de Blok}, W.~J.~G. and {Walter}, F.},
        title = "{A high-resolution rotation curve of NGC 6822: a test-case for cold dark matter}",
      journal = {\mnras},
         year = 2003,
        month = mar,
       volume = {340},
       number = {1},
        pages = {12-28},
          doi = {10.1046/j.1365-8711.2003.06170.x},
archivePrefix = {arXiv},
       eprint = {astro-ph/0210568},
 primaryClass = {astro-ph},
       adsurl = {https://ui.adsabs.harvard.edu/abs/2003MNRAS.340...12W}
}

@ARTICLE{Deger2022,
       author = {{Deger}, Sinan and {Lee}, Janice C. and {Whitmore}, Bradley C. and {Thilker}, David A. and {Boquien}, Mederic and {Chandar}, Rupali and {Dale}, Daniel A. and {Ubeda}, Leonardo and {White}, Rick and {Grasha}, Kathryn and {Glover}, Simon C.~O. and {Schruba}, Andreas and {Barnes}, Ashley T. and {Klessen}, Ralf and {Kruijssen}, J.~M. Diederik and {Rosolowsky}, Erik and {Williams}, Thomas G.},
        title = "{Bright, relatively isolated star clusters in PHANGS-HST galaxies: Aperture corrections, quantitative morphologies, and comparison with synthetic stellar population models}",
      journal = {\mnras},
         year = 2022,
        month = feb,
       volume = {510},
       number = {1},
        pages = {32-53},
          doi = {10.1093/mnras/stab3213},
archivePrefix = {arXiv},
       eprint = {2110.10708},
 primaryClass = {astro-ph.GA},
       adsurl = {https://ui.adsabs.harvard.edu/abs/2022MNRAS.510...32D}
}

@ARTICLE{Huxor2005,
       author = {{Huxor}, A.~P. and {Tanvir}, N.~R. and {Irwin}, M.~J. and {Ibata}, R. and {Collett}, J.~L. and {Ferguson}, A.~M.~N. and {Bridges}, T. and {Lewis}, G.~F.},
        title = "{A new population of extended, luminous star clusters in the halo of M31}",
      journal = {\mnras},
         year = 2005,
        month = jul,
       volume = {360},
       number = {3},
        pages = {1007-1012},
          doi = {10.1111/j.1365-2966.2005.09086.x},
archivePrefix = {arXiv},
       eprint = {astro-ph/0412223},
 primaryClass = {astro-ph},
       adsurl = {https://ui.adsabs.harvard.edu/abs/2005MNRAS.360.1007H}
}

@ARTICLE{Deason2014,
       author = {{Deason}, Alis and {Wetzel}, Andrew and {Garrison-Kimmel}, Shea},
        title = "{Satellite Dwarf Galaxies in a Hierarchical Universe: The Prevalence of Dwarf-Dwarf Major Mergers}",
      journal = {\apj},
         year = 2014,
        month = oct,
       volume = {794},
       number = {2},
          eid = {115},
        pages = {115},
          doi = {10.1088/0004-637X/794/2/115},
archivePrefix = {arXiv},
       eprint = {1406.3344},
 primaryClass = {astro-ph.GA},
       adsurl = {https://ui.adsabs.harvard.edu/abs/2014ApJ...794..115D}
}

@ARTICLE{Swan2016,
       author = {{Swan}, J. and {Cole}, A.~A. and {Tolstoy}, E. and {Irwin}, M.~J.},
        title = "{Ca II triplet spectroscopy of RGB stars in NGC 6822: kinematics and metallicities}",
      journal = {\mnras},
         year = 2016,
        month = mar,
       volume = {456},
       number = {4},
        pages = {4315-4327},
          doi = {10.1093/mnras/stv2774},
archivePrefix = {arXiv},
       eprint = {1602.01897},
 primaryClass = {astro-ph.GA},
       adsurl = {https://ui.adsabs.harvard.edu/abs/2016MNRAS.456.4315S}
}

@ARTICLE{Salaris1993,
       author = {{Salaris}, Maurizio and {Chieffi}, Alessandro and {Straniero}, Oscar},
        title = "{The alpha -enhanced Isochrones and Their Impact on the FITS to the Galactic Globular Cluster System}",
      journal = {\apj},
         year = 1993,
        month = sep,
       volume = {414},
        pages = {580},
          doi = {10.1086/173105},
       adsurl = {https://ui.adsabs.harvard.edu/abs/1993ApJ...414..580S}
}

@ARTICLE{Mackey2010,
       author = {{Mackey}, A.~D. and {Huxor}, A.~P. and {Ferguson}, A.~M.~N. and {Irwin}, M.~J. and {Tanvir}, N.~R. and {McConnachie}, A.~W. and {Ibata}, R.~A. and {Chapman}, S.~C. and {Lewis}, G.~F.},
        title = "{Evidence for an Accretion Origin for the Outer Halo Globular Cluster System of M31}",
      journal = {\apjl},
         year = 2010,
        month = jul,
       volume = {717},
       number = {1},
        pages = {L11-L16},
          doi = {10.1088/2041-8205/717/1/L11},
archivePrefix = {arXiv},
       eprint = {1005.3812},
 primaryClass = {astro-ph.GA},
       adsurl = {https://ui.adsabs.harvard.edu/abs/2010ApJ...717L..11M}
}

@ARTICLE{Fall2001,
       author = {{Fall}, S. Michael and {Zhang}, Qing},
        title = "{Dynamical Evolution of the Mass Function of Globular Star Clusters}",
      journal = {\apj},
         year = 2001,
        month = nov,
       volume = {561},
       number = {2},
        pages = {751-765},
          doi = {10.1086/323358},
archivePrefix = {arXiv},
       eprint = {astro-ph/0107298},
 primaryClass = {astro-ph},
       adsurl = {https://ui.adsabs.harvard.edu/abs/2001ApJ...561..751F}
}

@ARTICLE{Elmegreen1997,
       author = {{Elmegreen}, Bruce G. and {Efremov}, Yuri N.},
        title = "{A Universal Formation Mechanism for Open and Globular Clusters in Turbulent Gas}",
      journal = {\apj},
         year = 1997,
        month = may,
       volume = {480},
       number = {1},
        pages = {235-245},
          doi = {10.1086/303966},
       adsurl = {https://ui.adsabs.harvard.edu/abs/1997ApJ...480..235E}
}

@article{vdVen2006, 
  year     = {2006}, 
  title    = {The dynamical distance and intrinsic structure of the globular cluster ω Centauri}, 
  author   = {van de Ven, G and van den Bosch, R C E and Verolme, E K and de Zeeuw, P T}, 
  journal  = {\aap}, 
  doi      = {10.1051/0004-6361:20053061}, 
  pages    = {513 -- 543}, 
  number   = {2}, 
  volume   = {445}, 
  language = {English}
}

@article{2003MNRAS346L11B, 
  year     = {2003}, 
  title    = {Formation of ω Centauri from an ancient nucleated dwarf galaxy in the young Galactic disc}, 
  author   = {Bekki, K and Freeman, K C}, 
  journal  = {\mnras}, 
  doi      = {10.1046/j.1365-2966.2003.07275.x}, 
  pages    = {L11 -- L15}, 
  number   = {2}, 
  volume   = {346}, 
  language = {English}
}

@ARTICLE{Meylan1997,
       author = {{Meylan}, G. and {Heggie}, D.~C.},
        title = "{Internal dynamics of globular clusters}",
      journal = {\aapr},
         year = 1997,
        month = jan,
       volume = {8},
        pages = {1-143},
          doi = {10.1007/s001590050008},
archivePrefix = {arXiv},
       eprint = {astro-ph/9610076},
 primaryClass = {astro-ph},
       adsurl = {https://ui.adsabs.harvard.edu/abs/1997A&ARv...8....1M}
}

@article{Bianchini2014bk, 
  year     = {2014}, 
  title    = {The inefficiency of satellite accretion in forming extended star clusters}, 
  author   = {Bianchini, P and Renaud, F and Gieles, M and Varri, A L}, 
  journal  = {\mnras}, 
  doi      = {10.1093/mnrasl/slu177}, 
  pages    = {L40 -- L44}, 
  number   = {1}, 
  volume   = {447}, 
  language = {English}, 
  month    = {11}
}

@article{2010MNRAS4041157M, 
  year    = {2010}, 
  title   = {A faint extended cluster in the outskirts of {NGC} 5128: evidence of a low mass accretion}, 
  author  = {Mouhcine, M and Harris, W E and Ibata, R and Rejkuba, M}, 
  journal = {\mnras}, 
  doi     = {10.1111/j.1365-2966.2010.16363.x}, 
  pages   = {1157 -- 1164}, 
  number  = {3}, 
  volume  = {404}
}

@ARTICLE{Weisz2016,
       author = {{Weisz}, Daniel R. and {Koposov}, Sergey E. and {Dolphin}, Andrew E. and {Belokurov}, Vasily and {Gieles}, Mark and {Mateo}, Mario L. and {Olszewski}, Edward W. and {Sills}, Alison and {Walker}, Matthew G.},
        title = "{A Hubble Space Telescope Study of the Enigmatic Milky Way Halo Globular Cluster Crater*}",
      journal = {\apj},
         year = 2016,
        month = may,
       volume = {822},
       number = {1},
          eid = {32},
        pages = {32},
          doi = {10.3847/0004-637X/822/1/32},
archivePrefix = {arXiv},
       eprint = {1510.08533},
 primaryClass = {astro-ph.GA},
       adsurl = {https://ui.adsabs.harvard.edu/abs/2016ApJ...822...32W}
}

@ARTICLE{Jones2023,
       author = {{Jones}, Michael G. and {Karunakaran}, Ananthan and {Bennet}, Paul and {Sand}, David J. and {Spekkens}, Kristine and {Mutlu-Pakdil}, Bur{\c{c}}in and {Crnojevi{\'c}}, Denija and {Janowiecki}, Steven and {Leisman}, Lukas and {Fielder}, Catherine E.},
        title = "{Gas-rich, Field Ultra-diffuse Galaxies Host Few Globular Clusters}",
      journal = {\apjl},
         year = 2023,
        month = jan,
       volume = {942},
       number = {1},
          eid = {L5},
        pages = {L5},
          doi = {10.3847/2041-8213/acaaab},
archivePrefix = {arXiv},
       eprint = {2211.00651},
 primaryClass = {astro-ph.GA},
       adsurl = {https://ui.adsabs.harvard.edu/abs/2023ApJ...942L...5J}
}

@ARTICLE{Eadie2022,
       author = {{Eadie}, Gwendolyn M. and {Harris}, William E. and {Springford}, Aaron},
        title = "{Clearing the Hurdle: The Mass of Globular Cluster Systems as a Function of Host Galaxy Mass}",
      journal = {\apj},
         year = 2022,
        month = feb,
       volume = {926},
       number = {2},
          eid = {162},
        pages = {162},
          doi = {10.3847/1538-4357/ac33b0},
archivePrefix = {arXiv},
       eprint = {2110.15376},
 primaryClass = {astro-ph.GA},
       adsurl = {https://ui.adsabs.harvard.edu/abs/2022ApJ...926..162E}
}

@ARTICLE{Jarrett2003,
       author = {{Jarrett}, T.~H. and {Chester}, T. and {Cutri}, R. and {Schneider}, S.~E. and {Huchra}, J.~P.},
        title = "{The 2MASS Large Galaxy Atlas}",
      journal = {\aj},
         year = 2003,
        month = feb,
       volume = {125},
       number = {2},
        pages = {525-554},
          doi = {10.1086/345794},
       adsurl = {https://ui.adsabs.harvard.edu/abs/2003AJ....125..525J}
}

@article{carlsten2022elves-c85, 
  year    = {2022}, 
  title   = {{ELVES} {II}: Globular Clusters and Nuclear Star Clusters of Dwarf Galaxies: the Importance of Environment}, 
  author  = {Carlsten, Scott G. and Greene, Jenny E. and Beaton, Rachael L. and Greco, Johnny P.}, 
  journal = {\apj}, 
  issn    = {0004-637X}, 
  doi     = {10.3847/1538-4357/ac457e}, 
  eprint  = {2105.03440}, 
  pages   = {44}, 
  number  = {1}, 
  volume  = {927}
}

@ARTICLE{Lim2025,
       author = {{Lim}, Sungsoon and {Peng}, Eric W. and {C{\^o}t{\'e}}, Patrick and {Ferrarese}, Laura and {Roediger}, Joel C. and {Liu}, Chengze and {Spengler}, Chelsea and {Sola}, Elisabeth and {Duc}, Pierre-Alain and {Sales}, Laura V. and {Blakeslee}, John P. and {Cuillandre}, Jean-Charles and {Durrell}, Patrick R. and {Emsellem}, Eric and {Gwyn}, Stephen D.~J. and {Lan{\c{c}}on}, Ariane and {Marleau}, Francine R. and {Mihos}, J. Christopher and {M{\"u}ller}, Oliver and {Puzia}, Thomas H. and {S{\'a}nchez-Janssen}, Rub{\'e}n},
        title = "{The Spatial Distribution of Globular Cluster Systems in Early-type Galaxies: Estimation Procedure and Catalog of Properties for Globular Cluster Systems Observed with Deep Imaging Surveys}",
      journal = {\apjs},
         year = 2025,
        month = feb,
       volume = {276},
       number = {2},
          eid = {34},
        pages = {34},
          doi = {10.3847/1538-4365/ad97b7},
archivePrefix = {arXiv},
       eprint = {2411.17049},
 primaryClass = {astro-ph.GA},
       adsurl = {https://ui.adsabs.harvard.edu/abs/2025ApJS..276...34L}
}

@article{floyd2024phangs-11d, 
  year    = {2024}, 
  title   = {{PHANGS} Hubble Space Telescope Treasury Survey: Globular Cluster Systems in 17 Nearby Spiral Galaxies}, 
  author  = {Floyd, Matthew and Chandar, Rupali and Whitmore, Bradley C. and Thilker, David A. and Lee, Janice C. and Pauline, Rachel E. and Thomas, Zion L. and Berschback, William J. and Henny, Kiana F. and Dale, Daniel A. and Klessen, Ralf S. and Schinnerer, Eva and Grasha, Kathryn and Boquien, Médéric and Larson, Kirsten L. and Deger, Sinan and Barnes, Ashley T. and Leroy, Adam K. and Rosolowsky, Erik and Williams, Thomas G. and Úbeda, Leonardo}, 
  journal = {\aj}, 
  issn    = {0004-6256}, 
  doi     = {10.3847/1538-3881/ad1889}, 
  eprint  = {2403.13908}, 
  pages   = {95}, 
  number  = {3}, 
  volume  = {167}
}

@ARTICLE{Glatt2008,
       author = {{Glatt}, Katharina and {Grebel}, Eva K. and {Sabbi}, Elena and {Gallagher}, III, John S. and {Nota}, Antonella and {Sirianni}, Marco and {Clementini}, Gisella and {Tosi}, Monica and {Harbeck}, Daniel and {Koch}, Andreas and {Kayser}, Andrea and {Da Costa}, Gary},
        title = "{Age Determination of Six Intermediate-Age Small Magellanic Cloud Star Clusters with HST/ACS}",
      journal = {\aj},
         year = 2008,
        month = oct,
       volume = {136},
       number = {4},
        pages = {1703-1727},
          doi = {10.1088/0004-6256/136/4/1703},
archivePrefix = {arXiv},
       eprint = {0807.3744},
 primaryClass = {astro-ph},
       adsurl = {https://ui.adsabs.harvard.edu/abs/2008AJ....136.1703G}
}

@article{2019AARv278G, 
  year     = {2019}, 
  title    = {What is a globular cluster? An observational perspective}, 
  author   = {Gratton, Raffaele and Bragaglia, Angela and Carretta, Eugenio and D'Orazi, Valentina and Lucatello, Sara and Sollima, Antonio}, 
  journal  = {A\&ARv}, 
  doi      = {10.1007/s00159-019-0119-3}, 
  pages    = {8 -- 136}, 
  number   = {1}, 
  volume   = {27}, 
  language = {English}
}

@article{cerny2023delve-f41, 
  year    = {2023}, 
  title   = {{DELVE} 6: An Ancient, Ultra-faint Star Cluster on the Outskirts of the Magellanic Clouds}, 
  author  = {Cerny, W. and Drlica-Wagner, A. and Li, T. S. and Pace, A. B. and Olsen, K. A. G. and Noël, N. E. D. and Marel, R. P. van der and Carlin, J. L. and Choi, Y. and Erkal, D. and Geha, M. and James, D. J. and Martínez-Vázquez, C. E. and Massana, P. and Medina, G. E. and Miller, A. E. and Mutlu-Pakdil, B. and Nidever, D. L. and Sakowska, J. D. and Stringfellow, G. S. and Carballo-Bello, J. A. and Ferguson, P. S. and Kuropatkin, N. and Mau, S. and Tollerud, E. J. and Vivas, A. K. and Collaboration, {DELVE}}, 
  journal = {\apj}, 
  issn    = {2041-8205}, 
  doi     = {10.3847/2041-8213/aced84}, 
  pages   = {L21}, 
  number  = {2}, 
  volume  = {953}
}

@ARTICLE{Cerny2021,
       author = {{Cerny}, W. and {Pace}, A.~B. and {Drlica-Wagner}, A. and {Ferguson}, P.~S. and {Mau}, S. and {Adam{\'o}w}, M. and {Carlin}, J.~L. and {Choi}, Y. and {Erkal}, D. and {Johnson}, L.~C. and {Li}, T.~S. and {Mart{\'\i}nez-V{\'a}zquez}, C.~E. and {Mutlu-Pakdil}, B. and {Nidever}, D.~L. and {Olsen}, K.~A.~G. and {Pieres}, A. and {Tollerud}, E.~J. and {Simon}, J.~D. and {Vivas}, A.~K. and {James}, D.~J. and {Kuropatkin}, N. and {Majewski}, S. and {Mart{\'\i}nez-Delgado}, D. and {Massana}, P. and {Miller}, A.~E. and {Neilsen}, E.~H. and {No{\"e}l}, N.~E.~D. and {Riley}, A.~H. and {Sand}, D.~J. and {Santana-Silva}, L. and {Stringfellow}, G.~S. and {Tucker}, D.~L. and {Delve Collaboration}},
        title = "{Discovery of an Ultra-faint Stellar System near the Magellanic Clouds with the DECam Local Volume Exploration Survey}",
      journal = {\apj},
         year = 2021,
        month = mar,
       volume = {910},
       number = {1},
          eid = {18},
        pages = {18},
          doi = {10.3847/1538-4357/abe1af},
archivePrefix = {arXiv},
       eprint = {2009.08550},
 primaryClass = {astro-ph.GA},
       adsurl = {https://ui.adsabs.harvard.edu/abs/2021ApJ...910...18C}
}

@article{Cole2017, 
  year    = {2017}, 
  title   = {{DDO} 216-A1: A Central Globular Cluster in a Low-luminosity Transition-type Galaxy}, 
  author  = {Cole, Andrew A and Weisz, Daniel R and Skillman, Evan D and Leaman, Ryan and Williams, Benjamin F and Dolphin, Andrew E and Johnson, L Clifton and {McConnachie}, Alan W and Boylan-Kolchin, Michael and Dalcanton, Julianne and Governato, Fabio and Madau, Piero and Shen, Sijing and Vogelsberger, Mark}, 
  journal = {\apj}, 
  doi     = {10.3847/1538-4357/aa5df6}, 
  pages   = {54}, 
  number  = {1}, 
  volume  = {837}
}

@ARTICLE{Forbes2018a,
       author = {{Forbes}, Duncan A. and {Bastian}, Nate and {Gieles}, Mark and {Crain}, Robert A. and {Kruijssen}, J.~M. Diederik and {Larsen}, S{\o}ren S. and {Ploeckinger}, Sylvia and {Agertz}, Oscar and {Trenti}, Michele and {Ferguson}, Annette M.~N. and {Pfeffer}, Joel and {Gnedin}, Oleg Y.},
        title = "{Globular cluster formation and evolution in the context of cosmological galaxy assembly: open questions}",
      journal = {Proceedings of the Royal Society of London Series A},
         year = 2018,
        month = feb,
       volume = {474},
       number = {2210},
          eid = {20170616},
        pages = {20170616},
          doi = {10.1098/rspa.2017.0616},
archivePrefix = {arXiv},
       eprint = {1801.05818},
 primaryClass = {astro-ph.GA},
       adsurl = {https://ui.adsabs.harvard.edu/abs/2018RSPSA.47470616F}
}

@ARTICLE{Georgiev2010,
       author = {{Georgiev}, Iskren Y. and {Puzia}, Thomas H. and {Goudfrooij}, Paul and {Hilker}, Michael},
        title = "{Globular cluster systems in nearby dwarf galaxies - III. Formation efficiencies of old globular clusters}",
      journal = {\mnras},
         year = 2010,
        month = aug,
       volume = {406},
       number = {3},
        pages = {1967-1984},
          doi = {10.1111/j.1365-2966.2010.16802.x},
archivePrefix = {arXiv},
       eprint = {1004.2039},
 primaryClass = {astro-ph.CO},
       adsurl = {https://ui.adsabs.harvard.edu/abs/2010MNRAS.406.1967G}
}

@ARTICLE{Usher2023,
       author = {{Usher}, Christopher and {Dage}, Kristen C. and {Girardi}, L{\'e}o and {Barmby}, Pauline and {Bonatto}, Charles J. and {Chies-Santos}, Ana L. and {Clarkson}, William I. and {G{\'o}mez Camus}, Matias and {Hartmann}, Eduardo A. and {Ferguson}, Annette M.~N. and {Pieres}, Adriano and {Prisinzano}, Loredana and {Rhode}, Katherine L. and {Rich}, R. Michael and {Ripepi}, Vincenzo and {Santiago}, Basilio and {Stassun}, Keivan G. and {Street}, R.~A. and {Szab{\'o}}, R{\'o}bert and {Venuti}, Laura and {Zaggia}, Simone and {Canossa}, Marco and {Floriano}, Pedro and {Lopes}, Pedro and {Miranda}, Nicole L. and {Oliveira}, Raphael A.~P. and {Reina-Campos}, Marta and {Roman-Lopes}, A. and {Sobeck}, Jennifer},
        title = "{Rubin Observatory LSST Stars Milky Way and Local Volume Star Clusters Roadmap}",
      journal = {\pasp},
         year = 2023,
        month = jul,
       volume = {135},
       number = {1049},
          eid = {074201},
        pages = {074201},
          doi = {10.1088/1538-3873/ace3f7},
archivePrefix = {arXiv},
       eprint = {2306.17333},
 primaryClass = {astro-ph.GA},
       adsurl = {https://ui.adsabs.harvard.edu/abs/2023PASP..135g4201U}
}

@INPROCEEDINGS{DDP2022,
       author = {{Guy}, Leanne P. and {Cuillandre}, Jean-Charles and {Bachelet}, Etienne and {Banerji}, Manda and {Bauer}, Franz E. and {Collett}, Thomas and {Conselice}, Christopher J. and {Eggl}, Siegfried and {Ferguson}, Annette and {Fontana}, Adriano and {Heymans}, Catherine and {Hook}, Isobel M. and {Aubourg}, {\'E}ric and {Aussel}, Herv{\'e} and {Bosch}, James and {Carry}, Benoit and {Hoekstra}, Henk and {Kuijken}, Konrad and {Lanusse}, Francois and {Melchior}, Peter and {Mohr}, Joseph and {Moresco}, Michele and {Nakajima}, Reiko and {Paltani}, St{\'e}phane and {Troxel}, Michael and {Allevato}, Viola and {Amara}, Adam and {Andreon}, Stefano and {Anguita}, Timo and {Bardelli}, Sandro and {Bechtol}, Keith and {Birrer}, Simon and {Bisigello}, Laura and {Bolzonella}, Micol and {Botticella}, Maria Teresa and {Bouy}, Herv{\'e} and {Brinchmann}, Jarle and {Brough}, Sarah and {Camera}, Stefano and {Cantiello}, Michele and {Cappellaro}, Enrico and {Carlin}, Jeffrey L. and {Castander}, Francisco J. and {Castellano}, Marco and {Chari}, Ranga Ram and {Chisari}, Nora Elisa and {Collins}, Christopher and {Courbin}, Fr{\'e}d{\'e}ric and {Cuby}, Jean-Gabriel and {Cucciati}, Olga and {Daylan}, Tansu and {Diego}, Jose M. and {Duc}, Pierre-Alain and {Fotopoulou}, Sotiria and {Fouchez}, Dominique and {Gavazzi}, Rapha{\"e}l and {Gruen}, Daniel and {Hatfield}, Peter and {Hildebrandt}, Hendrik and {Landt}, Hermine and {Hunt}, Leslie K. and {Ibata}, Rodrigo and {Ilbert}, Olivier and {Jasche}, Jens and {Joachimi}, Benjamin and {Joseph}, R{\'e}my and {Knight}, Matthew and {Kotak}, Rubina and {Laigle}, Clotilde and {Lan{\c{c}}on}, Ariane and {Larsen}, S{\o}ren S. and {Lavaux}, Guilhem and {Leclercq}, Florent and {Leonard}, C. Danielle and {von der Linden}, Anja and {Liu}, Xin and {Longo}, Giuseppe and {Magliocchetti}, Manuela and {Maraston}, Claudia and {Marshall}, Phil and {Mart{\'\i}n}, Eduardo L. and {Mattila}, Seppo and {Maturi}, Matteo and {McCracken}, Henry Joy and {Metcalf}, R. Benton and {Montes}, Mireia and {Mortlock}, Daniel and {Moscardini}, Lauro and {Narayan}, Gautham and {Paolillo}, Maurizio and {Papaderos}, Polychronis and {Pello}, Roser and {Pozzetti}, Lucia and {Radovich}, Mario and {Rejkuba}, Marina and {Rom{\'a}n}, Javier and {S{\'a}nchez-Janssen}, Rub{\'e}n and {Sarpa}, Elena and {Sartoris}, Barbara and {Schrabback}, Tim and {Sluse}, Dominique and {Smartt}, Stephen J. and {Smith}, Graham P. and {Snodgrass}, Colin and {Talia}, Margherita and {Tao}, Charling and {Toft}, Sune and {Tortora}, Crescenzo and {Tutusaus}, Isaac and {Usher}, Christopher and {van Velzen}, Sjoert and {Verma}, Aprajita and {Vernardos}, Georgios and {Voggel}, Karina and {Wandelt}, Benjamin and {Watkins}, Aaron E. and {Weller}, Jochen and {Wright}, Angus H. and {Yoachim}, Peter and {Yoon}, Ilsang and {Zucca}, Elena},
        title = "{Rubin-Euclid Derived Data Products: Initial Recommendations}",
    booktitle = {Zenodo id. 5836022},
         year = 2022,
       volume = {58},
        month = jan,
          eid = {5836022},
        pages = {5836022},
          doi = {10.5281/zenodo.5836022},
archivePrefix = {arXiv},
       eprint = {2201.03862},
 primaryClass = {astro-ph.IM},
       adsurl = {https://ui.adsabs.harvard.edu/abs/2022zndo...5836022G}
}

@ARTICLE{Pace2021,
       author = {{Pace}, Andrew B. and {Walker}, Matthew G. and {Koposov}, Sergey E. and {Caldwell}, Nelson and {Mateo}, Mario and {Olszewski}, Edward W. and {Bailey}, III, John I. and {Wang}, Mei-Yu},
        title = "{Spectroscopic Confirmation of the Sixth Globular Cluster in the Fornax Dwarf Spheroidal Galaxy}",
      journal = {\apj},
         year = 2021,
        month = dec,
       volume = {923},
       number = {1},
          eid = {77},
        pages = {77},
          doi = {10.3847/1538-4357/ac2cd2},
archivePrefix = {arXiv},
       eprint = {2105.00064},
 primaryClass = {astro-ph.GA},
       adsurl = {https://ui.adsabs.harvard.edu/abs/2021ApJ...923...77P}
}

\begin{appendix}
\onecolumn
\section{The effects of stochasticity}
\label{stoch_effects}
Stochastic sampling of the IMF can affect the integrated colours of lower-mass clusters. \citet{Fouesneau2010} showed that using continuous population models to analyse low-mass clusters (around $10^4 \, M{_\odot}$ and below) often leads to age underestimation. This bias also affects mass estimates, as younger models are more luminous per unit mass than older ones. While underestimation was more common, some ages were overestimated, as clusters with an excess of luminous stars above the average predicted by continuous models tend to be very red. Additionally, \citet{Popescu2010} demonstrated that, at these low masses, clusters may appear much redder and older than they would if their IMF was fully sampled, primarily affecting clusters with ages less than ${\sim}10$\,Myr but still evident at all ages in clusters below $10^3\, M{_\odot}$. Offsets to bluer colours exist too, but their amplitudes are less significant.

To explore the contribution of stochasticity to our photometric uncertainties, we divided the circular aperture used for colour determination into halves, slicing it at five different angles. We then calculated the $B-V$ and \YE$-$\HE colours in each half (ten measurements in total for each cluster) and computed the standard deviation of these colour estimates for each cluster as a measure of internal colour variation. If stochastic fluctuations were dominant, we would expect to see significant variation in colour across a given aperture. In IC\,10, the variation in the $B-V$ colours increased significantly at \IE-band magnitudes below ${\sim}20.5$, indicative of stochastic effects, but no correlation was apparent in \YE$-$\HE or in NGC\,6822 at all. Furthermore, the median and interquartile ranges of our measured colour variations closely match the corresponding median photometric uncertainties and interquartile ranges in both NGC\,6822 and IC\,10. Taken together, this suggests that measurement uncertainties unassociated with stochastic sampling of the LF are the dominant source of the observed colour variations. 

We next proceeded to examine the impact of the colour variations measured above on the inferred masses and ages.  We adopt fiducial colour intervals of $B-V=0.7\pm0.1$ and $\YE-\HE=0.0\pm0.1$, where the widths reflect the maximum uncertainties typically seen in our measurements described above. We explore the properties of low-mass clusters ($\leq 10^4 \, M{_\odot}$) computed using stochastic population models with \texttt{pypegase} \citep{Fouesneau2010}, focusing on those that fall within our adopted colour intervals. By extracting the ages of all synthetic clusters within these colour ranges, we find that the age distributions remain largely independent of \IE-band magnitude, despite the stochastic effects. Furthermore, the spread in age estimates is comparable to the typical uncertainties from SED fitting.

To test the robustness of \texttt{BAGPIPES} in producing cluster property uncertainties commensurate with the photometric uncertainties, we drew 100 synthetic photometric realisations from the observed magnitudes and photometric uncertainties of each cluster in IC\,10, and re-fit each realisation with \texttt{BAGPIPES}. We found agreement between the standard deviation of the resulting fits for each cluster and the single-fit posterior uncertainties. Additionally, we investigated whether the discrepancies between our age measurements and those reported in \citetalias{Lim2015} (shown in Fig.~\ref{fig:Lim_comp_SED}) could arise from stochastic effects. To do this, we examined whether the age offsets between our results and those of \citetalias{Lim2015} correlate with the standard deviations derived from the above 100 realisations per cluster. No correlations were apparent. 

While we have not found strong evidence for stochasticity influencing our results, it will nevertheless be present. The tests that we have performed are quite limited and a more detailed exploration, including sampling across the full colour space, examining the effects of varying metallicity and invoking cluster luminosity as an additional age constraint, should be pursued in future work. Given their proximity, NGC\,6822 and IC\,10 provide excellent laboratories for testing and refining stochastic models at low metallicity, since many of the younger clusters can be resolved into individual stars with HST and hence also studied via CMD fitting.  As demonstrated by \citet{Johnson2022}, comparison of CMD-derived parameters with SED-fitted ones provides a valuable assessment of the performance of stochastic modelling of star clusters in the low-mass regime.  

\section{Tables of photometry and properties of star clusters}
\label{results_tables}
In these tables, we list the results of our photometry and SED-fitting analysis of the clusters (classifications 1--4). The tables include IDs, $UBVRI$ photometry, \Euclid \IE\YE\JE\HE photometry, $R_{\mathrm{h}}$ in pc, ages in Gyrs, $\mathrm{log_{10}}(M/{M}_{\odot})$ for both initial mass formed and living stellar mass, metallicities (expressed in terms of $Z/Z_{\odot}$, where $Z_{\odot}=0.02$) and $A_V$ values for each cluster. Total apparent magnitudes for {\it V} band and \YE band are provided alongside colours to convert to the remaining bands. We provide only the first six rows here; full tables are available at the CDS.

\begin{landscape}
\begin{table}
\caption{NGC\,6822 magnitudes, colours and half-light radii. The full table is available at the CDS.}
\label{Table:6822_phot}
\small
\setlength{\tabcolsep}{3.25pt}
\begin{tabular}{c c c c c c c c c c c c c c c c c c c c c}
\hline\hline
  \multicolumn{1}{c}{\tt ID} &
  \multicolumn{1}{c}{\tt Vmag} &
  \multicolumn{1}{c}{\tt e\_Vmag} &
  \multicolumn{1}{c}{\tt B-V} &
  \multicolumn{1}{c}{\tt e\_B-V} &
  \multicolumn{1}{c}{\tt U-B} &
  \multicolumn{1}{c}{\tt e\_U-B} &
  \multicolumn{1}{c}{\tt V-R} &
  \multicolumn{1}{c}{\tt e\_V-R} &
  \multicolumn{1}{c}{\tt R-I} &
  \multicolumn{1}{c}{\tt e\_R-I} &
  \multicolumn{1}{c}{\tt YEmag} &
  \multicolumn{1}{c}{\tt e\_YEmag} &
  \multicolumn{1}{c}{\tt YE-HE} &
  \multicolumn{1}{c}{\tt e\_YE-HE} &
  \multicolumn{1}{c}{\tt JE-HE} &
  \multicolumn{1}{c}{\tt e\_JE-HE} &
  \multicolumn{1}{c}{\tt IE-HE} &
  \multicolumn{1}{c}{\tt e\_IE-HE} &
  \multicolumn{1}{c}{\tt Rh} &
  \multicolumn{1}{c}{\tt e\_Rh} \\
\hline
  ESCC-NGC6822-01 & 16.09 & 0.02 & 0.42 & 0.04 & -0.16 & 0.07 & 0.36 & 0.03 & 0.45 & 0.04 & 15.95 & 0.03 & -0.04 & 0.05 & -0.06 & 0.06 & 0.07 & 0.05 & 3.09 & 0.14\\
  ESCC-NGC6822-02 & 15.57 & 0.02 & 0.87 & 0.01 & 0.1 & 0.01 & 0.61 & 0.01 & 0.63 & 0.01 & 14.57 & 0.01 & 0.1 & 0.01 & 0.0 & 0.01 & 0.66 & 0.01 & 2.27 & 0.37\\
  ESCC-NGC6822-03 & 17.43 & 0.02 & 0.72 & 0.01 & 0.25 & 0.02 & 0.49 & 0.01 & 0.52 & 0.01 & 16.34 & 0.07 & -0.06 & 0.03 & -0.05 & 0.03 & 0.4 & 0.03 & 5.69 & 0.77\\
  ESCC-NGC6822-04 & 17.91 & 0.03 & 0.63 & 0.01 & 0.23 & 0.01 & 0.44 & 0.01 & 0.47 & 0.02 & 17.01 & 0.14 & -0.23 & 0.08 & -0.14 & 0.08 & 0.07 & 0.06 & 3.09 & 0.54\\
  ESCC-NGC6822-05 & 18.06 & 0.02 & 0.89 & 0.01 & 0.27 & 0.02 & 0.57 & 0.01 & 0.55 & 0.01 & 16.91 & 0.03 & 0.0 & 0.01 & -0.03 & 0.01 & 0.54 & 0.01 & 7.15 & 0.17\\
  ESCC-NGC6822-06 & 20.48 & 0.06 & 0.99 & 0.09 & 0.37 & 0.19 & 0.61 & 0.07 & 0.47 & 0.07 & 18.66 & 0.02 & 0.02 & 0.03 & 0.0 & 0.03 & 0.65 & 0.02 & 9.49 & 0.37\\

\hline\end{tabular}
\newline
\vspace*{0.25 cm}
\newline
\caption{IC\,10 magnitudes, colours and half-light radii. The full table is available at the CDS.}
\label{Table:ic10_phot}
\begin{tabular}{c c c c c c c c c c c c c c c c c c c c c}
\hline\hline
  \multicolumn{1}{c}{\tt ID} &
  \multicolumn{1}{c}{\tt Vmag} &
  \multicolumn{1}{c}{\tt e\_Vmag} &
  \multicolumn{1}{c}{\tt B-V} &
  \multicolumn{1}{c}{\tt e\_B-V} &
  \multicolumn{1}{c}{\tt U-B} &
  \multicolumn{1}{c}{\tt e\_U-B} &
  \multicolumn{1}{c}{\tt V-R} &
  \multicolumn{1}{c}{\tt e\_V-R} &
  \multicolumn{1}{c}{\tt R-I} &
  \multicolumn{1}{c}{\tt e\_R-I} &
  \multicolumn{1}{c}{\tt YEmag} &
  \multicolumn{1}{c}{\tt e\_YEmag} &
  \multicolumn{1}{c}{\tt YE-HE} &
  \multicolumn{1}{c}{\tt e\_YE-HE} &
  \multicolumn{1}{c}{\tt JE-HE} &
  \multicolumn{1}{c}{\tt e\_JE-HE} &
  \multicolumn{1}{c}{\tt IE-HE} &
  \multicolumn{1}{c}{\tt e\_IE-HE} &
  \multicolumn{1}{c}{\tt Rh} &
  \multicolumn{1}{c}{\tt e\_Rh} \\
\hline
  ESCC-IC10-01 & 19.0 & 0.05 & 1.27 & 0.04 & 0.48 & 0.05 & 0.86 & 0.03 & 0.88 & 0.03 & 17.0 & 0.03 & 0.31 & 0.01 & 0.12 & 0.01 & 1.35 & 0.01 & 2.65 & 0.07\\
  ESCC-IC10-02* & 21.28 & 0.08 & 0.94 & 0.12 & 0.22 & 0.11 & 0.61 & 0.12 & 0.76 & 0.12 & 20.76 & 0.33 & -0.38 & 0.86 & -0.32 & 0.92 & 0.21 & 0.8 & 1.89 & 0.3\\
  TG3 & 21.69 & 0.07 & 1.02 & 0.1 & -0.1 & 0.1 & 0.76 & 0.09 & 1.12 & 0.09 & 19.29 & 0.04 & 0.8 & 0.05 & 0.42 & 0.05 & 2.2 & 0.04 & 1.05 & 0.31\\
  ESCC-IC10-03 & 19.87 & 0.07 & 1.16 & 0.05 & 0.61 & 0.07 & 0.85 & 0.04 & 0.9 & 0.04 & 18.12 & 0.13 & 0.45 & 0.05 & 0.25 & 0.05 & 1.53 & 0.04 & 2.58 & 0.11\\
  ESCC-IC10-04 & 20.42 & 0.09 & 1.01 & 0.06 & 0.19 & 0.07 & 0.67 & 0.05 & 0.8 & 0.06 & 19.86 & 0.45 & -0.02 & 0.27 & -0.02 & 0.3 & 0.74 & 0.25 & 2.13 & 0.12\\
  TG6 & 20.46 & 0.04 & 0.94 & 0.07 & 0.25 & 0.08 & 0.82 & 0.06 & 0.87 & 0.06 & 18.91 & 0.42 & 0.77 & 0.13 & 0.4 & 0.13 & 1.92 & 0.09 & 1.15 & 0.16\\

\hline\end{tabular}
\end{table}
\begin{table}

\caption{NGC\,6822 SED-fitting results. The full table is available at the CDS.}
\label{Table:6822_SED}
\begin{tabular}{c c c c c c c c c c c}
\hline
  \multicolumn{1}{c}{\tt ID} &
  \multicolumn{1}{c}{\tt Age} &
  \multicolumn{1}{c}{\tt e\_Age} &
  \multicolumn{1}{c}{\tt Mass} &
  \multicolumn{1}{c}{\tt e\_Mass} &
  \multicolumn{1}{c}{\tt Living Mass} &
  \multicolumn{1}{c}{\tt e\_Living Mass} &
  \multicolumn{1}{c}{\tt Z/Z\_0} &
  \multicolumn{1}{c}{\tt e\_Z/Z\_0} &
  \multicolumn{1}{c}{\tt AV} &
  \multicolumn{1}{c}{\tt e\_AV} \\
\hline\hline
  ESCC-NGC6822-01 & 0.1 & 0.05 & 4.44 & 0.03 & 4.30 & 0.02 & 0.03 & 0.01 & 0.73 & 0.09\\
  ESCC-NGC6822-02 & 6.21 & 0.82 & 5.81 & 0.04 & 5.45 & 0.03 & 0.04 & 0.01 & 0.78 & 0.03\\
  ESCC-NGC6822-03 & 0.7 & 0.18 & 4.46 & 0.03 & 4.24 & 0.02 &  0.24 & 0.05 & 1.02 & 0.15\\
  ESCC-NGC6822-04 & 0.55 & 0.11 & 4.15 & 0.05 & 
  3.95 & 0.04 &
  0.17 & 0.09 & 0.96 & 0.18\\
  ESCC-NGC6822-05 & 2.49 & 0.31 & 4.52 & 0.03 & 4.23 & 0.04 & 0.19 & 0.02 & 0.67 & 0.06\\
  ESCC-NGC6822-07 & 8.17 & 1.29 & 5.73 & 0.04 & 5.37 & 0.03 & 0.03 & 0.01 & 0.64 & 0.02\\
\hline\end{tabular}
\newline
\vspace*{0.25 cm}
\newline
\caption{IC\,10 SED-fitting results. The full table is available at the CDS.}
\label{Table:ic10_SED}
\begin{tabular}{c c c c c c c c c c c}
\hline
  \multicolumn{1}{c}{\tt ID} &
  \multicolumn{1}{c}{\tt Age} &
  \multicolumn{1}{c}{\tt e\_Age} &
  \multicolumn{1}{c}{\tt Mass} &
  \multicolumn{1}{c}{\tt e\_Mass} &
  \multicolumn{1}{c}{\tt Living Mass} &
  \multicolumn{1}{c}{\tt e\_Living Mass} &
  \multicolumn{1}{c}{\tt Z/Z\_0} &
  \multicolumn{1}{c}{\tt e\_Z/Z\_0} &
  \multicolumn{1}{c}{\tt AV} &
  \multicolumn{1}{c}{\tt e\_AV} \\
\hline\hline
  ESCC-IC10-01 & 7.5 & 4.54 & 5.36 & 0.24 & 4.85 & 0.22 & 0.07 & 0.02 & 2.12 & 0.14\\
  ESCC-IC10-02 & 0.07 & 0.01 & 3.23 & 0.04 & 3.05 & 0.02 & 0.06 & 0.01 & 3.61 & 0.08\\
  TG3 & 0.13 & 0.03 & 3.64 & 0.04 & 3.39 & 0.02 & 0.36 & 0.12 & 2.53 & 0.06\\
  ESCC-IC10-03 & 0.63 & 0.25 & 4.41 & 0.1 & 4.08 & 0.04 & 0.1 & 0.1 & 2.65 & 0.15\\
  ESCC-IC10-04 & 0.08 & 0.02 & 3.58 & 0.05 & 3.42 & 0.02 & 0.06 & 0.01 & 2.51 & 0.05\\
  TG6 & 0.15 & 0.07 & 3.81 & 0.07 & 3.58 & 0.04 & 0.26 & 0.16 & 2.72 & 0.16\\
\hline\end{tabular}

\end{table}
\end{landscape}

\section{Cutouts of clusters}
\label{appendix_thumbnails}
Here we present cutouts of all objects in the final catalogues (class 1--6), including newly-identified clusters, previously-known candidates and previously-identified candidates that do not appear as clusters in the \Euclid \IE images.

\begin{figure}[!htbp]
\includegraphics[width=\columnwidth]{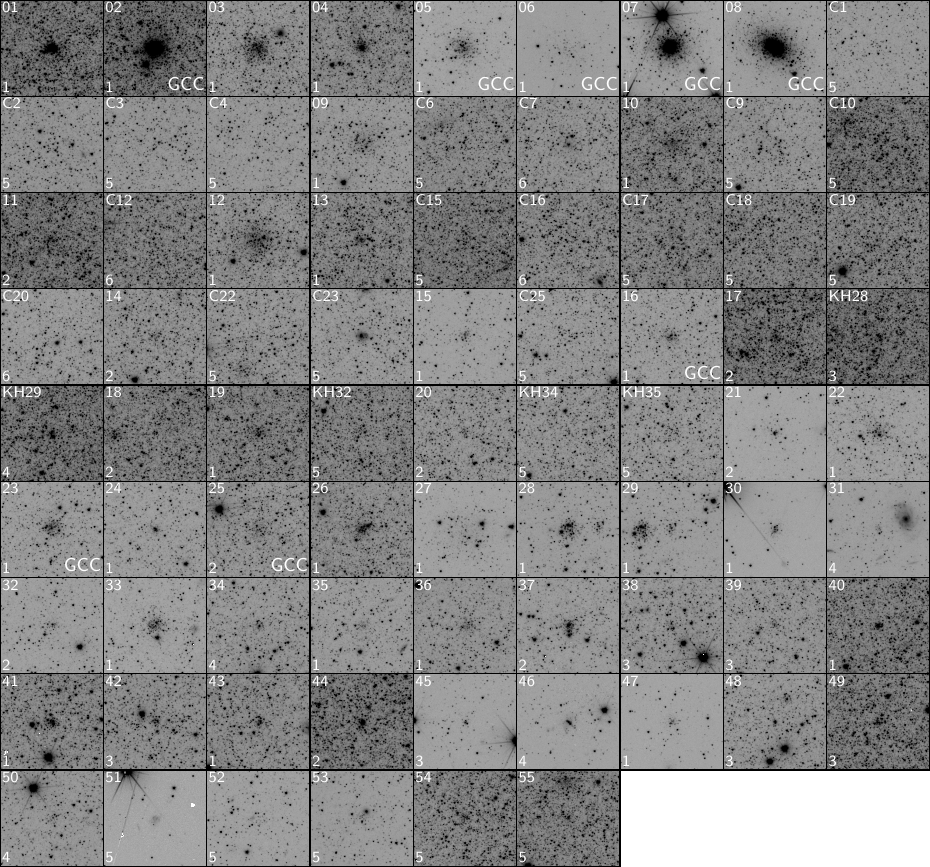}
\caption{Thumbnails of all clusters and candidate clusters in NGC\,6822 shown in asinh scaling. New and recovered candidates display their \Euclid ID in the top left corner and candidates that were not recovered in this work display their previous literature IDs. New clusters from this study start from \Euclid ID 21 onwards. Classifications are indicated in the bottom left of each subplot. Clusters satisfying our definition of GC are indicated by a {\lq{GCC}\rq} label in the bottom right of each subplot. Thumbnails are $28\farcs0$ on the side, corresponding to $\sim70$~pc at the distance of NGC\,6822.}
\label{fig:6822_compilation}
\end{figure}

\begin{figure}[!htbp]
\centering
\includegraphics[width=0.98\columnwidth]{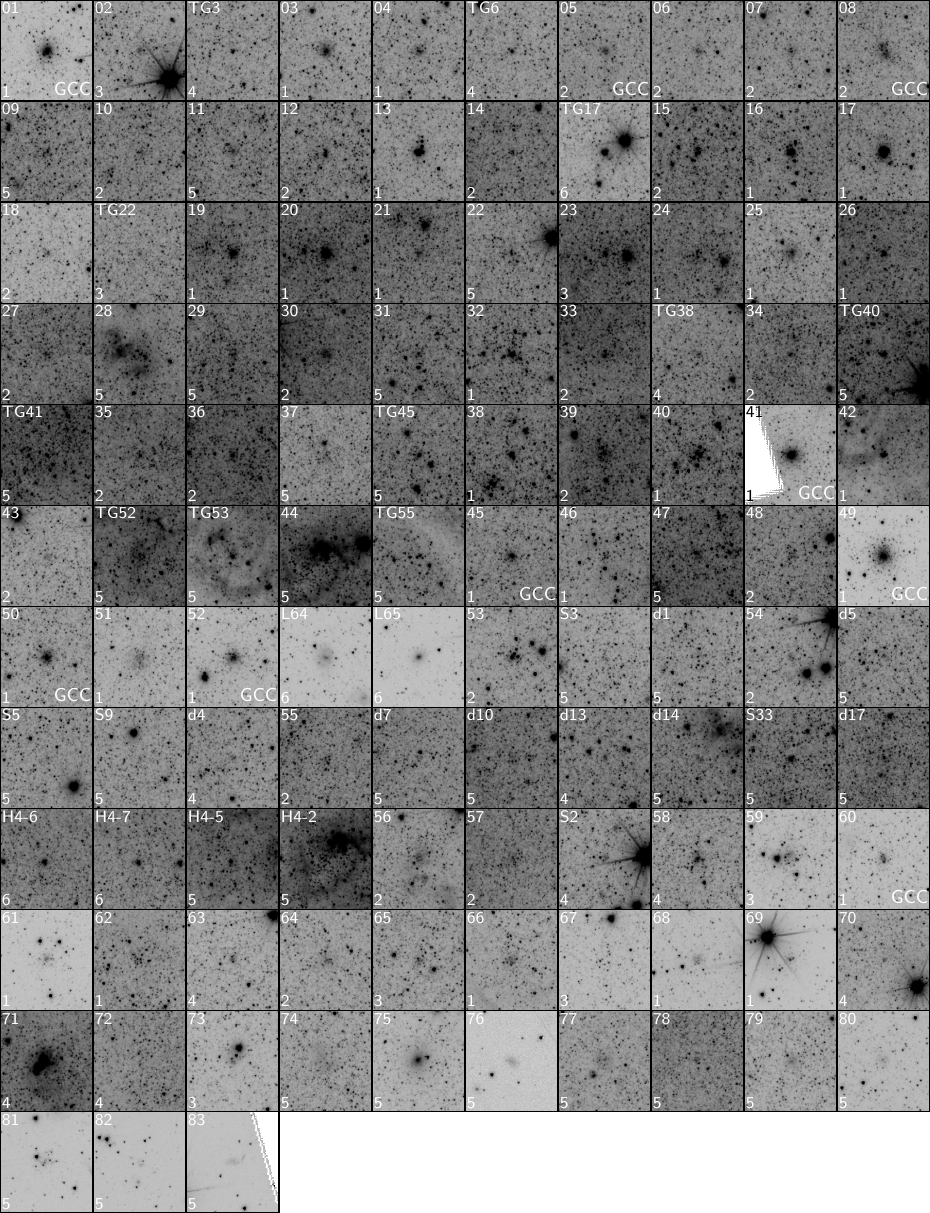}
\caption{Same as Fig.\,\ref{fig:6822_compilation} but for IC\,10. New clusters from this study start from \Euclid ID 58 onwards. Thumbnails are $18\farcs0$ on the side (corresponding to approximately the same physical size as the NGC\,6822 thumbnails).}
\label{fig:ic10_compilation}
\end{figure}

\end{appendix}

\end{document}